\documentclass{article}
\usepackage{arxiv}
\usepackage[utf8]{inputenc}
\usepackage[T1]{fontenc}
\usepackage[english]{babel}

\usepackage[numbers]{natbib}
\usepackage{doi}
\usepackage{url}
\usepackage{hyperref}

\usepackage{amsmath,amssymb,amsfonts}
\usepackage{amsthm}
\usepackage{mathrsfs}
\usepackage{nicefrac}

\usepackage{graphicx}
\usepackage{subcaption}
\usepackage{booktabs}
\usepackage{tabularx}
\usepackage{longtable}
\usepackage{array}
\usepackage{multirow}
\usepackage{ragged2e}
\usepackage{adjustbox}
\usepackage{pdflscape}
\usepackage{rotating}

\usepackage{algorithm}
\usepackage{algpseudocode}
\usepackage{listings}

\usepackage{xcolor}
\usepackage{textcomp}
\usepackage{microtype}
\usepackage{manyfoot}
\usepackage{float}
\usepackage[section]{placeins}

\usepackage{geometry}
\usepackage{titlesec}
\usepackage{etoolbox}
\usepackage{nameref}
\usepackage{lipsum}

\newcolumntype{L}{>{\raggedright\arraybackslash}l}

\title{Towards Explainability with a Multi-Branch Gated and Cross-Attention Hybrid Rating Predictor Recommender System}

\author{
	1$^{st}$ Sina Heydari \\
	Department of Computer Science and Information Technology \\ 
	Institude for Advanced Studies in Basic Sciences (IASBS) \\
	Zanjan, Iran, 45137-66731\\
	\texttt{sinaheydari@iasbs.ac.ir} \\
	\And
	2$^{nd}$ Amirreza Abbasi\\
	Department of Computer Science and Information Technology \\ 
	Institude for Advanced Studies in Basic Sciences (IASBS) \\
	Zanjan, Iran, 45137-66731\\
	\texttt{a.abbasi@iasbs.ac.ir} \\
	\And
	3$^{rd}$ Mohsen Hooshmand\\
	Department of Computer Science and Information Technology \\ 
	Institude for Advanced Studies in Basic Sciences (IASBS) \\
	Zanjan, Iran, 45137-66731\\
	\texttt{mohsen.hooshmand@iasbs.ac.ir} \\
	\And
	4$^{th}$ Amirhossein Dadashzadeh Taromi\\
	Department of Computer Science and Information Technology \\ 
	Institude for Advanced Studies in Basic Sciences (IASBS) \\
	Zanjan, Iran, 45137-66731\\
	\texttt{dadashzadeh@iasbs.ac.ir} \\
}

\begin{document}
	\maketitle
	\begin{abstract}
		While recommender systems that combine collaborative embeddings with side features have achieved strong results on rating prediction, there remains a gap in improving prediction accuracy and robustness to sparse rating matrices, along with a need to explain how much each source contributes to a given prediction. To address these challenges, we propose a two-stage approach evaluated on MovieLens 100K and 1M. In the first stage, a Collaborative Attention Autoencoder (CAAE) learns user and movie embeddings. In the second stage, a predictor builds four interaction branches from these embeddings and the side features, and we compare two branch-aggregation mechanisms, an independent softmax gate (AFIN) and self-attention across branches before pooling (AFINca), each evaluated with frozen and fine-tuned encoders. Fine-tuned AFIN achieves the lowest RMSE, 0.9073 on ML-100K and 0.8497 on ML-1M, and cross-branch attention yields no accuracy gain. Branch-zeroing ablations show that AFIN's weights do not track branch importance, since the collaborative branch receives a middling weight yet zeroing it raises RMSE by about 17\%. In AFINca, the weights align better with the ablation effects, but all effects are small and spread across branches. Branch weights should therefore be validated by intervention before being interpreted as importance. The source code is publicly available at: \url{https://github.com/ComputationIASBS/RecSys}.
	\end{abstract}


\section{Introduction}
\label{sec:intro}
Recommender systems facilitate the identification of relevant items across diverse domains. In entertainment, these systems recommend movies, music, and other media based on users’ preferences and interaction histories \cite{roy2022systematic}. In healthcare, recommender systems have been investigated for suggesting drugs, food, healthcare services, and medical professionals, thereby supporting both patients and clinicians \cite{tran2021healthcare}. Within movie recommendation, a prevalent approach is rating prediction, which involves estimating how a user would rate a movie they have not yet seen \cite{chen2019attribute}. This estimate quantifies user-item compatibility and forms the basis for ranking and filtering candidate movies \cite{koren2009matrix}. However, the task is challenging due to the sparsity and unevenness of user-movie rating matrices. Most users rate only a small fraction of the catalogue, and the number of ratings varies substantially across users and movies \cite{koren2009matrix}. Consequently, models must learn preference representations from incomplete histories while accounting for differences in user behavior, movie characteristics, and the amount of available evidence \cite{he2017neuralcol}.

Collaborative filtering (CF) addresses this challenge by leveraging patterns in observed interactions. Users with similar rating histories are likely to share preferences, and movies rated similarly may appeal to comparable audiences \cite{sarwar2001item}. Matrix factorization models these patterns using latent user and item vectors \cite{koren2009matrix, hafeez2026}. Neural collaborative filtering extends this approach by replacing the inner product with a learned nonlinear function, enabling the modeling of interactions that linear models cannot capture \cite{he2017neuralcol}.

Interaction histories are often insufficient for users and movies with limited ratings. Side information, including user demographics and movie attributes, provides additional context beyond the rating matrix, and hybrid models integrate this information with collaborative signals \cite{cheng2016wide, guo2017deepfm, taromi2025}. Collaborative representations summarize the information inferred from interaction data, while side features explicitly encode known attributes of users and movies. The primary challenge lies not in incorporating both sources, but in determining how their contributions should be represented and combined for each prediction \cite{dong2017hybrid}.

Representation learning provides an alternative approach to addressing sparsity. Rather than processing high-dimensional, incomplete rating vectors directly, models can compress each user’s ratings or each movie’s received ratings into dense vectors. Autoencoders achieve this by reconstructing observed ratings, as demonstrated by AutoRec, which utilized this principle to predict missing entries \cite{sedhain2015autorec, heydari2026}. These dense vectors serve as compact inputs for downstream predictors, although their effectiveness depends on the extent to which the encoder preserves collaborative information.

When multiple representations are available, their aggregation constitutes a key design decision. Independent gating assigns each representation an input-dependent weight, while attention-based contextualization enables representations to exchange information prior to pooling \cite{vaswani2017attention}. These approaches embody distinct modeling assumptions, and their relative effectiveness depends on the specific task and the structure of the representations \cite{he2023aain}.

The method of aggregation also influences the interpretability of model predictions. Explainable recommender systems aim to clarify the rationale behind predictions and facilitate model inspection by designers \cite{gao2019explainable}. Explanations may arise from inherently interpretable models or from post-hoc analyses of complex architectures \cite{zhang2020explainable}. In hybrid neural models, aggregation weights are often interpreted post-hoc, with branches assigned larger weights presumed to be more influential. This interpretation is prevalent in recommendation research. For example, attentional factorization machines present attention weights as interpretable \cite{xiao2017attentional}, and gated multi-expert models utilize gate weights to analyze task relationships \cite{ma2018modeling}. However, in language processing, the validity of this interpretation has been questioned. Attention weights may correlate weakly with other importance measures, and different weight configurations can yield identical predictions. The explanatory value of weights depends on how explanation is defined and empirically tested \cite{jain2019attention, wiegreffe2019attention, serrano2019attention}. In the context of recommender branch weights, such empirical validation through intervention is rare. This gap motivates the present study.

This study investigates the outlined questions within a movie rating prediction pipeline using the MovieLens 100K and 1M datasets \cite{harper2015movielens}. Both datasets associate ratings with user and movie attributes, enabling the simultaneous examination of collaborative interactions and side features. The pipeline consists of two stages. First, a Collaborative Attention Autoencoder (CAAE) learns dense user and movie embeddings from observed rating patterns. Second, these embeddings and the side features are combined into four \emph{interaction branches}, each representing a distinct perspective on user-movie compatibility. The collaborative branch pairs the user embedding with the movie embedding, while the feature-affinity branch pairs user attributes with movie attributes. The two cross branches pair a user embedding with movie attributes and a movie embedding with user attributes, thereby linking learned evidence on one side to explicit attributes on the other. These combinations are consistent with established feature-based and hybrid designs \cite{chen2019attribute, rendle2010factorizing}. The presence of four branches directly raises the aggregation question: how should their contributions be weighted, and should they interact prior to combination?

Two aggregation mechanisms are compared. The Attentive Factorization Interaction Network (AFIN) employs an input-conditioned softmax gate, which functions as attention over branches and assigns independent weights to each branch. Because the gate is conditioned on the current user-movie pair, the weights vary across predictions rather than assigning a fixed global importance to each branch. AFINca (AFIN with cross-branch attention) first applies self-attention across the branch representations, allowing each branch to be influenced by the others, and then pools them using a gate. As a result, the branches entering the pooling stage are no longer independent, which has implications for both predictive accuracy and the interpretation of pooling weights. Both architectures are evaluated with frozen pretrained embeddings and with encoder fine-tuning, where the rating loss also updates the encoder. This experimental design isolates the effect of the aggregation mechanism from the effect of adapting representations to the downstream task, and enables both predictors to utilize the same embeddings.

In addition to predictive accuracy, the internal behavior of the two aggregation mechanisms is analyzed. The study reports which branch dominates each prediction, the mean weight assigned to each branch, the variation of weights with user activity, and the change in prediction error when a branch is zeroed during inference. These analyses address two distinct questions: which branches receive large weights, and which branches the model relies on for prediction. These aspects may not coincide; a branch may receive a small weight yet contain essential information, while a heavily weighted branch may be removed with minimal impact. Thus, comparing weights with zeroing effects provides complementary insights into each branch’s role and offers a more nuanced basis for explanation. This comparison is particularly relevant for AFINca, where the attention step may transfer information between branches prior to pooling.

This work addresses the following research questions:

\begin{enumerate}
	\item[\textbf{RQ1}] \textbf{Do branch weights identify the branches a model depends on?} We test whether the branch weighted most heavily is also the branch whose removal degrades prediction most. For AFIN, we compare branch dominance and mean weights with zeroing results.
	\item[\textbf{RQ2}] \textbf{Does the mismatch between branch weights and branch dependence carry over to AFINca?} We apply the same dominance, weight, and zeroing analyses to AFINca, whose attention step mixes branch information before pooling.
	\item[\textbf{RQ3}] \textbf{Does self-attention across branches improve rating prediction over independent gating?} We compare AFIN and AFINca by test RMSE over 11 random seeds, and by MAE, on MovieLens 1M and 100K.
	\item[\textbf{RQ4}] \textbf{Does fine-tuning the pretrained encoder help?} We compare frozen and fine-tuned variants of both architectures, in accuracy and in which branch each model weights most.
\end{enumerate}

The primary contributions of this work are as follows. We introduce a two-stage recommender system that improves rating prediction by incorporating side information and evaluate it on the MovieLens 100K and 1M datasets.

I) Representation learning. The Collaborative Attention Autoencoder (CAAE) leverages existing user and movie ratings, along with side features, to learn four distinct types of embeddings.

II) Rating prediction. Two predictors, AFIN (gating) and AFINca (cross-branch attention), integrate the four embeddings to decompose each user-movie pair into four interaction branches, explicitly quantifying each branch's weight.

Within this framework, we conduct four analyses. First, we assess each branch's influence by comparing its learned weight with the increase in error when it is removed. In AFIN, these two measures diverge: the most heavily weighted branches have minimal impact when removed, while the collaborative branch, though not the most heavily weighted, accounts for the largest increase in error. In contrast, AFINca demonstrates closer alignment between weights and removal effects. However, because attention aggregates branch information before pooling, this agreement does not necessarily mean the weights are more representative. Second, we evaluate the two combination mechanisms, gating and cross-branch attention, on MovieLens 100K and 1M across 11 random seeds. AFIN outperforms AFINca with frozen encoders; after fine-tuning, their performance is comparable, and AFINca shows no accuracy improvements beyond seed variation. Third, we examine encoder fine-tuning, which consistently reduces mean RMSE by small margins and alters the branch receiving the largest weight. Finally, we compare our results with state-of-the-art methods. These findings reflect design choices within a single evaluated setting and do not establish the general superiority of either aggregation mechanism. Furthermore, as both datasets require at least 20 ratings per user, the experiments do not address unseen users.

The remainder of this paper is organized as follows. Section~\ref{sec:prelim} presents the preliminaries, and Section~\ref{sec:rel} reviews related work. Section~\ref{sec:method} details the datasets, the representation-learning stage, the two predictors, and the explainability analyses. Section~\ref{sec:res} reports the results, Section~\ref{sec:disc} addresses the research questions, and Section~\ref{sec:concl} concludes with limitations and directions for future work.

\section{Preliminaries}
\label{sec:prelim}

This section establishes the notation and concepts employed in the framework, including rating prediction and collaborative filtering, hybrid models with side information, autoencoders, and attention and gating mechanisms.

\subsection{Rating Prediction and Collaborative Filtering}
\label{subsec:rpcf}

Rating prediction is a supervised regression task: given observed ratings, predict the rating of a user-movie pair absent from the training data. Let $U$ and $M$ denote the sets of users and movies:
\begin{equation}
	U = \{u_1, \dots, u_{|U|}\}, \qquad
	M = \{m_1, \dots, m_{|M|}\}.
	\label{eq:umsets}
\end{equation}
Let $\Omega \subseteq U \times M$ be the set of observed user-movie pairs and $\Omega^{c}$ its complement. The observed ratings form a partially observed matrix, where $r_{u_i,m_j}$ is the rating that user $u_i$ gave movie $m_j$:
\begin{equation}
	R \in (\{1,\dots,5\} \cup \{\varnothing\})^{|U| \times |M|}, \qquad
	R_{ij} =
	\begin{cases}
		r_{u_i,m_j}, & (u_i,m_j) \in \Omega, \\
		\varnothing, & (u_i,m_j) \in \Omega^{c}.
	\end{cases}
	\label{eq:obrm}
\end{equation}
Most users rate only a small fraction of the available movies, resulting in a sparse matrix $R$, where unobserved entries correspond to missing ratings rather than zero-valued preferences \cite{koren2009matrix}.

Collaborative filtering exploits patterns in $R$ to estimate the missing entries \cite{sarwar2001item}. Latent-factor models represent users and items by low-dimensional vectors whose compatibility determines the predicted rating, and neural CF learns a nonlinear interaction function in place of the inner product \cite{he2017neuralcol}. Both capture collaborative signals, but their effectiveness depends on how many interactions are observed and how they are distributed.

\subsection{Hybrid Recommendation and Side Information}
\label{subsec:sideinfos}
Interaction histories alone do not fully characterize users and items. Side information encompasses additional attributes, such as demographics, item categories, or other metadata. Hybrid models integrate side information with collaborative signals to produce richer representations \cite{guo2017deepfm, cheng2016wide}.

Let $\mathbf{e}_{u_i}, \mathbf{e}_{m_j} \in \mathbb{R}^{d}$ denote learned representations of user $u_i$ and movie $m_j$, and let $\mathbf{f}_{u_i}$ and $\mathbf{f}_{m_j}$ denote their side-feature vectors. In this work, rating prediction is the problem of learning a parameterized function $F(\cdot;\theta)$ that estimates $\hat{r}_{u_i,m_j}$ from these four inputs, for pairs that are observed or held out. Hybrid architectures differ in how $F$ combines the learned representations with the explicit attributes. Section~\ref{subsec:caae} specifies how $\mathbf{e}$ is obtained, and Section~\ref{subsec:probform} gives the exact form of $F$.

\subsection{Autoencoders}
\label{AE and Attention}
An autoencoder learns a compact representation through an encoding-decoding process. The encoder maps an input vector $\mathbf{v}$ to a latent representation $\mathbf{l}$, and the decoder reconstructs the input from it:
\begin{equation}
	\mathbf{l} = g_\phi(\mathbf{v}), \qquad
	\hat{\mathbf{v}} = h_\psi(\mathbf{l}).
	\label{eq:encdec}
\end{equation}
In this context, $g_\phi$ and $h_\psi$ denote the encoder and decoder, respectively, which are trained by minimizing a reconstruction loss. In collaborative filtering, $\mathbf{v}$ represents a user’s rating row or a movie’s rating column, and $\mathbf{l}$ is the dense embedding utilized in subsequent prediction tasks.

\subsection{Attention and Gating}
\label{subsec:attgate}
Attention computes input-dependent relationships among representations. Instead of combining them with fixed coefficients, it derives weights from learned compatibility scores. Scaled dot-product attention \cite{vaswani2017attention} combines queries $Q$, keys $K$, and values $V$ as
\begin{equation}
	\operatorname{Attention}(Q, K, V) = \operatorname{softmax}\left(\frac{QK^\top}{\sqrt{d_k}}\right)V,
	\label{eq:attention}
\end{equation}
where $d_k$ is the key dimensionality and the softmax turns compatibility scores into weights over the value vectors. Self-attention draws $Q$, $K$, and $V$ from one set of representations, whereas cross-attention draws $Q$ from one set and $K$ and $V$ from another. A softmax gate constitutes a simpler weighting mechanism. It maps the input to a normalized weight vector and computes the weighted sum of a fixed set of component vectors, such that each component is weighted independently and no information is exchanged among components \cite{jacobs1991adaptive, shazeer2017outrageously}. Both attention and gating produce normalized, input-dependent weights that are often interpreted as indicators of importance. However, in line with the attention-as-explanation literature \cite{jain2019attention, wiegreffe2019attention, serrano2019attention}, this interpretation is treated as a hypothesis rather than an established fact. A weight reflects how the model allocates pooling, whereas a component’s true importance is assessed by the change in predictions when it is modified. Section~\ref{subsec:intpret} defines the intervention used in this study.

\section{Related Work}
\label{sec:rel}
This section reviews collaborative filtering and rating prediction, neighborhood methods, autoencoder-based representation learning, hybrid and attention-based feature interaction, and the ongoing debate regarding the explanatory power of attention weights. The section concludes by identifying the research gap addressed in this study.

\subsection{Recent Developments in Collaborative Filtering and Rating Prediction}
Collaborative filtering remains a foundational approach, using historical interactions to infer preferences for unobserved items. Recent studies investigate how neural models can better represent user preferences, model feature interactions, and cope with noisy or sparse ratings. The survey of \citet{gheewala2025depth} covers deep learning for both rating prediction and top-$N$ recommendation and stresses that the two tasks have different objectives and evaluation protocols. This matters when comparing studies, because better ranking quality does not imply lower rating error.

One line of work integrates ratings with item or user side information in a single representation-learning framework. \citet{nahta2025deep} proposed a deep collaborative neural generative embedding model for movie-rating prediction that combines ratings with item metadata. It is relevant to hybrid rating prediction because it shows how auxiliary information can complement interaction data, but its probabilistic embeddings and reliability mechanisms address questions different from ours.

Model-based approaches also remain strong on explicit ratings. \citet{hafeez2026} compared user-based filtering with NMF and SVD and found SVD best. \citet{wang2026} introduced BayesNCF-v2, which combines user and item embedding tables with an MLP feature extractor, a Bayesian linear layer for the mean rating, and a second output for the log-variance, so that it can quantify predictive uncertainty. Both methods use interaction data only, with no side features and no branch aggregation.

This study addresses a distinct question: how multiple learned representations should be combined for explicit rating prediction. No new collaborative objective is proposed. Instead, mechanisms for aggregating branches that encode collaborative signals and user-movie attributes are compared, and the extent to which the resulting weights reflect dependence is evaluated.

\subsection{Neighborhood-Based Collaborative Filtering}
Neighborhood methods predict ratings from the ratings of similar users or items, so accuracy depends on how similarity is measured and neighbors are selected \cite{sarwar2001item}. Recent work usually modifies one of these steps. \citet{du2021} label ratings as like, neutral, or dislike to build a concordance ratio and apply empirical Bayes smoothing to penalize user pairs with few co-rated items. \citet{belmessous2024} modeled each user pair with binomial subjective logic, which adds an explicit uncertainty term to similarity, and rank neighbors with TOPSIS. \citet{nguyen2023} let both the distance metric and the number of neighbors adapt to local data density.

Other approaches added context. \citet{jain2023} applied exponential or power-law decay to older interactions so that recent ratings weigh more. \citet{cheng2023} randomized only the ratings users mark as sensitive and estimate item-item similarity from the perturbed data. \citet{aljunid2021} combined user-based and item-based predictions by linear regression. \citet{alam2025} fed restricted Boltzmann machine representations into kNN, which linked neighborhood methods to the representation learning discussed next.

None of these methods use side features or learn how to aggregate across branches. They served as accuracy benchmarks, not as direct comparators for the aggregation mechanisms studied here.

\subsection{Autoencoder-Based Representation Learning for Recommendation}
Autoencoders encode an input into a lower-dimensional representation and reconstruct the input from it, which suits the partially observed interaction vectors of recommendation. \citet{sedhain2015autorec} proposed AutoRec, an autoencoder for collaborative filtering, and \citet{wu2016collaborative} and \citet{liang2018variational} extended the idea with denoising and variational formulations. These works are the methodological background for learning representations by reconstruction. Surveys continue to list autoencoders and other generative architectures as central to deep recommendation, and \citet{gheewala2025depth} distinguish them from neural interaction models by learning objective and evaluation setting. \citet{nahta2025deep} combined a generative embedding with rating side information for movie-rating prediction. These approaches share the aim of learning representations that integrate collaborative evidence with auxiliary information, although their latent-variable assumptions and prediction mechanisms differ.

Reconstruction quality alone does not determine how well the final predictor performs. The representation must retain information relevant to the downstream task, and the architecture that combines it with other signals also shapes the outcome. Whether to freeze a pretrained encoder or fine-tune it jointly with the predictor is therefore an experimental variable in its own right.

In this study, the CAAE provides user and movie representations to a multi-branch predictor. The primary focus is not on the overall effectiveness of autoencoders as recommenders, but rather on how their representations behave under different aggregation mechanisms. Comparing frozen and fine-tuned configurations reveals whether adapting the encoder to the rating objective improves prediction and alters the distribution of weights.

\subsection{Hybrid Recommendation and Attention-Based Feature Interaction}
Hybrid recommenders combine collaborative signals with auxiliary information such as demographics and item attributes, aiming to capture facets of compatibility that interaction histories miss. Recent works study how to model interactions among these features and how to adapt their contributions to the input \cite{dong2017hybrid, he2017neuralfac}.

\citet{xiao2017attentional} introduced Attentional Factorization Machines (AFM), which weight pairwise feature-interaction vectors with a learned softmax attention before summing them and present the weights as indicators of each interaction’s contribution. AFIN applies a similar principle at the branch level: an input-conditioned softmax weights four interaction vectors before pooling. The difference lies in evaluation. AFM treats its weights as interpretable without an intervention test, whereas we compare ours with branch-zeroing effects. The gate of AFIN also resembles the multi-gate mixture of experts of \citet{ma2018modeling}, where input-conditioned softmax gates weight expert outputs and gate weights are read to understand task relationships. Those weights are not validated by removing an expert, and the same limitation applies to AFIN’s branch weights unless tested.

\citet{qian2020} introduced HGAR, which merges explicit and implicit feedback by granular computing: an SVD models explicit ratings, an MLP models implicit preferences, and the two are combined by sum pooling under a joint loss. In contrast to this fixed pooling, we compare input-conditioned gating with cross-branch attention over four branches and ask whether the resulting weights reflect branch dependence.

\citet{he2023aain} introduced the Attentional Aggregative Interaction Network (AAIN) and a deep variant, DAAIN, which use attention to reorganize input features and capture higher-order interactions through an aggregative mechanism. It is relevant because it shows how attention can support interaction modeling, but it concerns feature interactions, whereas we contrast distinct user-movie interaction types.

\citet{song2024multi} examined how feature interactions vary across scenarios and tasks and proposed a multi-task feature-interaction approach, on the finding that the importance of an interaction depends on context. Multi-scenario and multi-task settings differ from our single rating-prediction task, but the work supports learning interaction-specific weights over a uniform combination rule.

These studies provide the context for comparing the two aggregation mechanisms examined here. AFIN assigns weights to available branches through an input-conditioned gate, resulting in contributions that vary across examples. In contrast, AFINca enables branches to exchange information via attention prior to pooling. The key distinction lies in the modeling of dependencies: gating determines relative contributions, while attention contextualizes branch representations using information from other branches. These mechanisms represent alternative modeling choices rather than a hierarchical progression. Although more elaborate mechanisms offer increased flexibility, their impact on prediction error requires empirical evaluation. Both mechanisms are compared within a single pipeline to avoid confounding effects arising from differences in inputs or objectives.

\subsection{Attention and Contextualized Representation Interaction}
Attention is widely used in recommendation, especially where user behavior is sequential or feature relationships must be modeled explicitly \cite{kang2018self}. \citet{zhu2024feature} proposed a feature interaction dual self-attention network that applies separate self-attention to feature interactions and to sequential transitions in session-based recommendation. Sequential recommendation predicts future interactions from the temporal order of past activity, whereas rating prediction estimates a numerical preference for a given movie. Sequential models are therefore relevant to us architecturally, as examples of contextualizing representations, and their results are not directly comparable to this work.

The general attention operation comes from the Transformer \cite{vaswani2017attention}. In self-attention, queries, keys, and values derive from the same set of representations, and in cross-attention, queries attend to a separate set.

Self-attention over feature tokens is well established for tabular and feature-interaction models. AutoInt applies multi-head self-attention over embedded feature fields and interprets the weights as interaction importance \cite{song2019autoint}. FT-Transformer applies a Transformer over feature tokens with an added summary token and is a strong tabular baseline \cite{gorishniy2021revisiting}. AFINca applies the same operation to four branch tokens and a context token, so its architecture is not novel. In AFINca, queries, keys, and values all derive from the branch tokens, which makes the operation self-attention over branches. We call it cross-branch attention because it runs across interaction branches, not between separate source and target sets. The empirical question is whether contextualizing the branches improves rating prediction over weighting them independently. Recent attention-based architectures motivate the question but do not show that attention-based aggregation outperforms gating in a static rating-prediction setting.

\subsection{Explaining Attention Weights and Branch Contributions}
Explainability is a significant concern for neural recommenders, particularly when a model combines representations whose roles vary across predictions \cite{zhang2020explainable}. Attention weights are often inspected to see which features, items, or representations receive more emphasis, but a weight does not necessarily reflect a component’s influence on the final prediction.

The debate on attention as explanation is directly relevant. \citet{jain2019attention} showed that attention distributions in some language models correlate weakly with other importance measures and that very different distributions can yield similar predictions. \citet{wiegreffe2019attention} argued that whether attention explains depends on how explanation is defined and evaluated. \citet{serrano2019attention} compared attention weights with the output change caused by altering them, an intervention we adapt to branch-level gate weights. Surveys such as \citet{gheewala2025depth} document the growing range of neural recommenders, but the presence of attention weights does not by itself make predictions explainable. The relationship between internal weights and outputs must be tested.

Branch weights are examined alongside branch-zeroing experiments. Weights indicate how the model allocates its learned aggregation, while zeroing demonstrates how predictions change when a branch’s weight is set to zero and the remaining weights are renormalized, without retraining. These analyses provide complementary evidence, as a branch with a small weight may still contain information that is not captured by the other branches.

\setlength{\tabcolsep}{2.5pt}
\begin{longtable}{@{}p{1.5cm} p{3cm} p{3cm} p{4cm} p{3cm} p{1cm}@{}}
	\caption{Overview of related studies including their methodological approach, category relative to this study, datasets, treatment of explainability, and publication year.}
	\label{tab:relatedwork}\\
	\toprule
	\textbf{Reference} & \textbf{Method} & \textbf{Category} & \textbf{Data} & \textbf{Explainability} & \textbf{Year} \\
	\midrule
	\endfirsthead
	\multicolumn{6}{c}{\tablename~\thetable{} (continued)}\\
	\toprule
	\textbf{Reference} & \textbf{Method} & \textbf{Category} & \textbf{Data} & \textbf{Explainability} & \textbf{Year} \\
	\midrule
	\endhead
	\midrule
	\multicolumn{6}{r}{\emph{Continued on next page}}\\
	\endfoot
	\bottomrule
	\endlastfoot
	\citet{koren2009matrix}       & Latent Factor Models              & Collaborative Filtering                     & Netflix Prize                                                                   & Does not address explainability.                                         & 2009 \\
	\citet{rendle2010factorizing} & Latent Factor Models              & Factorization Machines                      & Anonymized purchase
	data of an online drug store                                                          & Does not address explainability.                                         & 2010 \\
	\citet{sedhain2015autorec}    & Autoencoder-Based Models          & Autoencoder-Based Recommendation            & MovieLens 1M, MovieLens 100K, Netflix                                           & Does not address explainability.                                         & 2015 \\
	\citet{cheng2016wide}         & Hybrid Deep Learning Models       & Hybrid Recommendation                       & Google Play app-store interaction data (industrial)                             & Does not address explainability.                                         & 2016 \\
	\citet{wu2016collaborative}   & Autoencoder-Based Models          & Autoencoder-Based Recommendation            & MovieLens, Netflix                                                              & Does not address explainability.                                         & 2016 \\
	\citet{he2017neuralcol}       & Neural Collaborative Filtering    & Neural Collaborative Filtering              & MovieLens 1M, Pinterest                                                         & Does not address explainability.                                         & 2017 \\
	\citet{he2017neuralfac}       & Neural Feature Interaction Models & Feature Interaction                         & MovieLens 1M, Frappe                                                            & Does not address explainability.                                         & 2017 \\
	\citet{guo2017deepfm}         & Hybrid Deep Learning Models       & Hybrid Recommendation / Feature Interaction & Criteo, industrial app-store data                                               & Explainability is not a central focus.                                   & 2017 \\
	\citet{xiao2017attentional}   & Attentional Factorization Machines & Feature Interaction / Attention-Based      & Frappe, MovieLens (tag recommendation)                                          & Attention weights presented as interpretable; no intervention test.      & 2017 \\
	\citet{ma2018modeling}        & Multi-Task Learning with Gating   & Gated Mixture of Experts                    & Synthetic data, Census-Income, large-scale content recommendation (industrial)  & Gate weights used to study task relatedness; no intervention test.       & 2018 \\
	\citet{liang2018variational}  & Autoencoder-Based Models          & Variational Autoencoder Recommendation      & MovieLens 20M, Netflix Prize, Million Song Dataset                              & Does not address explainability.                                         & 2018 \\
	\citet{kang2018self}          & Attention-Based Recommendation    & Attention-Based Recommendation              & Amazon Beauty, Amazon Games, Steam, MovieLens-1M                                & Explainability is not a central focus.                                   & 2018 \\
	\citet{jain2019attention}     & Explainability Studies            & Attention Interpretability                  & Multiple NLP datasets and tasks                                                 & Explainability is a central research focus.                              & 2019 \\
	\citet{wiegreffe2019attention} & Explainability Studies           & Attention Interpretability                  & NLP tasks, including sentiment analysis and textual inference                   & Explainability is a central research focus.                              & 2019 \\
	\citet{serrano2019attention}  & Explainability Studies            & Attention Interpretability                  & NLP classification and question-answering tasks                                 & Explainability is a central research focus.                              & 2019 \\
	\citet{song2019autoint}       & Self-Attentive Feature Interaction & Feature Interaction / Attention-Based      & Criteo, Avazu, KDD12, MovieLens-1M                                              & Attention weights read as interaction importance; no intervention test.  & 2019 \\
	\citet{qian2020}              & SVD + MLP Hybrid                  & Hybrid Recommendation / Explicit-Implicit Feedback & MovieLens 100K, 1M                                                       & Does not address explainability.                                         & 2020 \\
	\citet{gorishniy2021revisiting} & Transformer for Tabular Data    & Attention-Based Feature Tokens              & Tabular benchmarks (e.g., California Housing, Adult, Higgs); not recommendation & Explainability is not a central focus.                                   & 2021 \\
	\citet{du2021}                & Similarity Measure                & Neighborhood-Based CF                       & MovieLens 100K, 1M                                                              & Does not address explainability.                                         & 2021 \\
	\citet{aljunid2021}           & Combined User/Item KNN            & Neighborhood-Based CF                       & MovieLens 100K, 1M                                                              & Does not address explainability.                                         & 2021 \\
	\citet{cheng2023}             & Privacy-Preserving KNN            & Neighborhood-Based CF                       & MovieLens 1M                                                                    & Does not address explainability.                                         & 2023 \\
	\citet{jain2023}              & Temporal Decay KNN                & Neighborhood-Based CF                       & MovieLens 100K                                                                  & Does not address explainability.                                         & 2023 \\
	\citet{nguyen2023}            & Adaptive KNN                      & Neighborhood-Based CF                       & MovieLens 100K, 1M                                                              & Does not address explainability.                                         & 2023 \\
	\citet{belmessous2024}        & Subjective-Logic Similarity       & Neighborhood-Based CF                       & MovieLens 100K, 1M                                                              & Does not address explainability.                                         & 2024 \\
	\citet{song2024multi}         & Neural Feature Interaction Models & Feature Interaction                         & AliExpress, Tenrec                                                              & Explainability is not a central focus.                                   & 2024 \\
	\citet{zhu2024feature}        & Attention-Based Recommendation    & Attention-Based Sequential Recommendation   & Tmall, MovieLens                                                                & Explainability is not a central focus.                                   & 2024 \\
	\citet{alam2025}              & RBM + KNN                         & Neighborhood-Based CF / Neural              & MovieLens 1M                                                                    & Does not address explainability.                                         & 2025 \\
	\citet{nahta2025deep}         & Neural Collaborative Filtering    & Movie Rating Prediction                     & Indian Regional Movies (IRM), MovieLens 100K, MovieLens 1M                      & Explainability is not a central focus.                                   & 2025 \\
	\citet{hafeez2026}            & User-Based CF, NMF, SVD           & Collaborative Filtering                     & MovieLens 1M                                                                    & Does not address explainability.                                         & 2026 \\
	\citet{wang2026}              & Bayesian NCF                      & Neural Collaborative Filtering              & MovieLens 100K, 1M                                                              & Does not address explainability.                                         & 2026 \\
\end{longtable}

\subsection{Research Gap and Contribution}
Table \ref{tab:relatedwork} summarizes recent recommender research on rating reliability, metadata integration, higher-order feature interaction, and attention-based representation learning. Surveys from 2024 to 2025 stress the difference between rating prediction and ranking. Within this body of work, models commonly weight their components with attention or gates and present the resulting weights as explanations \cite{xiao2017attentional, ma2018modeling, song2019autoint}. Whether those weights reflect what the model actually relies on is rarely tested. In language processing, intervention studies show that attention weights can diverge from feature importance \cite{jain2019attention, wiegreffe2019attention, serrano2019attention}, yet recommender work seldom applies the same test. In particular, no study we know of compares learned weights with removal effects across both an independent-gating and an attention-based aggregation of the same interaction branches. Without this comparison, it remains unclear whether weight-based explanations in recommenders can be trusted, and whether the choice between gating and attention changes their faithfulness.

We address this gap within a movie-rating pipeline. We compare independent gating (AFIN) with cross-branch attention (AFINca) on shared interaction branches, under both frozen and fine-tuned configurations of the CAAE encoder, and we compare the learned branch weights with the effects of removing individual branches. The proposed method also achieves lower RMSE than the baselines on MovieLens 100K and 1M.

\section{Methodology}
\label{sec:method}

\begin{figure}[htbp]
	\centering
	\includegraphics[width=\textwidth]{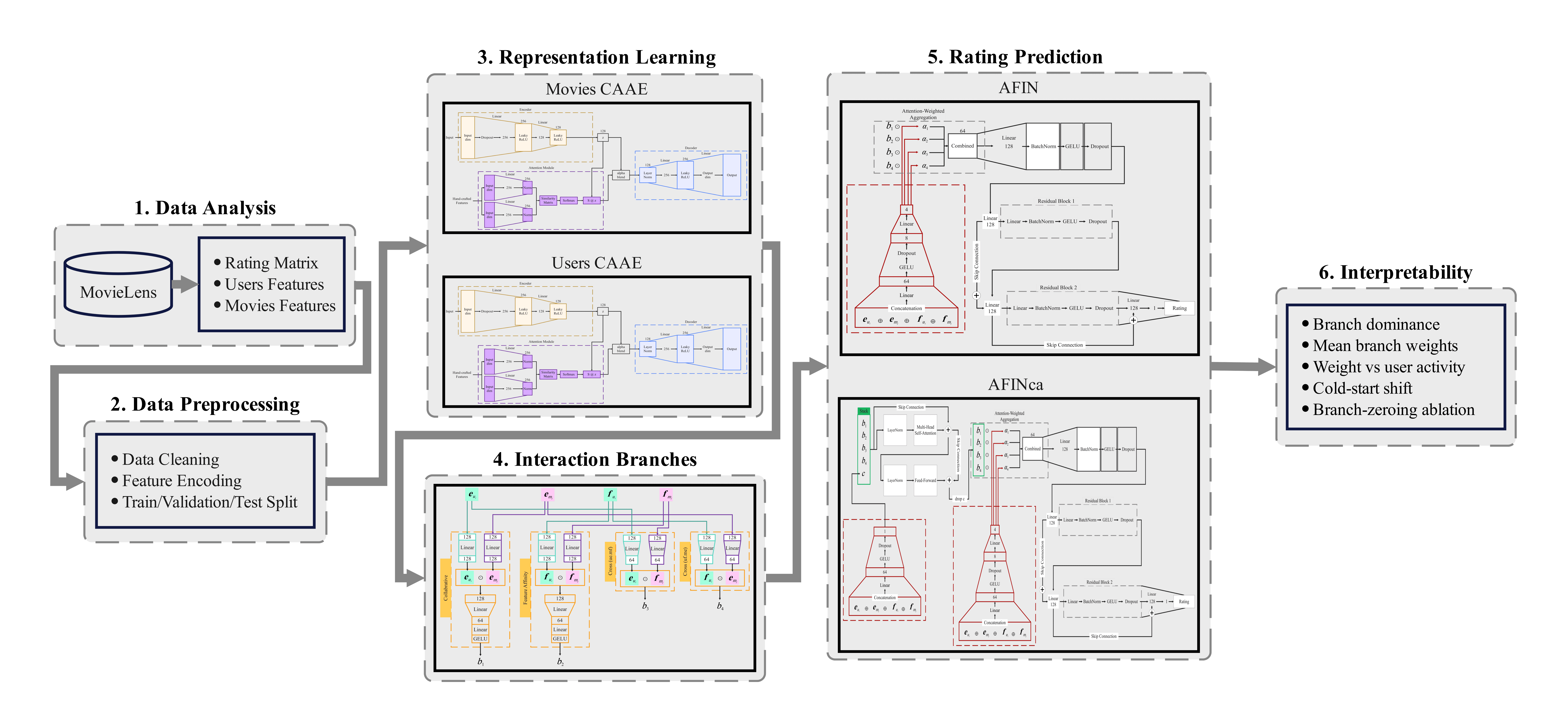}
	\caption{Overview of the pipeline. MovieLens ratings and user and movie features are cleaned, encoded, and split into training, validation, and test sets. Separate CAAE models for movies and users learn embeddings from the rating matrix. These embeddings are fed together with the side features into four interaction branches. AFIN or AFINca aggregates the branches and predicts the rating, and the trained models are then analyzed through branch dominance, mean branch weights, weight versus user activity, cold-start shift, and branch zeroing.}
	\label{fig:overalstructure}
\end{figure}

Fig.~\ref{fig:overalstructure} shows the proposed pipeline, starting by extracting the rating matrix, and user and movie features from MovieLens datasets, which are then cleaned, encoded, and split into training, validation, and test sets. The core methodology has three parts: representation learning (Section~\ref{subsec:caae}), rating prediction (Section~\ref{subsec:ratepred}), and explanation (Section~\ref{subsec:intpret}).

In representation learning, separate CAAE models are trained to learn dense user and movie embeddings from the sparse rating matrix and the side features. These embeddings are meant to give useful representations to users and movies with few ratings. Users and movies with no ratings are not evaluated. In rating prediction, each user-movie pair is processed through four interaction branches that capture complementary relationships between embeddings and side features. Two mechanisms separately aggregate the branches into a single rating: an Attentive Factorization Interaction Network (AFIN) and an Attentive Factorization Interaction Network with Cross-Attention (AFINca). Each is trained with frozen and fine-tuned encoders. For explainability, we examine the internal behavior of both models through branch dominance, mean branch weights, the relationship between branch weights and user activity, and the cold-start shift. A branch-zeroing ablation then measures each branch's effect on the predictions.

The rest of this section is organized as follows. Section~\ref{subse:data} presents the datasets and the feature-engineering decisions drawn from their statistics. Section~\ref{subsec:probform} formulates rating prediction as masked regression over a sparse rating matrix. Section~\ref{subsec:caae} describes CAAE pretraining. Section~\ref{subsec:ratepred} introduces the four interaction branches, and Sections~\ref{subsubsec:afin} and~\ref{subsubsec:afinca} describe AFIN and AFINca. Section~\ref{subsec:intpret} defines the explainability analyses and the branch-zeroing ablation.

\subsection{Dataset}
\label{subse:data}

We use MovieLens 100K (ML-100K) and MovieLens 1M (ML-1M), which are standard variants of the MovieLens benchmark widely used for recommender systems. Both datasets include integer ratings, given by users to movies, along with supplementary information tables for users and movies.

\begin{table}[hbp]
	\centering
	\caption{Descriptive statistics of the MovieLens 100K and MovieLens 1M datasets, including rating distribution, per-user and per-movie activity, sparsity, and user demographics.}
	\label{tab:mlstats}
	\begin{tabular*}{\textwidth}{@{\extracolsep{\fill}}LLL@{}}
		\toprule
		\textbf{Property} & \textbf{ML-100K} & \textbf{ML-1M} \\ 
		\midrule
		users & 943 & 6040 \\ 
		movies & 1682 & 3883 \\ 
		ratings & 100000 & 1000209\\ 
		sparsity & 93.70\% & 95.74\% \\ 
		rating mean & 3.53 & 3.58 \\ 
		rating std & 1.12  & 1.11 \\ 
		ratings per user median & 65 & 96\\ 
		ratings per user min & 20 & 20 \\ 
		ratings per user max & 737 & 2314 \\ 
		ratings per movie median & 27 & 123 \\ 
		ratings per movie min& 1 & 1 \\ 
		ratings per movie max & 583 & 3428 \\ 
		number of occupations & 21 & 21 \\ 
		number of males & 71.05\% & 71.71\% \\ 
		number of females & 28.95\% & 28.29\% \\
		\bottomrule
	\end{tabular*}
\end{table}

Table \ref{tab:mlstats} reports the statistical properties of two MovieLens datasets (ML-100K and ML-1M). The ML-100K contains 100000 ratings from 943 users on 1682 movies, and ML-1M contains 1000209 ratings from 6040 users on 3883 movies. Both datasets yield a high percentage of sparsity, with 93.70\% and 95.74\% for ML-100K and ML-1M, respectively, meaning most users did not rate most of the movies. ML-1M is about ten times larger than ML-100K and 2.04 percentage points sparser. Both datasets show a similar rating distribution, with the mean ratings of 3.53 for ML-100K, and 3.58 for ML-1M, which were demonstrated on a 1-to-5 scale, and nearly identical standard deviations of  1.12 for ML-100K and 1.11 for ML-1M, suggesting the growth in size of the dataset did not introduce a meaningfully different rating bias. While both datasets show a minimum of 20 ratings per user, the median user has rated 65 in ML-100K compared to ML-1M where the median user has rated 96 movies, and the most active user in ML-100K has 737 ratings versus 2314 in ML-1M. A similar pattern occurs as the median movie in ML-100K received 27 ratings compared to 123 in ML-1M. However, both datasets contain movies with only a single rating, reflecting a heavy-tailed distribution in popularity. Demographic composition is nearly identical across both datasets, with 21 categories and a gender split of roughly 72\% male and  28\% female.

\begin{figure}[ht]
	\centering
	\begin{subfigure}[t]{0.3\textwidth}
		\includegraphics[width=\textwidth, height=3cm]{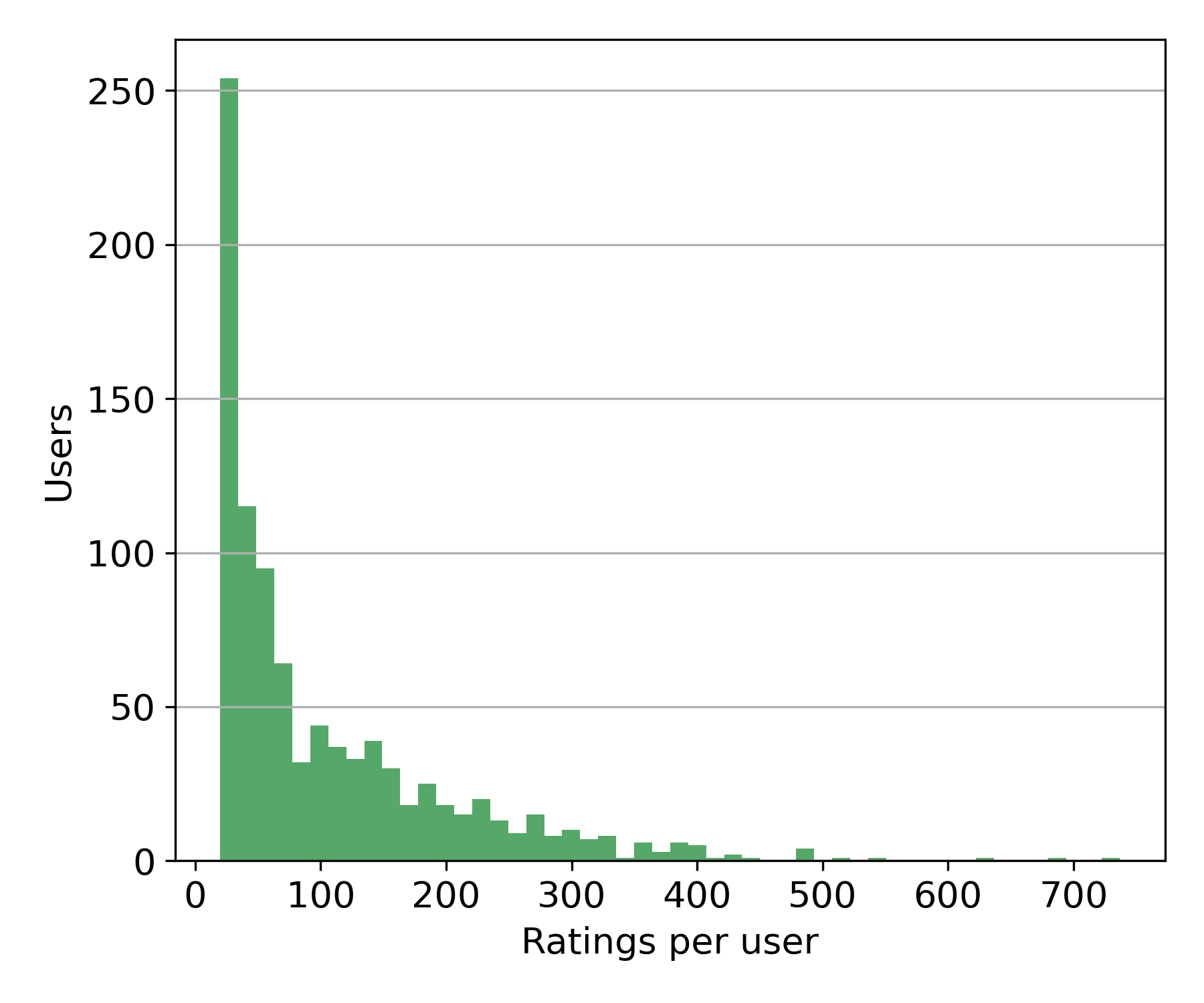}
		\caption{ML-100K: ratings per user}
		\label{subfig:100kuserrate}
	\end{subfigure}
	\hfill
	\begin{subfigure}[t]{0.3\textwidth}
		\includegraphics[width=\textwidth, height=3cm]{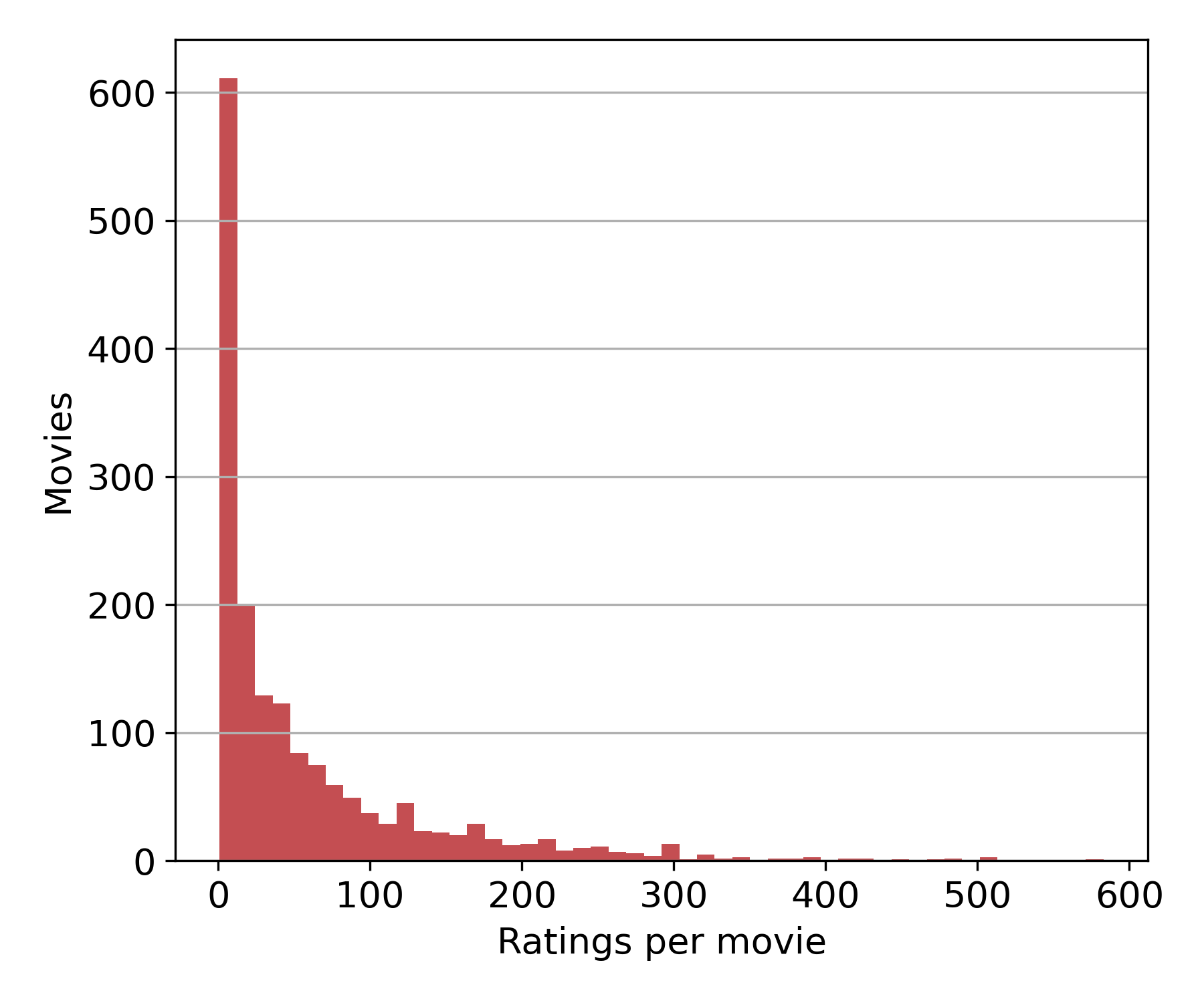}
		\caption{ML-100K: ratings per movie}
		\label{subfig:100kmovierate}
	\end{subfigure}
	\hfill
	\begin{subfigure}[t]{0.3\textwidth}
		\includegraphics[width=\textwidth, height=3cm]{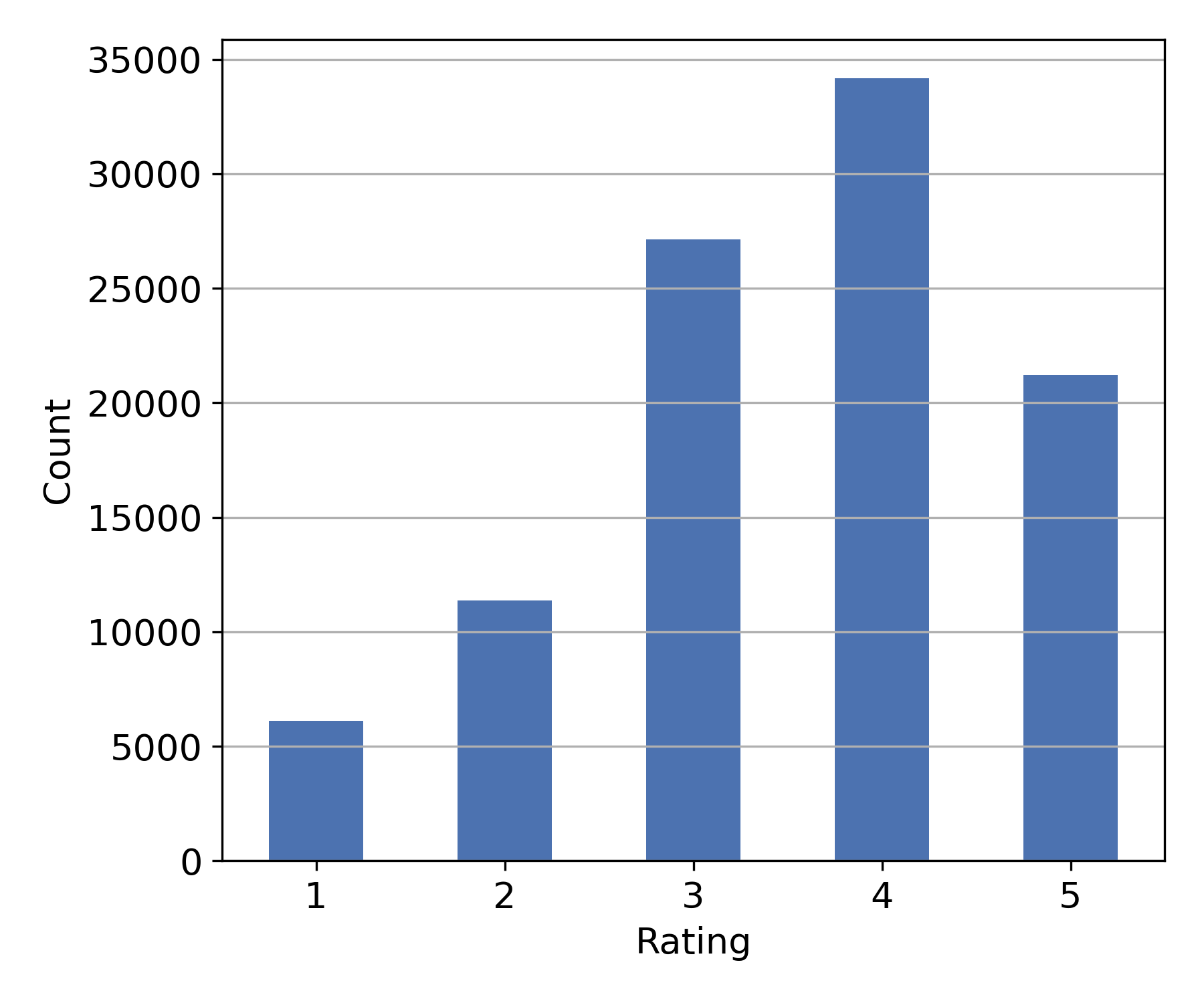}
		\caption{ML-100K: rating distribution}
		\label{subfig:100kratedist}
	\end{subfigure}
	\hfill
	\begin{subfigure}[t]{0.3\textwidth}
		\includegraphics[width=\textwidth, height=3cm]{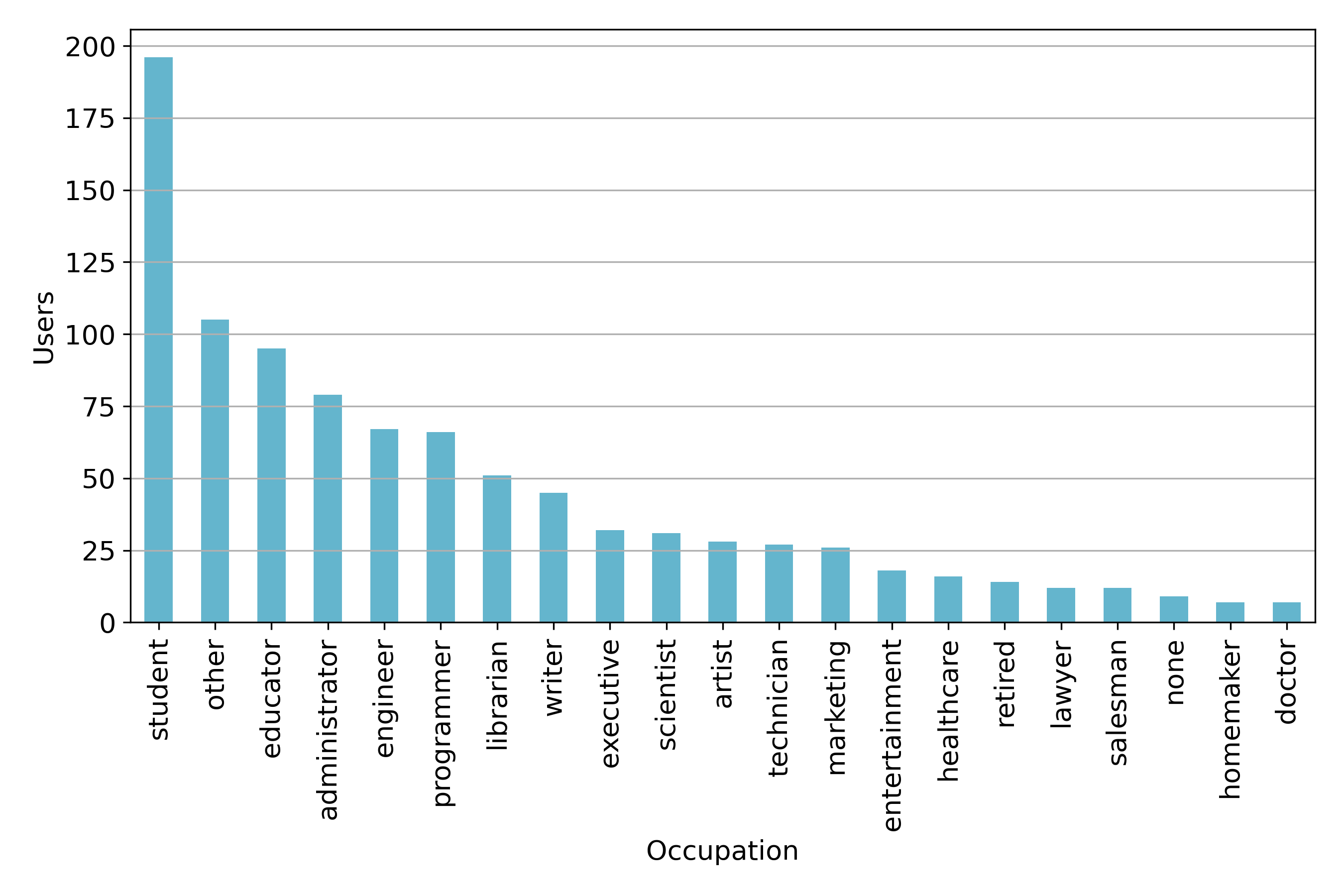}
		\caption{ML-100K: occupation distribution}
		\label{subfig:100koccupation}
	\end{subfigure}
	\hfill
	\begin{subfigure}[t]{0.3\textwidth}
		\includegraphics[width=\textwidth, height=3cm]{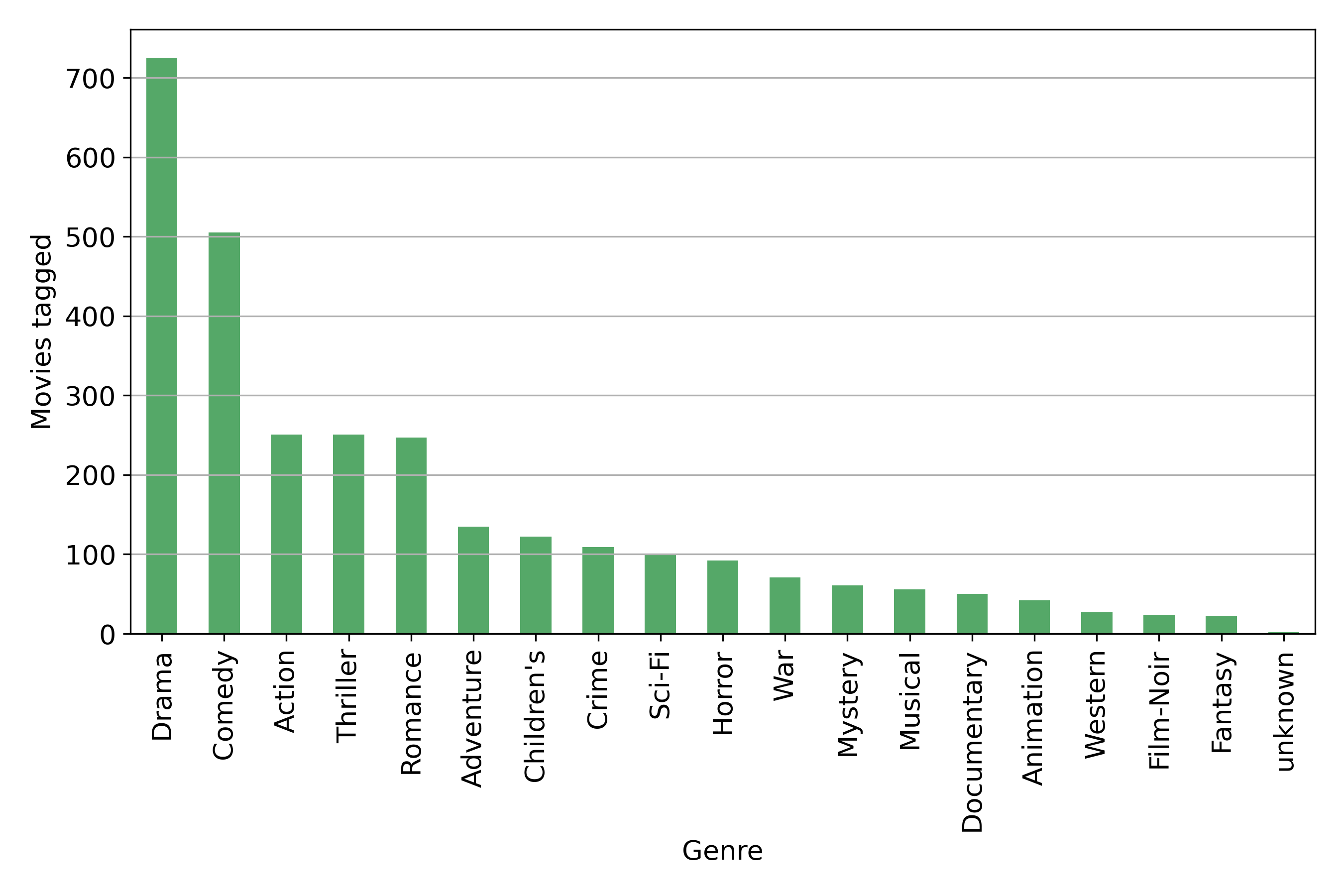}
		\caption{ML-100K: genre distribution}
		\label{subfig:100kgenre}
	\end{subfigure}
	\hfill
	\begin{subfigure}[t]{0.3\textwidth}
		\includegraphics[width=\textwidth, height=3cm]{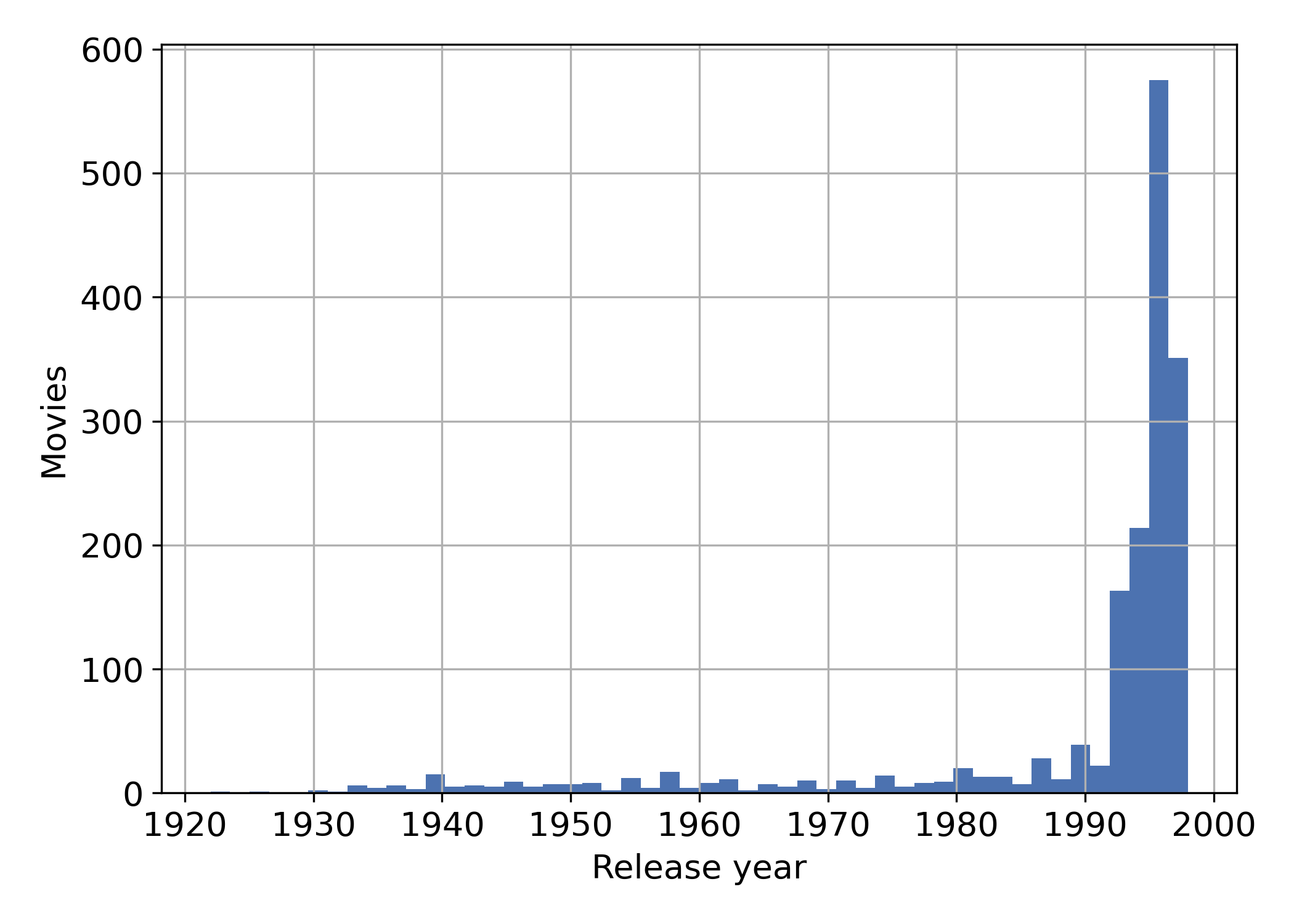}
		\caption{ML-100K: release year distribution}
		\label{subfig:100krelyear}
	\end{subfigure}
	\hfill
	\begin{subfigure}[t]{0.3\textwidth}
		\includegraphics[width=\textwidth, height=3cm]{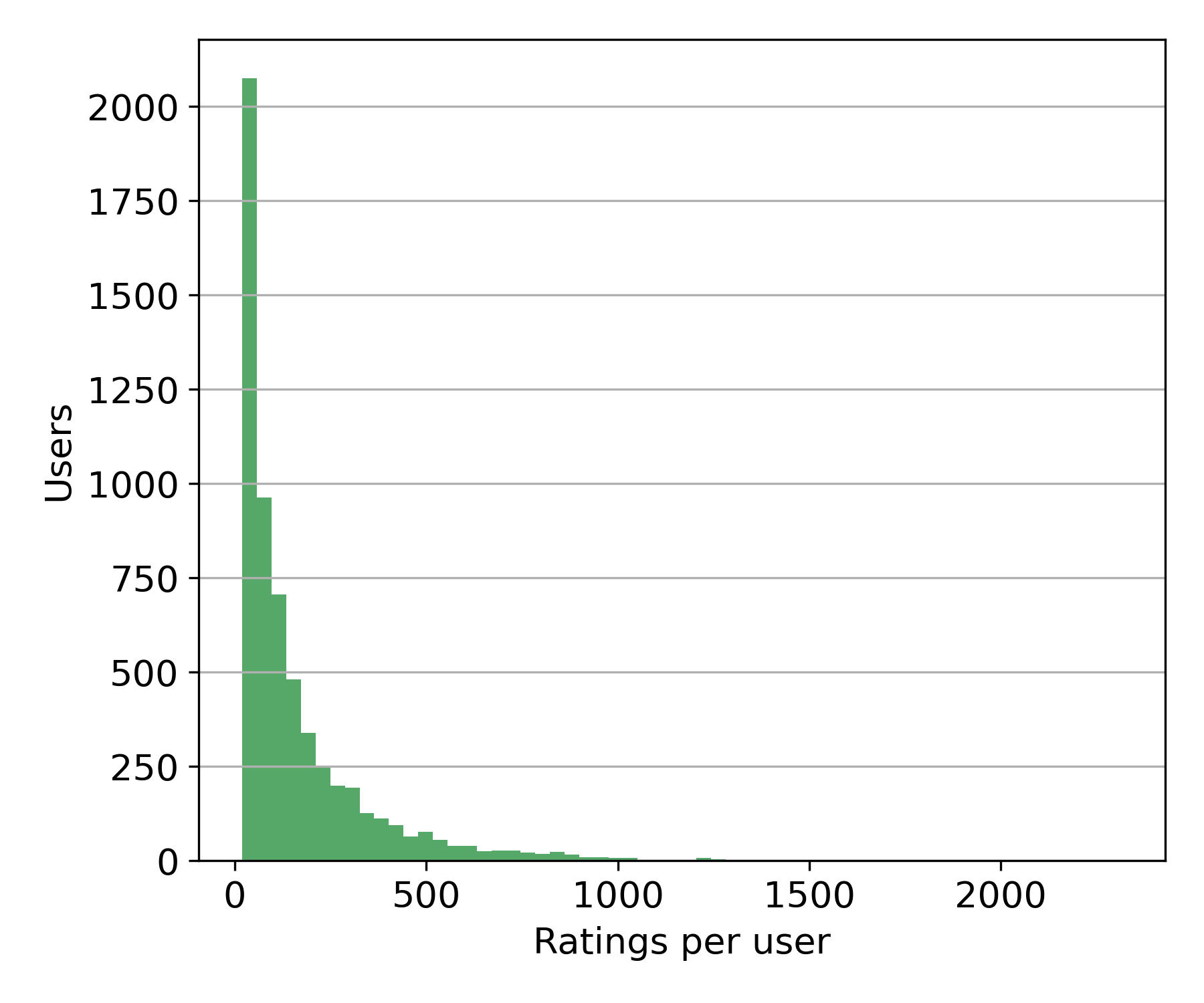}
		\caption{ML-1M: ratings per user}
		\label{subfig:1muserrate}
	\end{subfigure}
	\hfill
	\begin{subfigure}[t]{0.3\textwidth}
		\includegraphics[width=\textwidth, height=3cm]{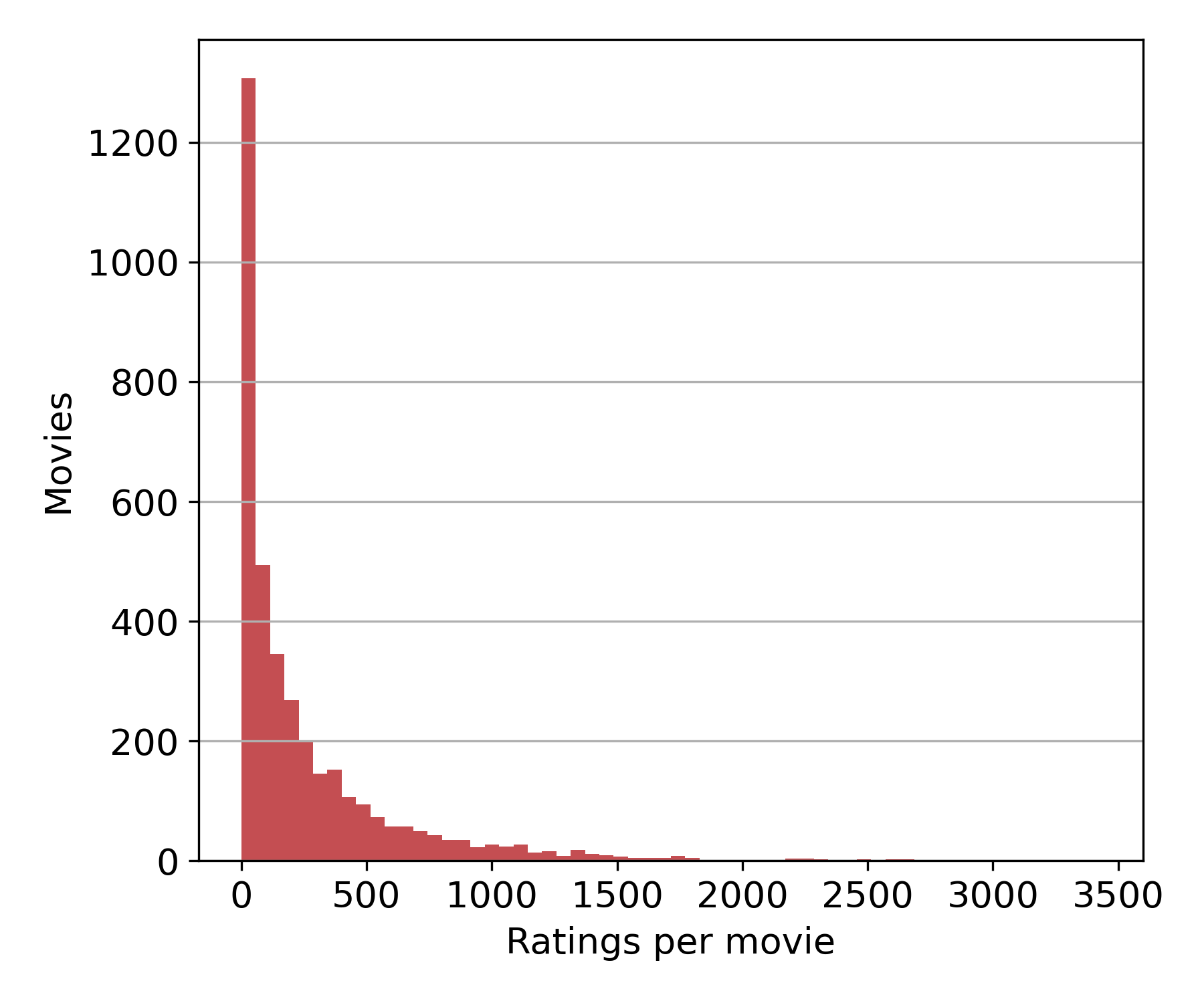}
		\caption{ML-1M: ratings per movie}
		\label{subfig:1mmovierate}
	\end{subfigure}
	\hfill
	\begin{subfigure}[t]{0.3\textwidth}
		\includegraphics[width=\textwidth, height=3cm]{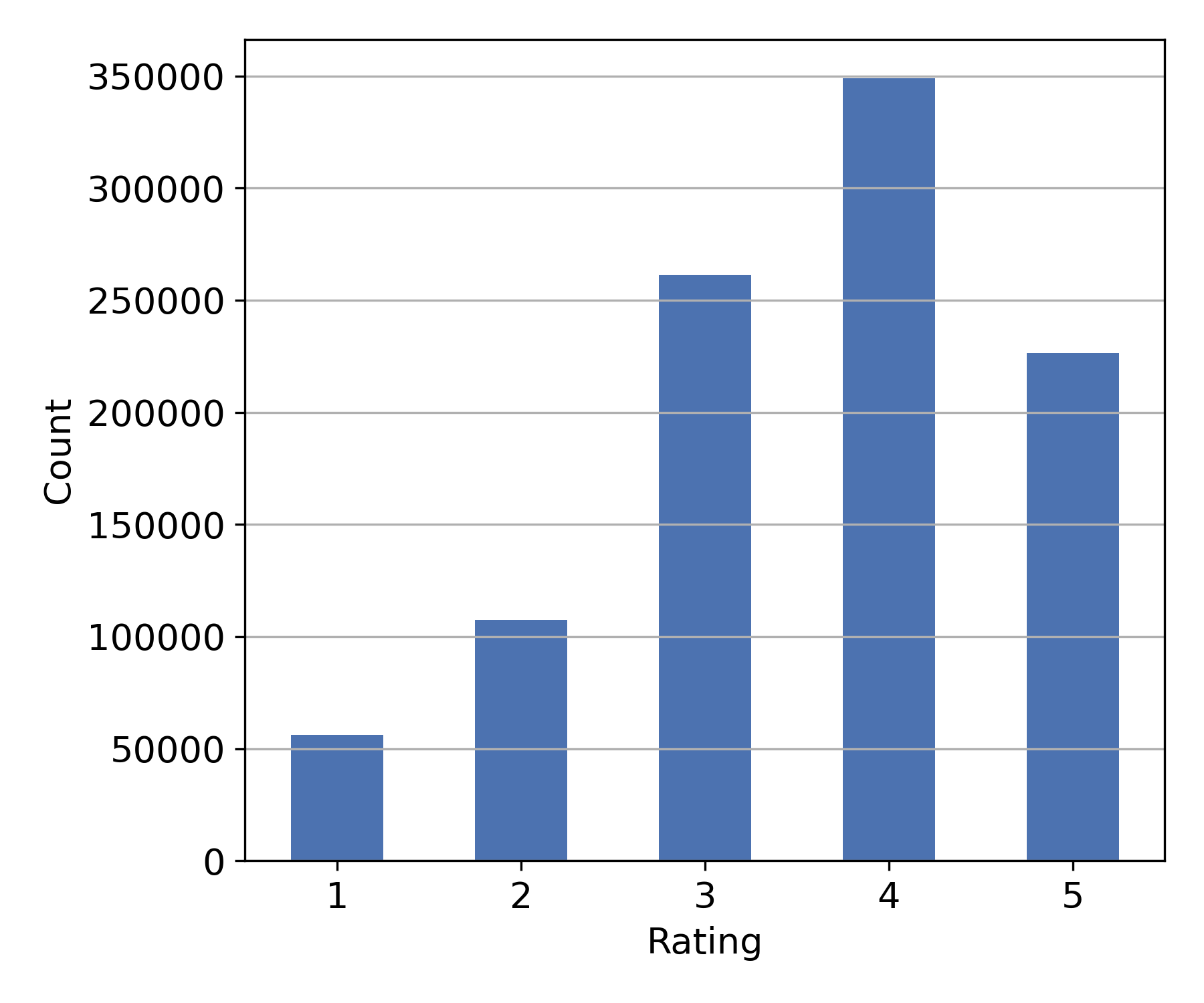}
		\caption{ML-1M: rating distribution}
		\label{subfig:1mratedist}
	\end{subfigure}
	\hfill
	\begin{subfigure}[t]{0.3\textwidth}
		\includegraphics[width=\textwidth, height=3cm]{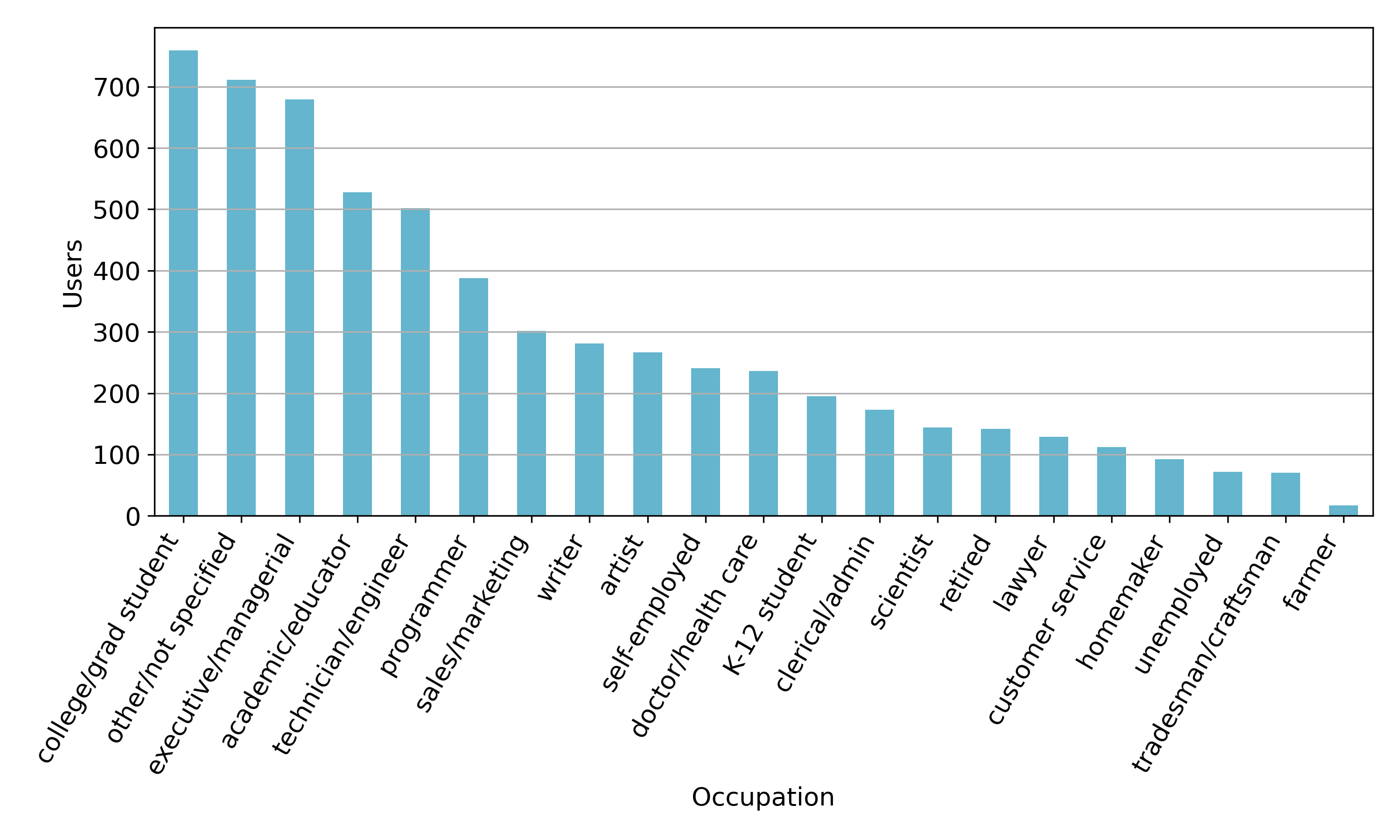}
		\caption{ML-1M: occupation distribution}
		\label{subfig:1moccupation}
	\end{subfigure}
	\hfill
	\begin{subfigure}[t]{0.3\textwidth}
		\includegraphics[width=\textwidth, height=3cm]{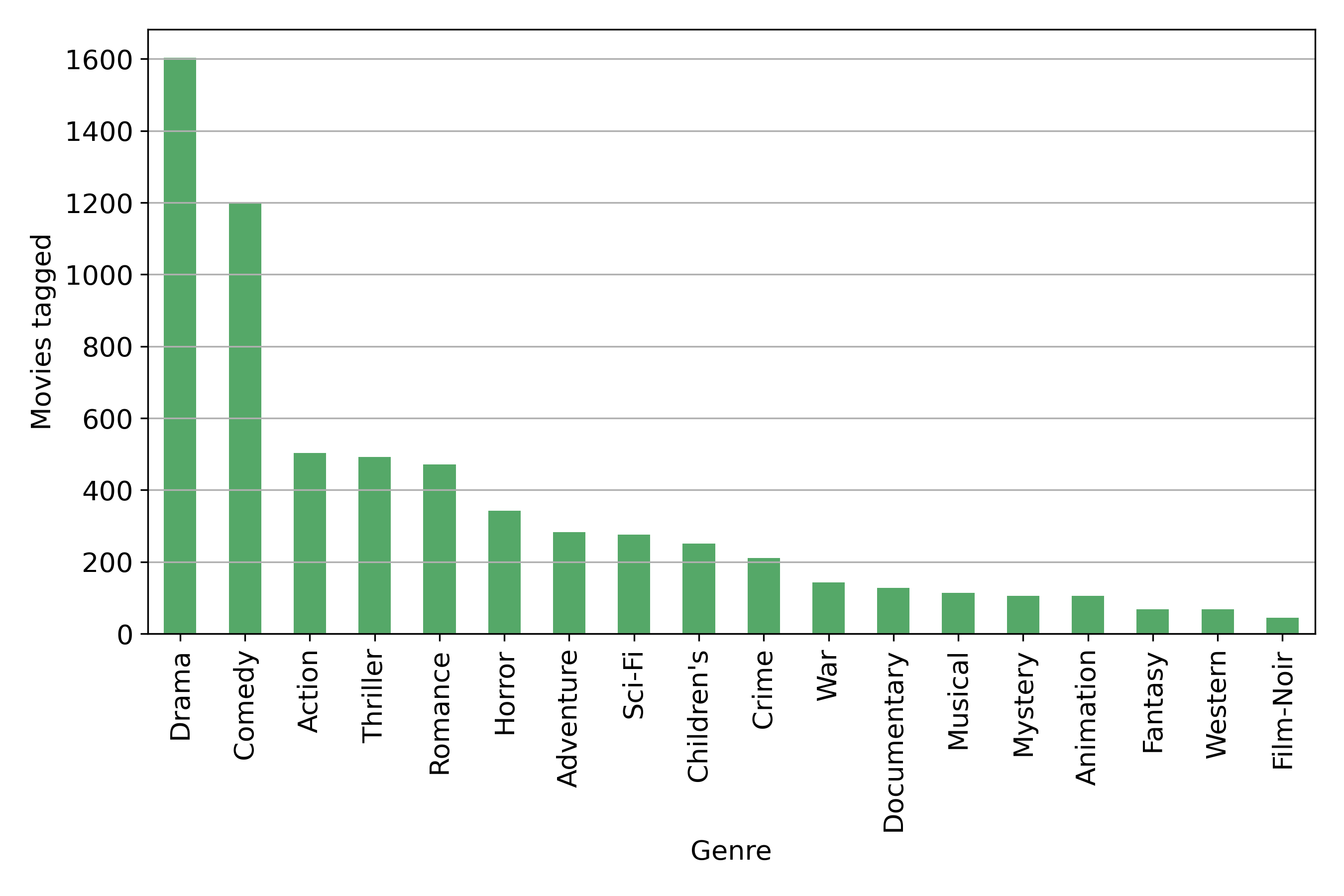}
		\caption{ML-1M: genre distribution}
		\label{subfig:1mgenre}
	\end{subfigure}
	\hfill
	\begin{subfigure}[t]{0.3\textwidth}
		\includegraphics[width=\textwidth, height=3cm]{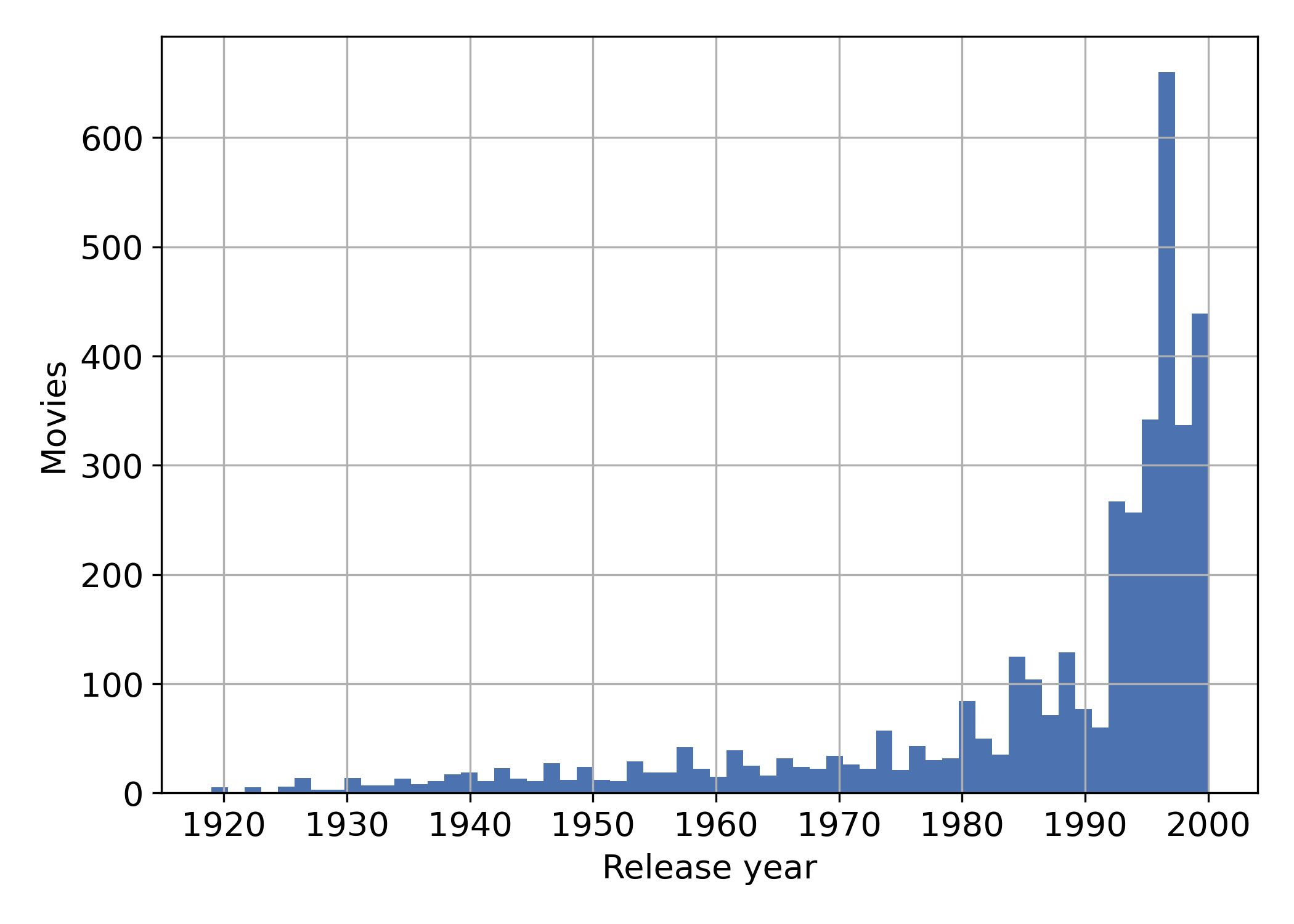}
		\caption{ML-1M: release year distribution}
		\label{subfig:1mrelyear}
	\end{subfigure}
	\caption{Exploratory statistics of ML-100K (panels a, b, c, d, e, and f) and ML-1M (panels g, h, i, j, k, and l): ratings per user, ratings per movie, rating distribution, user occupation, movie genre, and movie release year. Both datasets show heavy-tailed user and movie activity, right-skewed ratings concentrated at 3 and 4, imbalanced occupations and genres, and a skew toward recent releases.}
	\label{fig:ml}
\end{figure}

Fig.~\ref{fig:ml} presents exploratory visualizations of six features for ML-100K and ML-1M, showing matching distributional patterns across all six dimensions. Ratings per user (\ref{subfig:100kuserrate}, \ref{subfig:1muserrate}) and ratings per movie (\ref{subfig:100kmovierate}, \ref{subfig:1mmovierate}) are heavy-tailed, showing the characteristic sparsity, indicating that the majority of users and movies have relatively few ratings, while a small subset accounts for a disproportionate share of interactions. The rating distributions (\ref{subfig:100kratedist}, \ref{subfig:1mratedist}) are similarly right-skewed in both datasets, with ratings of 3 and 4 dominating and the ratings of 1 and 5 comprising a smaller share. Occupation distributions (\ref{subfig:100koccupation}, \ref{subfig:1moccupation}) show imbalance across categories in both datasets, with a handful of occupations (student, other, educator, and administrator in ML-100k, and college, other, executive, and academic in ML-1M) accounting for a large fraction of users. Genre distributions (\ref{subfig:100kgenre}, \ref{subfig:1mgenre}) are likewise consistent between datasets, with Drama and Comedy dominating, which forms an imbalance. Finally, release year distributions (\ref{subfig:100krelyear}, \ref{subfig:1mrelyear}) show the same right-skew toward more recent films in both datasets.

Collectively, these distributional properties informed the feature engineering decisions. ZIP code was excluded due to its high cardinality, which would result in most categories containing few users. In contrast, occupation, with 21 categories and a consistent imbalance across datasets, was retained as a more tractable categorical variable. The skew in release year toward recent films indicated that using raw year values or fine-grained bins would underpopulate early decades; therefore, era-based bins were used to ensure each temporal category had sufficient representation for meaningful pattern learning. The genre distribution's imbalance was preserved through multi-valued binary encoding, as genre co-occurrence was considered informative. Title and URL columns were omitted, as they provided no generalizable information beyond movie identity, which was already represented by the collaborative embedding. The pronounced sparsity in ratings per user and per movie motivated the use of dense embeddings via the attention autoencoder, as sparse identity vectors would convey minimal information for most users and movies.

\subsection{Problem Formulation}
\label{subsec:probform}

We treat rating prediction as a supervised regression task over the sparse rating matrix $R$ of Eq.~\ref{eq:obrm}, with $U$, $M$ (Eq.~\ref{eq:umsets}) and observed pairs $\Omega$ as defined in Section~\ref{subsec:rpcf}. For ML-1M, $|\Omega|$ covers roughly 4.3\% of all possible pairs, and the remaining entries of $R$ are missing, not zero. This sparsity is the constraint the rest of the methodology is built around.

Each user $u_i$ has an associated feature vector $\mathbf{f}_{u_i} = (\text{age}_{u_i}, \text{gender}_{u_i}, \text{occupation}_{u_i})$, and each movie $m_j$ has a feature vector $\mathbf{f}_{m_j} = (\text{releaseyear}_{m_j}, \text{genre}_{m_j})$. Separately, each user and movie is assigned a dense latent embedding, $\mathbf{e}_{u_i} \in \mathbb{R}^{d}$ and $\mathbf{e}_{m_j} \in \mathbb{R}^{d}$, learned by the collaborative filtering stage described in Section~\ref{subsec:caae} rather than derived from $\mathbf{f}_{u_i}$ or $\mathbf{f}_{m_j}$ directly. Eq.~\ref{eq:predictiontask} formulates the prediction task, which is to learn a function.
\begin{equation}
	\hat{r}_{u_i ,m_j} = F(\mathbf{e}_{u_i}, \mathbf{e}_{m_j}, \mathbf{f}_{u_i}, \mathbf{f}_{m_j}; \theta),
	\label{eq:predictiontask}
\end{equation}
that approximates $r_{u_i ,m_j}$ for $(u_i, m_j) \in \Omega$ and generalizes to held-out pairs. The predictor is trained by minimizing the mean squared error over the training pairs. We report RMSE and MAE on the held-out test pairs, where $\Omega_{\mathrm{tr}}, \Omega_{\mathrm{va}}, \Omega_{\mathrm{te}} \subset \Omega$ denote the training, validation, and test pairs (see the data-splits paragraph below). Because $\Omega$ is sparse and unevenly distributed across users and movies, a model that relies only on $\mathbf{f}_{u_i}, \mathbf{f}_{m_j}$ underuses the collaborative signal in $R$, while a model that relies only on identity-based embeddings $\mathbf{e}_{u_i}, \mathbf{e}_{m_j}$ struggles on cold-start users and movies with few or no observed ratings.

\paragraph{Features and data splits}
User features are a gender flag, a one-hot age bin (five bins), and a one-hot occupation (21 categories), giving 27 dimensions on ML-100K. Movie features are a one-hot release-era bin (four bins) and multi-hot genre indicators (19 on ML-100K), giving 23 dimensions. For ML-100K we use the official u1 split (u1.base and u1.test). For ML-1M, the test set contains about 100{,}000 ratings. On both datasets, 15\% of the training ratings are held out for validation. Let $R_{\mathrm{tr}}$ and $R_{\mathrm{va}}$ denote copies of $R$ that keep only the entries in $\Omega_{\mathrm{tr}}$ and $\Omega_{\mathrm{va}}$, respectively, with all other entries set to zero.

\subsection{Representation Learning}
\label{subsec:caae}

\begin{algorithm}[htbp]
	\caption{CAAE pretraining (users: input $R$, features $\mathbf{F}_U$; movies: input $R^{\top}$, features $\mathbf{F}_M$)}
	\label{alg:caae}
	\begin{algorithmic}[1]
		\Require Rating matrices $R_{\mathrm{tr}}, R_{\mathrm{va}}$, side features $\mathbf{F}$, dimension $d$, input dropout $p$, number of epochs $N_{\mathrm{ep}}$
		\Ensure Embeddings $\mathbf{E}$, encoder weights $\phi^\ast$
		\State Initialize encoder $g_\phi$, decoder $h_\psi$, attention projections $W_1, W_2$, blend scalar $\omega$, LayerNorm
		\State $\mathcal{L}^\ast \gets \infty$
		\For{$t = 1$ \textbf{to} $N_{\mathrm{ep}}$}
		\State \textbf{Training step on $R_{\mathrm{tr}}$:}
		\State $\mathbf{E} \gets g_\phi(\mathrm{Dropout}_p(R_{\mathrm{tr}}))$ \Comment{encoder, Eq.~\ref{eq:encdec}}
		\State $\mathbf{a} \gets \mathbf{F} \oplus \mathrm{detach}(\mathbf{E})$
		\State $S \gets \mathrm{softmax}\!\big(\mathrm{norm}(\mathbf{a}W_1)\,\mathrm{norm}(\mathbf{a}W_2)^{\top}/\sqrt{d_k}\big)$ \Comment{Eq.~\ref{eq:attention}}
		\State $\tilde{\mathbf{E}} \gets \mathrm{LN}\big(\omega\, S\mathbf{E} + (1-\omega)\,\mathbf{E}\big)$ \Comment{Eq.~\ref{eq:caaed-blend}, then LN}
		\State $\hat{\mathbf{v}} \gets h_\psi(\tilde{\mathbf{E}})$ \Comment{decoder, Eq.~\ref{eq:encdec}}
		\State $\mathcal{L} \gets \mathrm{RMSE}$ over $\Omega_{\mathrm{tr}}$ \Comment{Eq.~\ref{eq:rmse}, masked}
		\State Update parameters with $\nabla\mathcal{L}$
		\State \textbf{Validation (eval mode):}
		\State Forward pass on $R_{\mathrm{tr}}+R_{\mathrm{va}}$; $\mathcal{L}_{\mathrm{va}} \gets \mathrm{RMSE}$ over $\Omega_{\mathrm{va}}$ \Comment{Eq.~\ref{eq:rmse}, masked}
		\If{$\mathcal{L}_{\mathrm{va}} \le \mathcal{L}^\ast$}
		\State $\mathcal{L}^\ast \gets \mathcal{L}_{\mathrm{va}}$; \; save weights as best
		\EndIf
		\EndFor
		\State Restore best weights
		\State $\mathbf{E} \gets g_{\phi^\ast}(R_{\mathrm{tr}}+R_{\mathrm{va}})$ \Comment{eval mode, no dropout}
		\State \Return $\mathbf{E}$, $\phi^\ast$
	\end{algorithmic}
\end{algorithm}

In the first part of the pipeline, called representation learning, each user or movie is reduced to a dense embedding in $\mathbb{R}^{d}$ with $d = 128$. We use CAAE (Collaborative Attention Autoencoder), acting as a pretraining step. We separately train two CAAEs on users and movies to learn the dense embeddings on each data set. Eq.~\ref{eq:embtrain} formulates the process as:

\begin{equation}
	\begin{aligned}
		\mathbf{E}_M &= \mathrm{CAAE}(R^{\top}, \mathbf{F}_M) \in \mathbb{R}^{|M| \times d},\\
		\mathbf{E}_U &= \mathrm{CAAE}(R, \mathbf{F}_U) \in \mathbb{R}^{|U| \times d}.
	\end{aligned}
	\label{eq:embtrain}
\end{equation}

\begin{figure}[htbp]
	\centering
	\includegraphics[width=\textwidth]{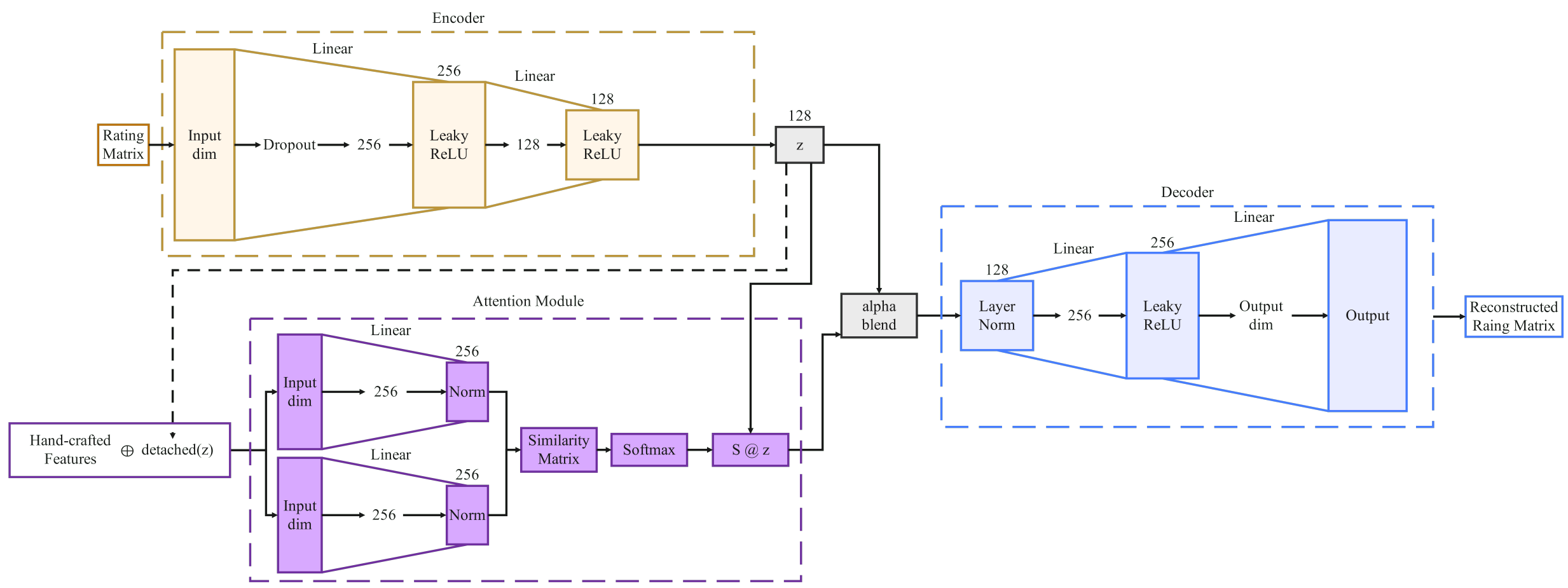}
	\caption{Architecture of the Collaborative Attention Autoencoder (CAAE). The encoder (yellow) compresses a dropout-corrupted rating row into a 128-dimensional embedding $\mathbf{e}$ (labeled $z$ in the figure). In the attention module (purple), the side features are concatenated with the detached embedding, projected by two linear layers with normalization, and passed through a softmax to give the similarity matrix $S$. The product $S\mathbf{E}$ is blended with $\mathbf{E}$ through the learned scalar $\omega$ (labeled $\alpha$ in the figure), layer-normalized, and decoded (blue) into the reconstructed rating row. The dashed line marks the detach: no gradient flows from the attention input into the encoder.}
	\label{fig:caae}
\end{figure}

Fig.~\ref{fig:caae} depicts the architecture of CAAE, where an encoder compresses the sparse rating row for a given input (user or movie) down to a $d$-dimensional embedding $\mathbf{e}$, and a decoder reconstructs the row from $\mathbf{e}$. The encoder uses two linear layers (256 and 128 units) with LeakyReLU, and the decoder uses a layer normalization followed by a 256-unit hidden layer and the output layer. Missing entries are encoded as zeros at the input. Dropout is applied to the input before encoding, and the reconstruction loss is masked RMSE, computed only over entries with an observed rating. The missing entries in $\Omega^c$ contribute nothing to the gradient, since there is no ground truth to reconstruct them against. The final decoder layer is linear, not passed through a nonlinearity, since the reconstruction target is a continuous rating rather than a bounded or categorical value.

What makes CAAE more than a plain autoencoder is the attention mechanism sitting between the encoder and the decoder. Each input (user or movie) also has a side-feature vector $\mathbf{f}_{u_i} \in \mathbf{F}_U$ for $user_i$ and $\mathbf{f}_{m_j} \in \mathbf{F}_M$ for $movie_j$, and CAAE uses this, together with the collaborative embedding itself, to build a similarity matrix over all users (or all movies), since training is full-batch. Two users with similar demographics and similar collaborative embeddings get a high attention weight, bringing them closer to each other. Concretely, the side features are concatenated ($\oplus$) with the encoder output, $\mathrm{attention\_module\_input} = \mathbf{f}_{u_i} \oplus \mathbf{e}_{u_i}$ for users and $\mathrm{attention\_module\_input} = \mathbf{f}_{m_j} \oplus \mathbf{e}_{m_j}$, projected through two separate linear layers, normalized, multiplied, and passed through a softmax to produce a weighted similarity matrix $S$. We detach $\mathbf{e}$ from the computation graph when forming the attention input, so the similarity scores are computed from the collaborative embedding without sending gradients from the similarity computation back into the encoder. Gradient still reaches the encoder through the value path $S\mathbf{E}$ and the blend in Eq.~\ref{eq:caaed-blend}, so the encoder is shaped by the reconstruction loss alone. The detach keeps the similarity computation from reshaping the embedding space through its own optimization dynamics.

$S$ is then applied to the batch of encoder outputs, giving each user's embedding a contribution pooled from demographically and collaboratively similar users:
\begin{equation}
	\tilde{\mathbf{E}} = \omega \, (S \cdot \mathbf{E}) + (1-\omega)\, \mathbf{E},
	\label{eq:caaed-blend}
\end{equation}
where $\mathbf{E}$ is the raw encoder output for the batch, $S$ is the weighted similarity matrix, and $\omega$ is a learned scalar controlling how much the attention-pooled signal is blended back in. $\tilde{\mathbf{E}}$ is then layer-normalized before being passed to the decoder.

During training, the side features let the attention step borrow strength across similar users (or movies): the reconstruction of a low-activity row is computed from an embedding blended with those of demographically and collaboratively similar rows, which regularizes the encoder on sparse rows. The extracted embeddings $\mathbf{e}_{u_i}$ and $\mathbf{e}_{m_j}$ are the raw encoder outputs $g_{\phi^\ast}(\cdot)$, without the attention blend, and they are the inputs to the rating predictor (Section~\ref{subsec:ratepred}). We do not evaluate users or movies without any observed ratings.

Algorithm~\ref{alg:caae} summarizes CAAE pretraining. The same procedure is run twice: for users with input $R$ and features $\mathbf{F}_U$, and for movies with input $R^{\top}$ and features $\mathbf{F}_M$. In the notation of Eq.~\ref{eq:encdec}, $\mathbf{v}$ is a rating row and $\mathbf{l}$ is the embedding $\mathbf{e}$. Each epoch encodes the dropout-corrupted training matrix, computes the similarity matrix $S$ from the side features and the detached embedding (Eq.~\ref{eq:attention}), blends $S\mathbf{E}$ with $\mathbf{E}$ (Eq.~\ref{eq:caaed-blend}), and decodes. The loss is the masked RMSE over $\Omega_{\mathrm{tr}}$ (Eq.~\ref{eq:rmse}). The detach means the similarity computation sends no gradient into the encoder. The checkpoint with the lowest masked validation RMSE on $\Omega_{\mathrm{va}}$ is kept, and the final embeddings are extracted from it with dropout disabled and $R_{\mathrm{tr}}+R_{\mathrm{va}}$ as input.

\subsection{Rating Prediction}
\label{subsec:ratepred}

The second part of the pipeline decomposes a user-movie pair $(u_i, m_j)$ into four interaction branches. It combines them, using the pretrained embeddings from CAAE alongside the hand-crafted side features, to produce $\hat{r}_{u_i,m_j}$.

\subsection{Interaction Branches}
\label{subsec:intbr}
\begin{figure}[htbp]
	\centering
	\includegraphics[width=\textwidth]{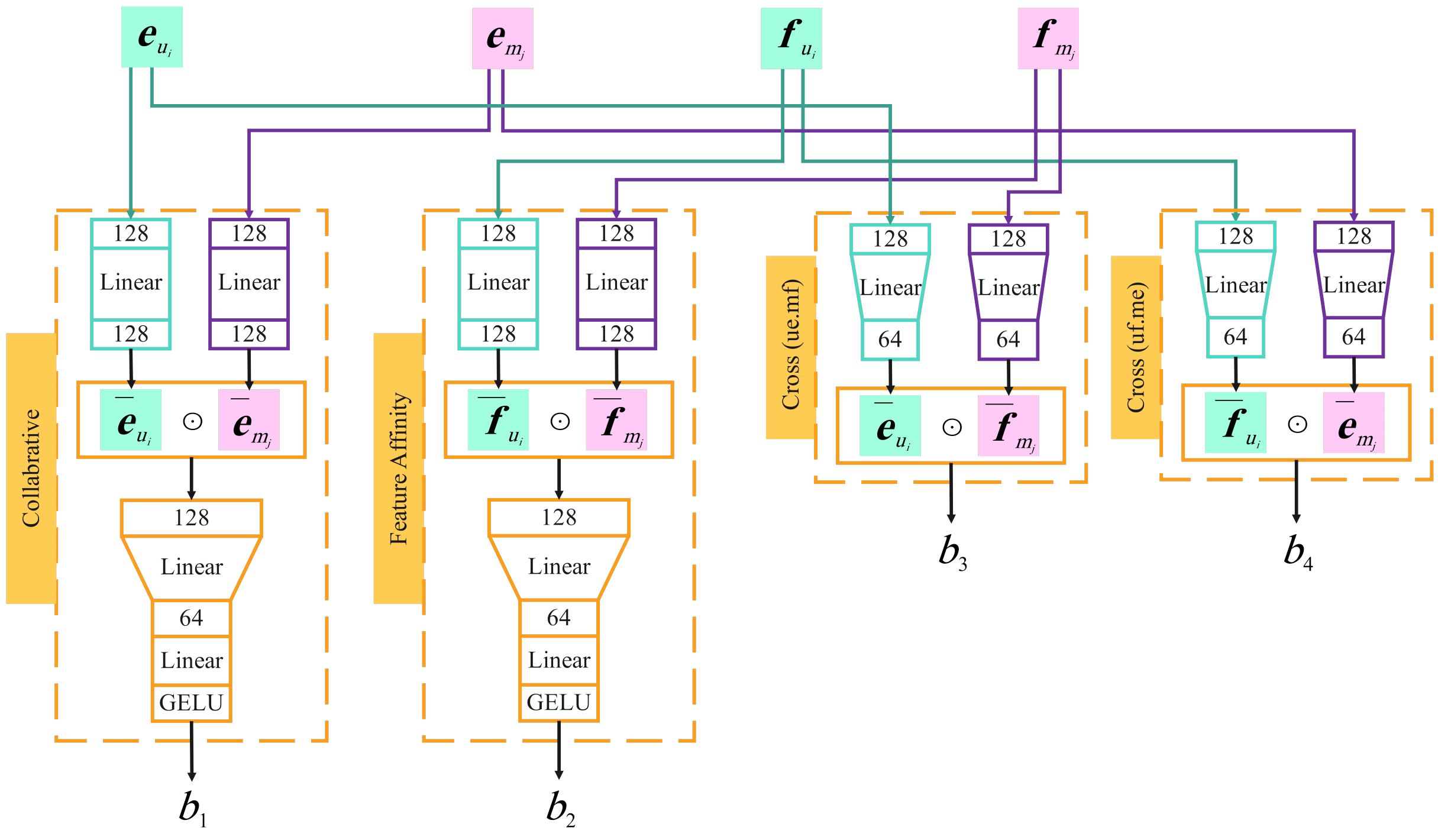}
	\caption{The four interaction branches built from the embeddings $\mathbf{e}_{u_i}$, $\mathbf{e}_{m_j}$ and the side features $\mathbf{f}_{u_i}$, $\mathbf{f}_{m_j}$. Green marks user-side inputs and pink marks movie-side inputs. Each branch projects its two inputs with separate affine projections and combines them by elementwise product. The collaborative ($\mathbf{b}_1$) and feature-affinity ($\mathbf{b}_2$) branches then apply a further linear layer with GELU. The cross branches pair a user embedding with movie features (ue.mf, $\mathbf{b}_3$) and a movie embedding with user features (uf.me, $\mathbf{b}_4$).}
	\label{fig:branches}
\end{figure}

Given a user embedding $\mathbf{e}_{u_i}$ and movie embedding $\mathbf{e}_{m_j}$ from CAAE, and the feature vectors $\mathbf{f}_{u_i}, \mathbf{f}_{m_j}$, this stage instantiates $F(\cdot; \theta)$ from Eq.~\ref{eq:predictiontask}; Fig.~\ref{fig:branches} illustrates it. The pair is decomposed into four branches, each representing a different notion of similarity:
\begin{equation}
	\begin{aligned}
		\mathbf{b}_1 &= \rho_1\big(\mathbf{P}_1\mathbf{e}_{u_i} \odot \mathbf{Q}_1\mathbf{e}_{m_j}\big), &&\text{(collaborative)}, \\
		\mathbf{b}_2 &= \rho_2\big(\mathbf{P}_2\mathbf{f}_{u_i} \odot \mathbf{Q}_2\mathbf{f}_{m_j}\big), &&\text{(feature affinity)}, \\
		\mathbf{b}_3 &= \mathbf{P}_3\mathbf{e}_{u_i} \odot \mathbf{Q}_3\mathbf{f}_{m_j}, &&\text{(cross, user embedding} \rightarrow \text{movie features)}, \\
		\mathbf{b}_4 &= \mathbf{P}_4\mathbf{e}_{m_j} \odot \mathbf{Q}_4\mathbf{f}_{u_i}, &&\text{(cross, movie embedding} \rightarrow \text{user features)}.
	\end{aligned}
	\label{eq:branches}
\end{equation}
Here $\mathbf{P}_k$ and $\mathbf{Q}_k$ are branch-specific affine projections of the first and second factor, $\odot$ is elementwise multiplication, and $\rho_1,\rho_2$ are linear layers with a GELU that map to the interaction dimension $D$ ($\rho_3,\rho_4$ are identities, since $\mathbf{P}_3,\mathbf{Q}_3,\mathbf{P}_4,\mathbf{Q}_4$ already map to $D$). For $\mathbf{b}_1$ and $\mathbf{b}_2$, the projections map both factors to the smaller of their two input dimensions so that the product is defined. $\mathbf{b}_1$ captures what the two identity embeddings alone imply about compatibility. $\mathbf{b}_2$ captures demographic and content fit, independent of collaborative history. The cross branches, $\mathbf{b}_3$ and $\mathbf{b}_4$, pick up asymmetric signals: a user's latent taste profile lining up with a movie's declared genre, or a movie's latent identity lining up with a user's declared demographics.

Both AFIN and AFINca are built on these same four branches. They differ only in how the branches are combined into a single vector $\mathbf{z}$ before the final residual MLP. AFIN gates each branch independently with a softmax weight, whereas AFINca lets the branches attend to each other first. Sections~\ref{subsubsec:afin} and~\ref{subsubsec:afinca} cover each in turn, and both are trained frozen and fine-tuned, as described in their respective subsections.

\subsubsection{AFIN}
\label{subsubsec:afin}

AFIN combines the four branches from Eq.~\ref{eq:branches} with a per-sample softmax gate $\boldsymbol{\alpha} = (\alpha_1, \alpha_2, \alpha_3, \alpha_4)$, conditioned on the raw inputs:
\begin{equation}
	\boldsymbol{\alpha} = \mathrm{softmax}\big(\gamma(\mathbf{e}_{u_i} \oplus \mathbf{e}_{m_j} \oplus \mathbf{f}_{u_i} \oplus \mathbf{f}_{m_j})\big),
	\label{eq:afin-gate}
\end{equation}
where $\oplus$ denotes vector concatenation and $\gamma(\cdot)$ is a three layer feed-forward network. The branches are combined as a weighted sum, $\mathbf{z} = \sum_{k=1}^{4} \alpha_k \mathbf{b}_k$, and $\mathbf{z}$ goes through a residual MLP to produce $\hat{r}_{u_i, m_j}$. Because $\boldsymbol{\alpha}$ is computed fresh for every pair rather than learned as one global weighting, it gives a direct, per-sample readout of which branch the model leans on, the basis for the explainability analysis in Section~\ref{subsec:intpret}.

\begin{figure}[tbp]
	\centering
	\includegraphics[width=\textwidth, keepaspectratio]{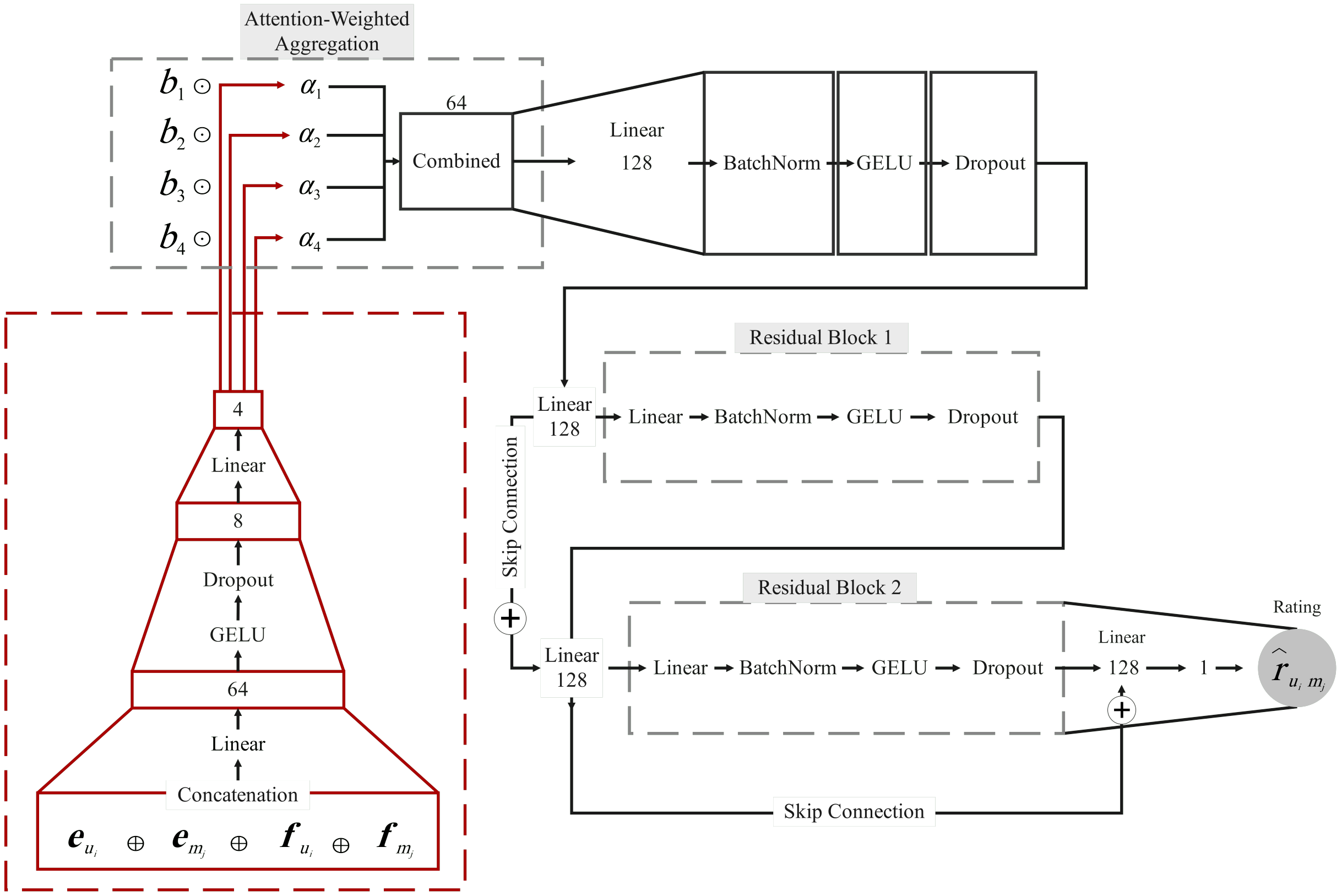}
	\caption{Architecture of AFIN. The concatenated inputs pass through a three-layer gating network (red) that outputs the four branch weights $\alpha_1,\dots,\alpha_4$. The branches $\mathbf{b}_1,\dots,\mathbf{b}_4$ are weighted by these values and summed into the pooled vector $\mathbf{z}$ (Combined). The pooled vector goes through a linear layer with batch normalization, GELU, and dropout, then through residual blocks with skip connections, and a final linear layer outputs the predicted rating $\hat{r}_{u_i,m_j}$.}
	\label{fig:afin}
\end{figure}

Under this gate, the branches do not interact. The $\mathbf{b}_1$ vector is computed without access to $\mathbf{b}_2$, $\mathbf{b}_3$, or $\mathbf{b}_4$ vectors, and the gate only reweights four fixed vectors, so no branch's content can shape another's. Fig.~\ref{fig:afin} shows the full architecture.

\paragraph{Frozen}
\label{par:afin-frozen}

Here $\mathbf{e}_{u_i}$ and $\mathbf{e}_{m_j}$ are fixed inputs. CAAE is trained once (Section~\ref{subsec:caae}); its encoder weights are then discarded, and only the resulting embedding matrices are kept. AFIN trains against them as it would against any precomputed feature, and no gradient reaches CAAE. The two stages stay fully decoupled: representation learning is optimized for reconstruction, rating prediction for downstream RMSE, and neither objective touches the other.

\paragraph{Fine-Tuned}
\label{par:afin-fine-tuned}

CAAE's encoder is loaded from its pretrained checkpoint but kept trainable, and gradients from AFIN's loss propagate back through $\mathbf{e}_{u_i}, \mathbf{e}_{m_j}$ into the encoder itself. The embeddings are then shaped by the rating-prediction loss as well as by the reconstruction pretraining. The encoder starts from a checkpoint that already reflects collaborative structure, so fine-tuning refines it rather than learning it from scratch.

\subsubsection{AFINca}
\label{subsubsec:afinca}

AFINca shares AFIN's four branches and overall pipeline. It changes only the aggregation: instead of an independent softmax over four fixed vectors, the branches first pass through a self-attention layer across branches, so each branch can be informed by the others before pooling.

The four branch vectors are treated as a sequence of tokens in $\mathbb{R}^{D}$, where $D$ is the shared interaction dimension. An optional context token $\mathbf{c} = \kappa(\mathbf{e}_{u_i} \oplus \mathbf{e}_{m_j} \oplus \mathbf{f}_{u_i} \oplus \mathbf{f}_{m_j})$, a projection of the same raw inputs that AFIN's gate uses, is prepended so that branches can attend directly to user and movie identity as well as to each other. All reported AFINca results use the context token. The five tokens pass through a standard multi-head self-attention Transformer encoder layer:
\begin{equation}
	(\mathbf{c}', \mathbf{b}'_1, \dots, \mathbf{b}'_4) = \mathrm{MHSA}\big(\mathbf{c}, \mathbf{b}_1, \dots, \mathbf{b}_4\big).
	\label{eq:afinca-attn}
\end{equation}
$\mathrm{MHSA}(\cdot)$ is a standard pre-norm Transformer encoder layer (multi-head self-attention plus a feed-forward block, both with residual connections), applied to five tokens, or four without the context token, instead of a sentence. The context token is discarded after attention, leaving four attended branch tokens $\mathbf{b}'_1,\dots,\mathbf{b}'_4 \in \mathbb{R}^{D}$. Each is a function of all four original branches and, with the context token, of the raw inputs directly, rather than of one branch alone.

\begin{figure}[tbp]
	\centering
	\includegraphics[width=\textwidth, keepaspectratio]{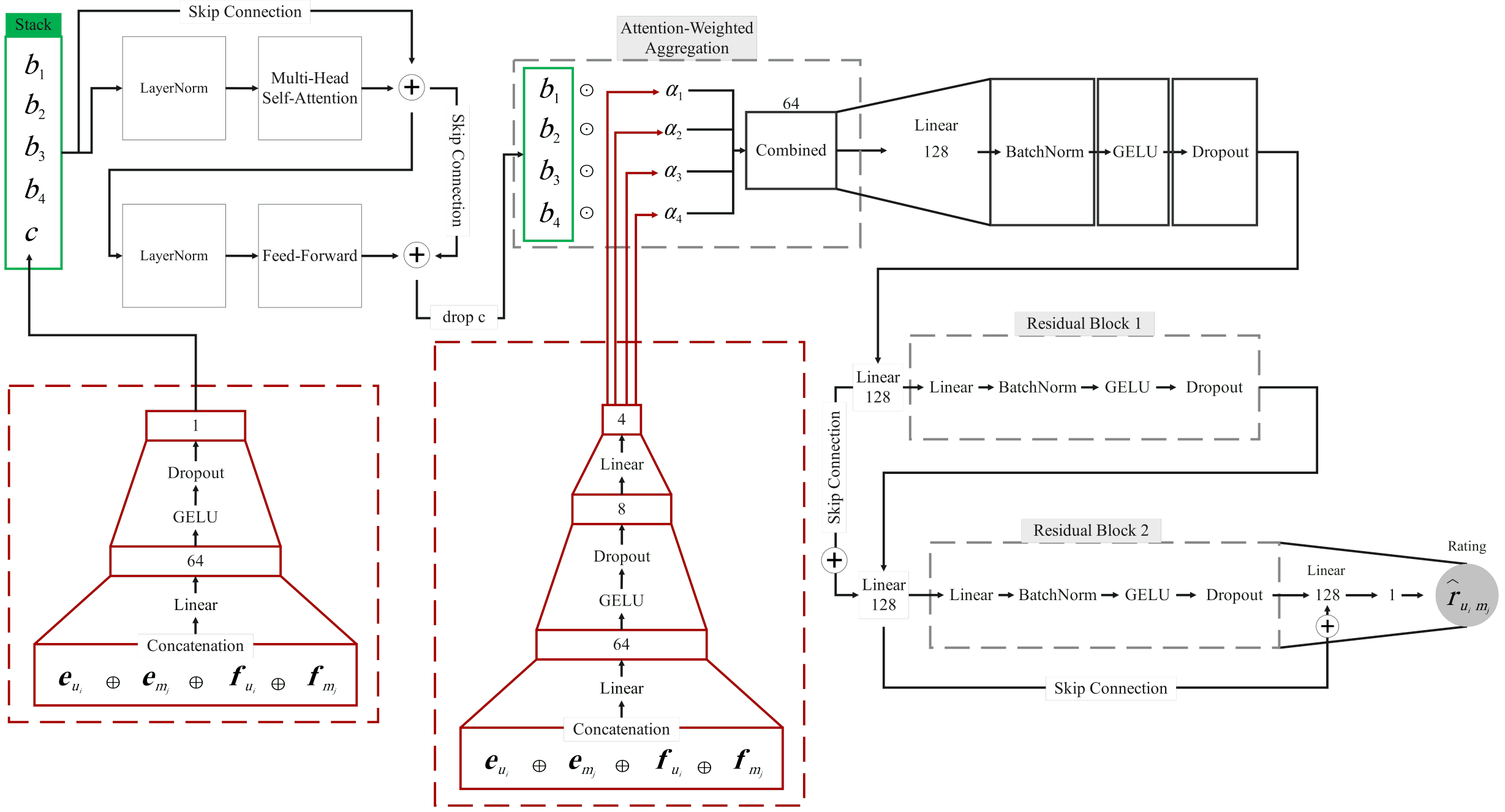}
	\caption{Architecture of AFINca. The four branches and a context token $\mathbf{c}$, produced by a feed-forward network from the concatenated inputs, are stacked (green) and passed through a pre-norm Transformer encoder layer: layer normalization, multi-head self-attention, layer normalization, and a feed-forward block, each with a skip connection. The context token is then dropped. The four attended tokens are pooled with weights $\beta_1,\dots,\beta_4$ from a gating network (red) conditioned on the context token. The pooled vector goes through the same residual MLP as in AFIN to give the predicted rating.}
	\label{fig:afinca}
\end{figure}

Pooling uses a gate like AFIN's, applied to the attended tokens and conditioned on the context vector:
\begin{equation}
	\boldsymbol{\beta} = \mathrm{softmax}\big(\gamma'(\mathbf{c})\big), \qquad \mathbf{z} = \sum_{k=1}^{4} \beta_k \mathbf{b}'_k ,
	\label{eq:afinca-gate}
\end{equation}
where $\mathbf{b}'_k$ is the $k$-th attended branch token and $\gamma'(\cdot)$ is a feed-forward network separate from AFIN's. The vector $\mathbf{z}$ goes through the same residual MLP as AFIN (Fig.~\ref{fig:afinca}).

The important difference from AFIN is what $\boldsymbol{\beta}$ gates. The pooled quantities are no longer four independent branches but branch representations that have already exchanged information through self-attention. A branch with a low pooling weight is therefore not necessarily uninformative, because its signal may already be present in another branch's representation after Eq.~\ref{eq:afinca-attn}. This complicates reading $\boldsymbol{\beta}$ the way one reads AFIN's $\boldsymbol{\alpha}$, as discussed in Section~\ref{subsec:intpret}.

\paragraph{Frozen}
\label{par:afinca-frozen}

Same setup as AFIN's frozen variant: $\mathbf{e}_{u_i}$ and $\mathbf{e}_{m_j}$ come from the pretrained CAAE checkpoint, fixed, with no gradient flowing back into the encoder. AFINca's attention layer and pooling gate train against these fixed embeddings, which isolates whatever the cross-branch attention itself contributes from any change in the upstream representation.

\paragraph{Fine-Tuned}
\label{par:afinca-fine-tuned}

CAAE's encoder starts from its pretrained checkpoint and stays trainable, with AFINca's loss propagating back into it, mirroring AFIN's fine-tuned setting. The point of running both is to see whether any RMSE gain from fine-tuning is specific to AFIN's simpler gate, or whether it carries over to AFINca's attention-based pooling too.

\begin{algorithm}[htbp]
	\caption{Forward pass of AFIN and AFINca for a pair $(u_i, m_j)$}
	\label{alg:forward}
	\begin{algorithmic}[1]
		\Require $\mathbf{e}_{u_i}, \mathbf{e}_{m_j}, \mathbf{f}_{u_i}, \mathbf{f}_{m_j}$; model $\in\{\text{AFIN},\text{AFINca}\}$
		\Ensure $\hat{r}_{u_i,m_j}$ and branch weights
		\State $\mathbf{b}_1,\dots,\mathbf{b}_4 \gets$ branches of Eq.~\ref{eq:branches}
		\If{model $=$ AFIN}
		\State $\boldsymbol{\alpha} \gets$ gate of Eq.~\ref{eq:afin-gate}
		\State $\mathbf{z} \gets \sum_{k=1}^{4}\alpha_k\mathbf{b}_k$
		\Else
		\State $\mathbf{c} \gets \kappa(\mathbf{e}_{u_i}\oplus\mathbf{e}_{m_j}\oplus\mathbf{f}_{u_i}\oplus\mathbf{f}_{m_j})$ \Comment{context token}
		\State $(\mathbf{c}',\mathbf{b}'_1,\dots,\mathbf{b}'_4) \gets \mathrm{MHSA}(\mathbf{c},\mathbf{b}_1,\dots,\mathbf{b}_4)$; discard $\mathbf{c}'$ \Comment{Eq.~\ref{eq:afinca-attn}}
		\State $\boldsymbol{\beta} \gets \mathrm{softmax}\big(\gamma'(\mathbf{c})\big)$
		\State $\mathbf{z} \gets \sum_{k=1}^{4}\beta_k\mathbf{b}'_k$ \Comment{Eq.~\ref{eq:afinca-gate}}
		\EndIf
		\State $\hat{r}_{u_i,m_j} \gets \mathrm{MLP}(\mathbf{z})$ \Comment{residual MLP}
		\State \Return $\hat{r}_{u_i,m_j}$, and $\boldsymbol{\alpha}$ (AFIN) or $\boldsymbol{\beta}$ (AFINca)
	\end{algorithmic}
\end{algorithm}

\begin{algorithm}[htbp]
	\caption{Rating-predictor training with frozen or fine-tuned CAAE encoders}
	\label{alg:ft}
	\begin{algorithmic}[1]
		\Require Pretrained encoders $g_{\phi_U}, g_{\phi_M}$; matrices $R_{\mathrm{tr}}, R_{\mathrm{va}}$; features $\mathbf{F}_U, \mathbf{F}_M$; pair sets $\Omega_{\mathrm{tr}}, \Omega_{\mathrm{va}}, \Omega_{\mathrm{te}}$; predictor $F(\cdot;\theta)$ (AFIN or AFINca); mode $\in\{\textsc{frozen},\textsc{finetune}\}$; learning rate $\eta$, encoder multiplier $\lambda$, epochs $N_{\mathrm{ep}}$, patience $\tau$
		\Ensure Trained $F(\cdot;\theta)$ (and encoders), test RMSE and MAE
		\If{mode $=$ \textsc{frozen}}
		\State $\mathbf{E}_U \gets g_{\phi_U}(R_{\mathrm{tr}}+R_{\mathrm{va}})$, \; $\mathbf{E}_M \gets g_{\phi_M}\big((R_{\mathrm{tr}}+R_{\mathrm{va}})^{\top}\big)$ \Comment{fixed, no gradient}
		\State Optimizer over $\theta$ with rate $\eta$
		\Else
		\State Optimizer over $\theta$ (rate $\eta$) and $\phi_U, \phi_M$ (rate $\lambda\eta$)
		\EndIf
		\For{epoch $=1$ \textbf{to} $N_{\mathrm{ep}}$}
		\ForAll{mini-batch $\{(u_i,m_j,r_{u_i,m_j})\}\subset\Omega_{\mathrm{tr}}$}
		\If{mode $=$ \textsc{finetune}}
		\State $\mathbf{E}_U \gets g_{\phi_U}(R_{\mathrm{tr}})$, \; $\mathbf{E}_M \gets g_{\phi_M}(R_{\mathrm{tr}}^{\top})$ \Comment{gradients enabled}
		\EndIf
		\State $\hat{r}_{u_i,m_j} \gets F(\mathbf{e}_{u_i}, \mathbf{e}_{m_j}, \mathbf{f}_{u_i}, \mathbf{f}_{m_j};\theta)$ \Comment{Eq.~\ref{eq:predictiontask}, Alg.~\ref{alg:forward}}
		\State Backpropagate MSE between $\hat{r}$ and $r$, clip gradient norm to 1, update
		\EndFor
		\State $\mathrm{RMSE}_{\mathrm{va}} \gets$ RMSE on $\Omega_{\mathrm{va}}$ \Comment{Eq.~\ref{eq:rmse}; eval mode, \textsc{finetune} uses $R_{\mathrm{tr}}$ as encoder input}
		\State Keep best checkpoint; stop if no improvement for $\tau$ epochs
		\EndFor
		\State Restore best checkpoint
		\If{mode $=$ \textsc{finetune}}
		\State $\mathbf{E}_U \gets g_{\phi_U}(R_{\mathrm{tr}}+R_{\mathrm{va}})$, \; $\mathbf{E}_M \gets g_{\phi_M}\big((R_{\mathrm{tr}}+R_{\mathrm{va}})^{\top}\big)$
		\EndIf
		\State \Return RMSE and MAE of $F(\cdot;\theta)$ on $\Omega_{\mathrm{te}}$ \Comment{Eqs.~\ref{eq:rmse}, \ref{eq:mae}}
	\end{algorithmic}
\end{algorithm}

\paragraph{Rating Prediction Forward pass}
\label{par:rpfp}
Algorithm~\ref{alg:forward} gives the forward pass of both models for a user-movie pair. Both compute the same four branches (Eq.~\ref{eq:branches}) and differ only in aggregation. AFIN weights the branches with the gate in Eq.~\ref{eq:afin-gate}. AFINca first applies self-attention over the branches and a context token (Eq.~\ref{eq:afinca-attn}), then pools the attended branches with the gate in Eq.~\ref{eq:afinca-gate}. In both cases the pooled vector $\mathbf{z}$ passes through the residual MLP, and the branch weights are returned for the analysis in Section~\ref{subsec:intpret}.

\paragraph{Rating Prediction with CAAE encoders (frozen and fine-tuned)}
\label{par:ftfrtrain}
Algorithm~\ref{alg:ft} describes how both predictors are trained in the frozen and fine-tuned settings. In the frozen setting, the embeddings are computed once from $R_{\mathrm{tr}}+R_{\mathrm{va}}$ and treated as fixed inputs. In the fine-tuned setting, the encoders are recomputed at every step from $R_{\mathrm{tr}}$ with gradients enabled, and their learning rate is scaled by $\lambda = 0.1$ relative to the predictor, since they start from a pretrained checkpoint. Model selection uses validation RMSE (Eq.~\ref{eq:rmse}) with early stopping in both settings. After the best checkpoint is restored, the fine-tuned embeddings are recomputed from $R_{\mathrm{tr}}+R_{\mathrm{va}}$ before test evaluation, matching the extraction convention of Algorithm~\ref{alg:caae}.

\subsection{Explainability}
\label{subsec:intpret}

AFIN's softmax gate gives a per-sample weight for each of the four interaction branches. AFINca, which applies cross-branch attention before pooling, gives a comparable per-sample weight vector through its pooling gate. We refer to both as branch weights. They are gate weights, not attention distributions over input tokens. A per-sample softmax over a small set of components resembles the gating of \citet{jacobs1991adaptive} and \citet{shazeer2017outrageously}. The attention-as-explanation debate of \citet{jain2019attention} and \citet{wiegreffe2019attention} therefore applies to them by analogy: both are normalized, input-dependent scores that are easily read as importance.

Throughout, $x=(u_i,m_j)\in\Omega_{\mathrm{te}}$ denotes a test pair, and $w_k(x)$ denotes the weight of branch $k\in\{1,\dots,4\}$ for that pair, with $w_k(x)=\alpha_k$ for AFIN (Eq.~\ref{eq:afin-gate}) and $w_k(x)=\beta_k$ for AFINca (Eq.~\ref{eq:afinca-gate}). In both models $w_k(x)\ge 0$ and $\sum_k w_k(x)=1$. The pooled vector is $\mathbf{z}(x)=\sum_{k=1}^{4} w_k(x)\,\bar{\mathbf{b}}_k(x)$, where $\bar{\mathbf{b}}_k=\mathbf{b}_k$ for AFIN and $\bar{\mathbf{b}}_k=\mathbf{b}'_k$ for AFINca. RMSE and MAE below are Eq.~\ref{eq:rmse} and~\ref{eq:mae} evaluated on the stated predictions. We analyze both models from five complementary angles:

\paragraph{Branch dominance}\label{par:branchdom} (\ref{subsubsec:resbranchdom}): For every test pair, the dominant branch is the one with the highest weight,
\begin{equation}
	k^\ast(x) = \arg\max_{k}\, w_k(x), \qquad
	\pi_k = \frac{1}{|\Omega_{\mathrm{te}}|}\sum_{x\in\Omega_{\mathrm{te}}} \mathbb{1}\big[k^\ast(x)=k\big],
	\label{eq:dominance}
\end{equation}
where $\pi_k$ is the fraction of test pairs on which branch $k$ is dominant. The counts $|\Omega_{\mathrm{te}}|\,\pi_k$ show which interaction type each model favors most often, for both the frozen and fine-tuned variants.

\paragraph{Branch weights}\label{par:branchwei} (\ref{subsubsec:resbranchweights}): The weight of each branch is averaged over the test set,
\begin{equation}
	\bar{w}_k = \frac{1}{|\Omega_{\mathrm{te}}|}\sum_{x\in\Omega_{\mathrm{te}}} w_k(x),
	\label{eq:meanweight}
\end{equation}
giving one mean weight per branch for each of the four model variants. Comparing frozen and fine-tuned variants shows how joint training redistributes weight across the four interactions. Dominance and mean weight can differ: a branch can win few pairs yet carry a sizable average weight, and the reverse.

\paragraph{Branch weights vs. user activity}\label{par:bwu} (\ref{subsubsec:resweightuser}): Let $n(x)$ be the number of training ratings of the user in pair $x$. Test pairs are sorted by $n(x)$ and split into four equal-size quartiles $\mathcal{B}_1,\dots,\mathcal{B}_4$, with $\mathcal{B}_1$ the coldest and $\mathcal{B}_4$ the warmest. The mean weight of branch $k$ in quartile $g$ is
\begin{equation}
	\bar{w}_k^{(g)} = \frac{1}{|\mathcal{B}_g|}\sum_{x\in \mathcal{B}_g} w_k(x).
	\label{eq:quartileweight}
\end{equation}
Because the bins are over test pairs, users with more ratings contribute more pairs. This tests whether the model's reliance on each branch changes with the amount of available user history.

\paragraph{Cold-start shift}\label{par:colds} (\ref{subsubsec:rescolsdstart}): For each branch, the shift is the difference between the coldest and warmest quartile means from Eq.~\ref{eq:quartileweight},
\begin{equation}
	\Delta_k^{\mathrm{cold}} = \bar{w}_k^{(1)} - \bar{w}_k^{(4)}.
	\label{eq:coldshift}
\end{equation}
A positive $\Delta_k^{\mathrm{cold}}$ means branch $k$ is weighted more heavily for users with little history, and a value near zero means the gate ignores activity. Every user has at least 20 ratings in the full dataset, so this is a low-activity proxy for the cold-start setting studied by \citet{schein2002methods}, not a test on unseen users.

\paragraph{Branch-zeroing ablation}\label{par:branchzero} (\ref{subsubsec:resbranchzeroing}): Following the intervention protocol of \citet{serrano2019attention}, who compare attention weights against the output change caused by altering them, we apply the test to branch-level gate weights. On a fixed, trained checkpoint, the post-softmax weight of branch $\ell$ is set to zero and the remaining three weights are renormalized to sum to one, with no retraining:
\begin{equation}
	w_k^{(-\ell)}(x) =
	\begin{cases}
		0, & k = \ell,\\[3pt]
		\dfrac{w_k(x)}{1 - w_\ell(x)}, & k \neq \ell,
	\end{cases}
	\qquad
	\mathbf{z}^{(-\ell)}(x) = \sum_{k=1}^{4} w_k^{(-\ell)}(x)\,\bar{\mathbf{b}}_k(x).
	\label{eq:zeroing}
\end{equation}
Softmax outputs are strictly below one, so the denominator is positive. Let $\hat{r}^{(-\ell)}_{x}$ be the prediction obtained by passing $\mathbf{z}^{(-\ell)}(x)$ through the unchanged residual MLP, and $\hat{r}_x$ the unmodified prediction. The importance of branch $\ell$ is the resulting degradation,
\begin{equation}
	\Delta\mathrm{RMSE}_\ell = \mathrm{RMSE}\big(\{\hat{r}^{(-\ell)}_x\}\big) - \mathrm{RMSE}\big(\{\hat{r}_x\}\big), \qquad
	\Delta\mathrm{RMSE}^{\%}_\ell = 100\,\frac{\Delta\mathrm{RMSE}_\ell}{\mathrm{RMSE}\big(\{\hat{r}_x\}\big)},
	\label{eq:zeroingdelta}
\end{equation}
both computed over $\Omega_{\mathrm{te}}$. $\Delta\mathrm{MAE}_\ell$ is defined identically with MAE in place of RMSE, and the ``$\Delta$RMSE (\%)'' column in Table~\ref{tab:branchzeroingabres} is $\Delta\mathrm{RMSE}^{\%}_\ell$. This quantity measures the predictive importance of the branch independent of how much weight it receives. Ablating components of a trained model without retraining follows \citet{michel2019sixteen}, and measuring the effect of occluding part of the input follows the general logic of \citet{zeiler2014visualizing}. For AFINca, $\bar{\mathbf b}_k=\mathbf b'_k$ are the tokens after self-attention, so zeroing $\beta_\ell$ removes only the pooling contribution of $\mathbf b'_\ell$. Branch $\ell$ may still influence the output through the other attended tokens, which already contain information from it.

\section{Results}
\label{sec:res}
\subsection{Configuration}\label{subsec:conf}
All experiments were implemented in Python 3.10.12 using PyTorch 2.12 \cite{paszke2019pytorch} (CUDA 13.0, cuDNN 9.20). Experiments were conducted on an NVIDIA GeForce RTX 4060 Laptop GPU (8GB VRAM, driver 595.99.02) and a 13th Gen Intel Core CPU, with both training and inference performed on the GPU. The user and movie CAAEs were trained for 1000 epochs using AdamW (learning rate $1\times10^{-3}$, weight decay $1\times10^{-5}$), input dropout of 0.5, and embedding dimension $d=128$. We trained AFIN and AFINca with RMSE loss using AdamW (learning rate $3\times10^{-4}$, weight decay 0.1), cosine learning-rate annealing, gradient-norm clipping at 1.0, a batch size of 16384, and up to 200 epochs. Early stopping based on validation RMSE was applied with a patience of 20. Both models utilise a dropout rate of 0.2. In the fine-tuned setting, the encoder learning rate was set to $0.1\times$ the predictor's learning rate. Validation RMSE was used to select checkpoints in all settings.

\subsection{Evaluation Metrics}\label{eval}
The objective of rating prediction is to minimize the discrepancy between predicted and observed ratings on data not used for parameter fitting. Let $\Omega' \subseteq \Omega$ denote the set of observed user-movie pairs on which a metric is computed and $N=|\Omega'|$. Unless stated otherwise, $\Omega'=\Omega_{\mathrm{te}}$, the held-out test set (Section~\ref{subsec:probform}); with $\Omega'=\Omega_{\mathrm{tr}}$ or $\Omega_{\mathrm{va}}$ the same definitions give the training and validation values used for the loss and for model selection. Two common measures of rating-prediction accuracy are root mean squared error ($\mathrm{RMSE}$), shown in Eq.~\ref{eq:rmse}, and mean absolute error ($\mathrm{MAE}$) shown in Eq.~\ref{eq:mae}. The $\mathrm{RMSE}$ is defined as:

\begin{equation}
	\mathrm{RMSE} = \sqrt{\frac{1}{N}\sum_{(u_i,m_j)\in\Omega'}\big(\hat{r}_{u_i,m_j}-r_{u_i,m_j}\big)^2},
	\label{eq:rmse}
\end{equation}
whereas $\mathrm{MAE}$ is defined as:
\begin{equation}
	\mathrm{MAE} = \frac{1}{N}\sum_{(u_i,m_j)\in\Omega'}\big|\hat{r}_{u_i,m_j}-r_{u_i,m_j}\big|.
	\label{eq:mae}
\end{equation}

$\mathrm{RMSE}$ penalizes larger prediction errors more strongly because the numbers are squared before averaging, while $\mathrm{MAE}$ measures their average absolute magnitude. Lower values indicate better rating-prediction accuracy under the corresponding metric. In this study, these metrics are used to compare model configurations on the MovieLens datasets. All models are evaluated on the same data split and the same set of observed target ratings.

\subsection{Experimental Setup}
\label{subsec:setup}
Both CAAE models (users and movies) are trained full-batch for 1000 epochs with AdamW (learning rate $10^{-3}$, weight decay $10^{-5}$), input dropout 0.5, and $d=128$. The embeddings were extracted form the model's checkpoint with the lowest validation RMSE. The rating predictors minimize RMSE with AdamW and cosine learning-rate annealing, gradient-norm clipping at 1.0, batch size 16384, and at most 200 epochs, with early stopping on validation RMSE (patience 20). AFIN uses an interaction dimension of 1536. AFINca uses interaction dimension 256 with 8 heads, one attention layer, and the context token, so the two models are not matched in parameter count. Both use a residual MLP with hidden width 128, five residual blocks, dropout 0.2, and weight decay 0.1. In the fine-tuned setting, the encoders use a learning rate $0.1\times$ that of the predictor (Algorithm~\ref{alg:ft}).

The reported test RMSE and MAE are computed on the test set using the model from checkpoint with the lowest validation RMSE, and MAE is not optimized separately. Each configuration was trained with 11 random seeds to ensure stable performance and reproducibility. 

Two details of the encoder inputs matter when comparing the settings. In the frozen setting, embeddings are computed once from $R_{\mathrm{tr}}+R_{\mathrm{va}}$, so validation ratings are part of the encoder input when validation RMSE is used for model selection. In the fine-tuned setting, validation uses $R_{\mathrm{tr}}$ only. In both settings, a training pair's own rating is part of its encoder input row during training, which is not the case at test time. To prevent data leakage, test ratings are never used as encoder input.

\subsection{Model Results}

\begin{table}[htbp]
	\centering
	\small
	\caption{Test RMSE and MAE (mean over 11 random seeds), lower is better, of AFIN and AFINca on MovieLens 100K (u1 split) and MovieLens 1M, with frozen and fine-tuned CAAE encoders. Bold marks the best result in each column, with ties in bold. Per-seed statistics are in Table~\ref{tab:seedres}.}
	\label{tab:modelsresults}
	\begin{tabular*}{\textwidth}{@{\extracolsep{\fill}} l rr rr}
		\toprule
		\multirow{2}{*}{\bfseries Model} & \multicolumn{2}{c}{\bfseries MovieLens 100K (u1 split)} & \multicolumn{2}{c}{\bfseries MovieLens 1M} \\
		\cmidrule(lr){2-3} \cmidrule(lr){4-5}
		& {\bfseries RMSE ($\downarrow$)} & {\bfseries MAE ($\downarrow$)} & {\bfseries RMSE ($\downarrow$)} & {\bfseries MAE ($\downarrow$)} \\
		\midrule
		AFIN frozen       & 0.9176 & 0.7211 & 0.8579 & 0.6749 \\
		AFIN fine-tuned   & 0.9109 & 0.7151 & \bfseries 0.8513 & \bfseries 0.6685 \\
		AFINca frozen     & 0.9230 & 0.7235 & 0.8589 & 0.6759 \\
		AFINca fine-tuned & \bfseries 0.9105 & \bfseries 0.7148 & \bfseries 0.8513 & 0.6693 \\
		\bottomrule
	\end{tabular*}
\end{table}

Table~\ref{tab:modelsresults} presents the mean test RMSE and MAE over 11 seeds for both architectures and embedding types. Fine-tuned AFIN and fine-tuned AFINca reach the lowest mean RMSE: 0.8513 for both on ML-1M, and 0.9109 and 0.9105, respectively, on ML-100K. Fine-tuning improves AFIN by 0.0066 RMSE on ML-1M and 0.0067 on ML-100K, with MAE improving by 0.0064 and 0.0060, respectively. These results indicate that enabling the rating loss to update the encoder yields embeddings more suitable for prediction than relying solely on reconstruction pretraining.

\begin{figure}[tbp]
	\centering
	\begin{subfigure}[t]{0.49\textwidth}
		\includegraphics[width=\textwidth]{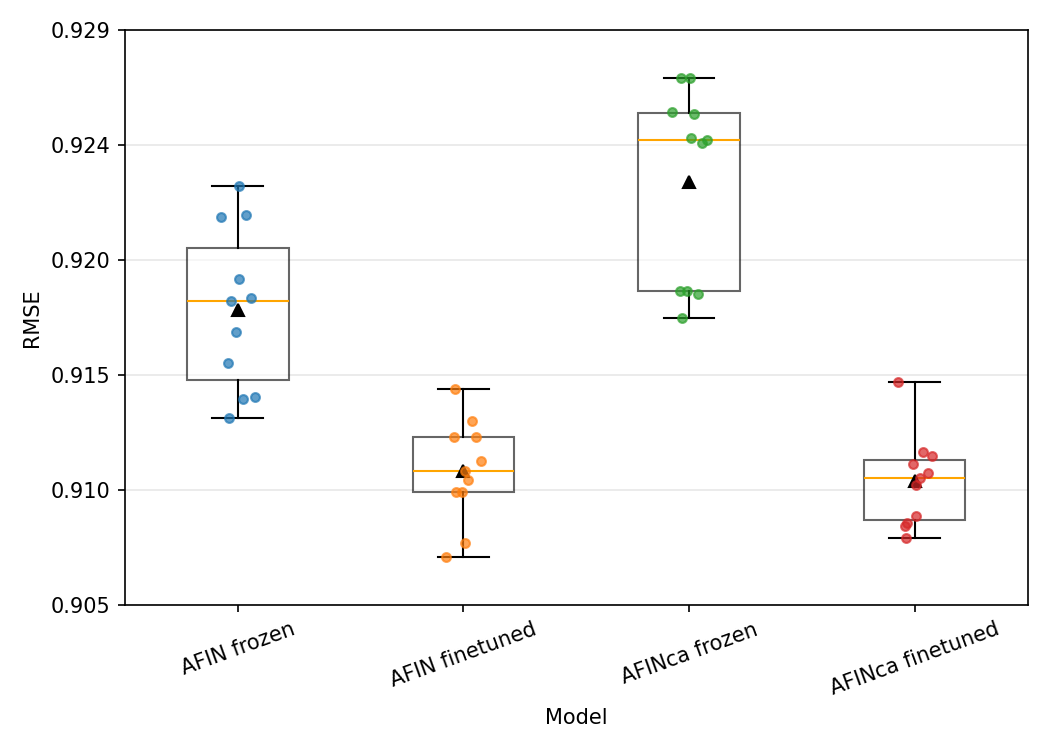}
		\caption{ML-100K: test RMSE}
		\label{subfig:testml100k}
	\end{subfigure}
	\hfill
	\begin{subfigure}[t]{0.49\textwidth}
		\includegraphics[width=\textwidth]{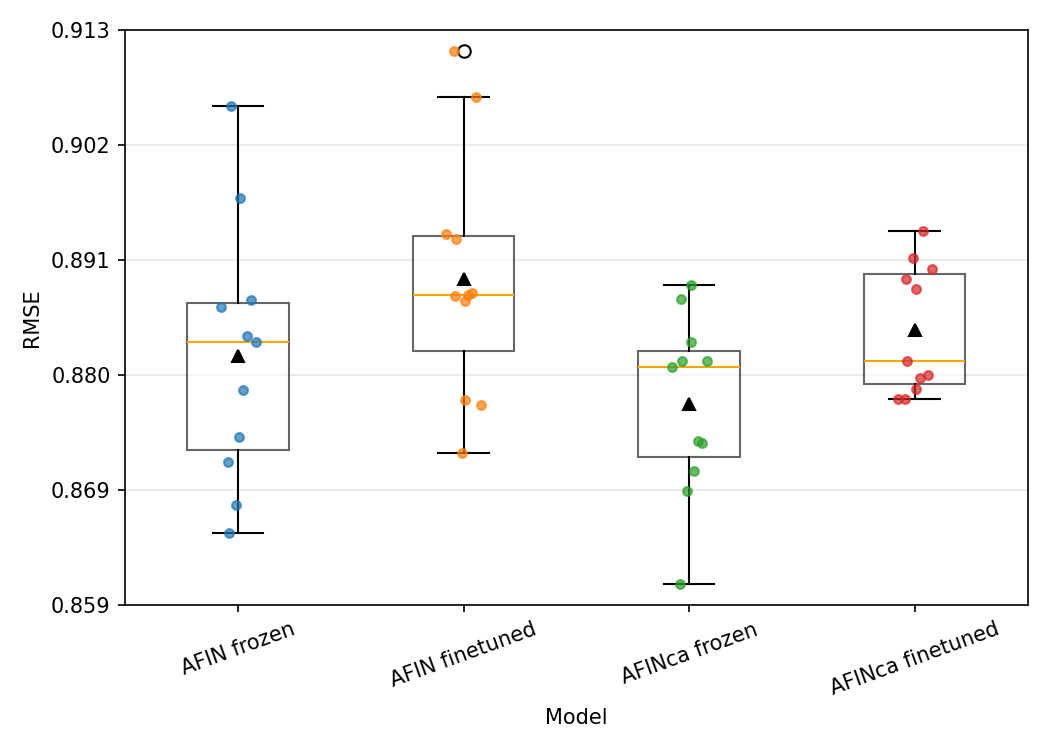}
		\caption{ML-100K: validation RMSE}
		\label{subfig:valml100k}
	\end{subfigure}
	\hfill
	\begin{subfigure}[t]{0.49\textwidth}
		\includegraphics[width=\textwidth]{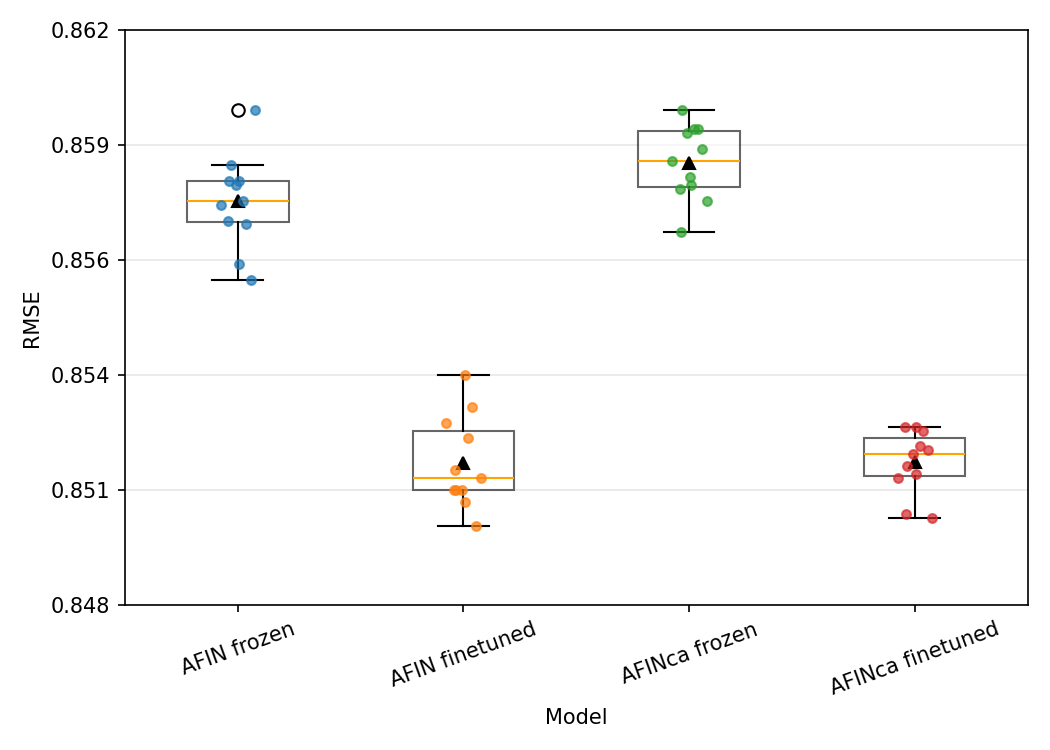}
		\caption{ML-1M: test RMSE}
		\label{subfig:testml1m}
	\end{subfigure}
	\hfill
	\begin{subfigure}[t]{0.49\textwidth}
		\includegraphics[width=\textwidth]{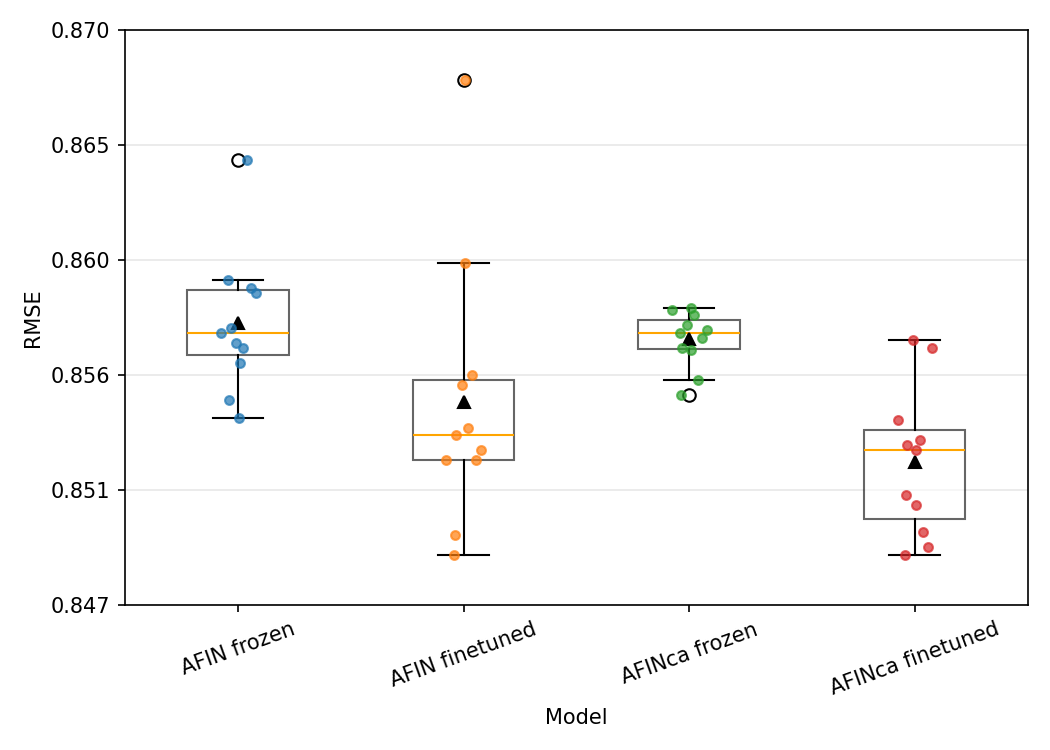}
		\caption{ML-1M: validation RMSE}
		\label{subfig:valml1m}
	\end{subfigure}
	\caption{Test and validation RMSE across 11 random seeds for each model variant on MovieLens 100K (panels a and b) and MovieLens 1M (panels c and d). Dots are individual seeds, boxes span the interquartile range with the median as a line and the mean as a triangle, whiskers follow the 1.5 interquartile-range rule, and circles mark outliers.}
	\label{fig:seedbox}
\end{figure}

AFINca also benefits from fine-tuning, with a larger improvement on ML-100K (0.0125 RMSE) than on ML-1M (0.0076 RMSE). With frozen embeddings, AFIN outperforms AFINca by 0.0010 RMSE on ML-1M and 0.0054 on ML-100K. After fine-tuning, the two are tied on ML-1M and AFINca is ahead by 0.0004 on ML-100K. Incorporating cross-branch attention therefore does not improve accuracy over the simpler branch-level gate. Table~\ref{tab:seedres} and Fig.~\ref{fig:seedbox} show that the fine-tuning gains exceed the seed standard deviations (0.0008 to 0.0040), whereas the differences between the two architectures after fine-tuning are well within them. We therefore read the results as showing that fine-tuning helps both architectures and that cross-branch attention gives no gain.

\begin{table}[htbp]
	\centering
	\small
	\caption{Test RMSE and MAE over 11 random seeds (mean, standard deviation, minimum, and maximum) of AFIN and AFINca with frozen and fine-tuned CAAE encoders on MovieLens 100K (u1 split) and MovieLens 1M.}
	\label{tab:seedres}
	\begin{tabular*}{\textwidth}{@{\extracolsep{\fill}} l rrrr rrrr}
		\toprule
		\multirow{2}{*}{\bfseries Model} & \multicolumn{4}{c}{\bfseries MovieLens 100K (u1 split)} & \multicolumn{4}{c}{\bfseries MovieLens 1M} \\
		\cmidrule(lr){2-5} \cmidrule(lr){6-9}
		& Mean   & Std    & Min    & Max    & Mean   & Std    & Min    & Max    \\ \midrule
		\multicolumn{9}{l}{\textbf{RMSE}} \\
		\midrule
		AFIN finetuned         & 0.9109 & 0.0021 & 0.9073 & 0.9143 & 0.8513 & 0.0011 & 0.8497 & 0.8535 \\
		AFIN frozen            & 0.9176 & 0.0033 & 0.9131 & 0.9228 & 0.8579 & 0.0012 & 0.8559 & 0.8602 \\
		AFINca finetuned       & 0.9105 & 0.0019 & 0.9081 & 0.9146 & 0.8513 & 0.0008 & 0.8499 & 0.8522 \\
		AFINca frozen          & 0.9230 & 0.0040 & 0.9173 & 0.9273 & 0.8589 & 0.0009 & 0.8571 & 0.8602 \\
		\midrule
		\multicolumn{9}{l}{\textbf{MAE}} \\
		\midrule
		AFIN finetuned         & 0.7151 & 0.0023 & 0.7107 & 0.7183 & 0.6685 & 0.0017 & 0.6657 & 0.6715 \\
		AFIN frozen            & 0.7211 & 0.0044 & 0.7155 & 0.7300 & 0.6749 & 0.0018 & 0.6720 & 0.6779 \\
		AFINca finetuned       & 0.7148 & 0.0028 & 0.7113 & 0.7197 & 0.6693 & 0.0016 & 0.6658 & 0.6717 \\
		AFINca frozen          & 0.7235 & 0.0028 & 0.7201 & 0.7289 & 0.6759 & 0.0013 & 0.6737 & 0.6782 \\
		\bottomrule
	\end{tabular*}
\end{table}

Table~\ref{tab:seedres} reports the mean, standard deviation, minimum, and maximum of test RMSE and MAE over 11 random seeds for each model variant, and Fig.~\ref{fig:seedbox} plots the corresponding RMSE distributions. In the figure, panels (a) and (c) show test RMSE on ML-100K and ML-1M, and panels (b) and (d) show the final validation RMSE. Each dot is one seed, each box spans the interquartile range with the median as a line and the mean as a triangle, and circles mark outliers.

The performance of all four variants is stable across seeds. The standard deviation of test RMSE is between 0.0019 and 0.0040 on ML-100K and between 0.0008 and 0.0012 on ML-1M, which is below 0.5\% and 0.15\% of the respective means. The spread between the best and worst seed is at most 0.0100 RMSE on ML-100K and 0.0043 on ML-1M, and MAE shows the same behavior, with standard deviations of at most 0.0044 and 0.0018. The ordering of the settings also holds across seeds. On ML-1M, the worst fine-tuned seed (0.8535 for AFIN, 0.8522 for AFINca) is better than the best frozen seed (0.8559 and 0.8571), and the same holds for AFINca on ML-100K (0.9146 against 0.9173). The only overlap between a fine-tuned and a frozen variant of the same architecture is AFIN on ML-100K, where the ranges meet at 0.9131 to 0.9143. Fine-tuned variants also vary less than frozen ones on ML-100K, in the figure as narrower boxes and in the table as smaller standard deviations. The differences between AFIN and AFINca after fine-tuning (0.0000 on ML-1M and 0.0004 on ML-100K) are smaller than the standard deviation of either model, so the boxes of the two architectures overlap almost completely in panels (a) and (c).

\subsection{Comparative Results}
\label{subsec:comres}

\begin{table}[htbp]
	\centering
	\caption{Comparison of AFIN with fine-tuned embeddings (lowest-RMSE run, with the MAE of the same run) against published baselines on MovieLens 100K and MovieLens 1M (lower is better). Baseline values are taken from the cited publications; a dash indicates that the metric was not reported for that dataset.}
	\label{tab:basecompresults}
	\begin{tabular*}{\textwidth}{@{\extracolsep{\fill}} l rr rr}
		\toprule
		\multirow{2}{*}{\bfseries Method} & \multicolumn{2}{c}{\bfseries MovieLens 100K} & \multicolumn{2}{c}{\bfseries MovieLens 1M} \\
		\cmidrule(lr){2-3} \cmidrule(lr){4-5}
		& {\bfseries RMSE ($\downarrow$)} & {\bfseries MAE ($\downarrow$)} & {\bfseries RMSE ($\downarrow$)} & {\bfseries MAE ($\downarrow$)} \\
		\midrule
		HGAR \citet{qian2020} & 0.9210 & 0.7170 & 0.8560 & 0.6680  \\
		$\Gamma$UICF \citet{aljunid2021} & 0.9229 & 0.7219 & 0.8506 & \textbf{0.6627}  \\
		EBCR \citet{du2021} & 0.9413 & 0.7348 & 0.9061 & 0.7052  \\
		PPPCF \citet{cheng2023} & -- & -- & 0.9207 & 0.7169  \\
		Time-based CF \citet{jain2023} & 1.0295 & 0.7481 & -- & --  \\
		ExtKNNCF \citet{nguyen2023} & 1.0370 & 0.8100 & 0.8530 & 0.7640  \\
		UASim \citet{belmessous2024} & 0.9484 & 0.7447 & 0.9372 & 0.7320  \\
		RBM+KNN \citet{alam2025} & -- & -- & 0.8736 & 0.6854  \\
		SVD \citet{hafeez2026} & -- & -- & 0.9397 & 0.7476  \\
		BayesNCF-v2 \citet{wang2026} & 0.9245 & 0.7249 & 0.8759 & 0.6853  \\
		\midrule
		\bfseries AFIN fine-tuned  &  \bfseries 0.9073 & \bfseries 0.7107 & \bfseries 0.8497 & 0.6657 \\
		\bottomrule
	\end{tabular*}
\end{table}

Table~\ref{tab:basecompresults} presents the performance of AFIN with fine-tuned embeddings, taking its lowest-RMSE run, in comparison to ten published methods drawn from the neighborhood-based, factorization-based, and hybrid collaborative filtering literature. The neural and attention-based methods of Section~\ref{sec:rel} are not included in this comparison. On MovieLens 100K, this run achieves the lowest RMSE (0.9073) and MAE (0.7107), outperforming the strongest baseline, HGAR, by 0.0137 RMSE and 0.0063 MAE. On MovieLens 1M, AFIN also attains the lowest RMSE (0.8497), although the margin over $\Gamma$UICF (0.8506) is only 0.0009. For MAE, AFIN (0.6657) ranks second to $\Gamma$UICF (0.6627) and ahead of HGAR (0.6680). Thus, AFIN leads on three out of four metrics. Methods employing neighborhood or similarity heuristics (EBCR, PPPCF, UASim, SVD) perform less effectively, with RMSE values exceeding 0.90 on ML-1M. ExtKNNCF reports a competitive ML-1M RMSE (0.8530) but a substantially higher MAE (0.7640). Its MAE-to-RMSE ratio is unusually high compared with the other methods, which may reflect a different evaluation protocol, so we do not draw conclusions from it. As the baseline results are sourced from the original publications, this comparison is indicative rather than definitive. Variations in data splits and preprocessing across studies, as well as the small ML-1M differences to the best baselines, suggest that the improvement should not be considered decisive.

\subsubsection{Baseline Methods}
\label{subsubsec:baselines}

\begin{itemize}
	\item \textbf{RBM+KNN \citet{alam2025}:} This method integrates a Restricted Boltzmann Machine (RBM) with k-Nearest Neighbors (kNN). Unrated items are assigned a value of zero in the rating matrix, and the RBM is trained using contrastive divergence. Cosine similarity identifies each user's neighbors, and missing ratings are inferred from these neighbors.
	\item \textbf{$\Gamma$UICF \citet{aljunid2021}:} This approach integrates user-based and item-based collaborative filtering within a single model. Similarities are computed offline using cosine similarity or the Pearson correlation coefficient (PCC) and restricted to the Top-K neighbors. For online prediction, linear regression combines the two predictions, with the parameter $\Gamma$ determining their respective contributions.
	\item \textbf{UASim \citet{belmessous2024}:} This method models user preferences using binomial subjective logic, decomposing each user pair into similarity, dissimilarity, and uncertainty based on co-rated items. Neighbor selection is formulated as a multi-criteria decision problem and ranked using the TOPSIS method. Predictions are generated through weighted averages.
	\item \textbf{PPPCF \citet{cheng2023}:} This privacy-preserving collaborative filtering method enables each user's device to label ratings as sensitive or weakly sensitive, randomizing only the sensitive ratings. The server applies iterative Bayesian updates to estimate item-item joint distributions, computes item similarities, and generates recommendations.
	\item \textbf{EBCR \citet{du2021}:} This method enhances the similarity measure in memory-based collaborative filtering. Ratings are categorized as like, neutral, or dislike to construct a concordance ratio. Empirical Bayes smoothing penalizes user pairs with limited co-rated items prior to KNN-based prediction.
	\item \textbf{SVD \citet{hafeez2026}:} This study compares two types of collaborative filtering. User-based filtering employs cosine and Euclidean similarity with weighted-average predictions. Model-based filtering factorizes the rating matrix using Non-negative Matrix Factorization (NMF) and Singular Value Decomposition (SVD). SVD demonstrates the best performance.
	\item \textbf{Time-based CF \citet{jain2023}:} This method incorporates temporal decay into memory-based collaborative filtering. Older interactions receive reduced weight through exponential or power decay, applied during similarity calculation or prediction. Similarity is measured using cosine similarity, Pearson correlation coefficient (PCC), or Gower’s coefficient, followed by KNN-based prediction.
	\item \textbf{ExtKNNCF \citet{nguyen2023}:} This approach integrates adaptive KNN with user cognition vectors, which are binary indicators of user interactions. A cognitive threshold and weight, determined by Euclidean distance and cosine similarity, establish an adaptive distance metric. The value of K varies between minimum and maximum limits according to local data density and user feedback.
	\item \textbf{HGAR \citet{qian2020}:} This method integrates explicit and implicit feedback using granular computing. Singular Value Decomposition (SVD) processes explicit ratings, while a Multi-Layer Perceptron (MLP) models implicit preferences. The two components are merged using sum pooling and jointly trained under a unified loss function.
	\item \textbf{BayesNCF-v2 \citet{wang2026}:} This model integrates deterministic user and item embedding tables with a Multi-Layer Perceptron (MLP) feature extractor. A Bayesian linear layer, constructed using Bayes-by-Backprop, predicts the mean rating, while a separate output predicts the log-variance. The model is trained end-to-end with Gaussian negative log-likelihood (NLL) and a Kullback-Leibler (KL) regularizer.
\end{itemize}

\subsection{Explainability Results and Analysis}
\label{subsec:intres}

\subsubsection{Branch dominance}
\label{subsubsec:resbranchdom}
\begin{figure}[tbp]
	\centering
	\begin{subfigure}[t]{0.49\textwidth}
		\includegraphics[width=\textwidth]{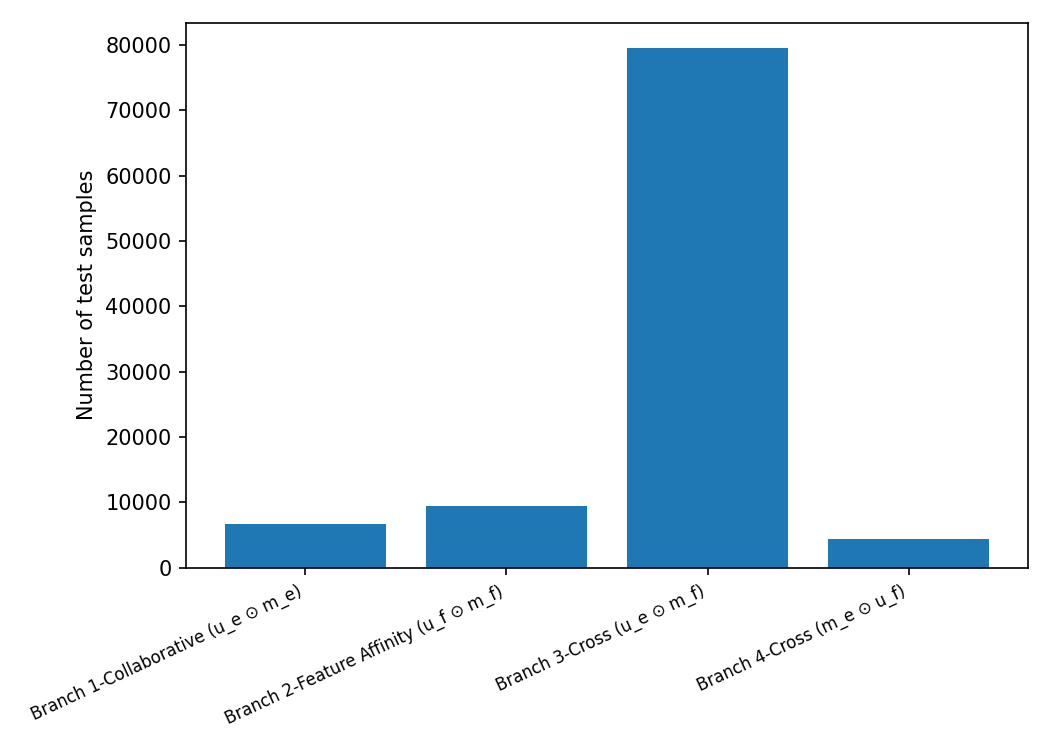}
		\caption{AFIN frozen}
		\label{subfig:afinbd}
	\end{subfigure}
	\hfill
	\begin{subfigure}[t]{0.49\textwidth}
		\includegraphics[width=\textwidth]{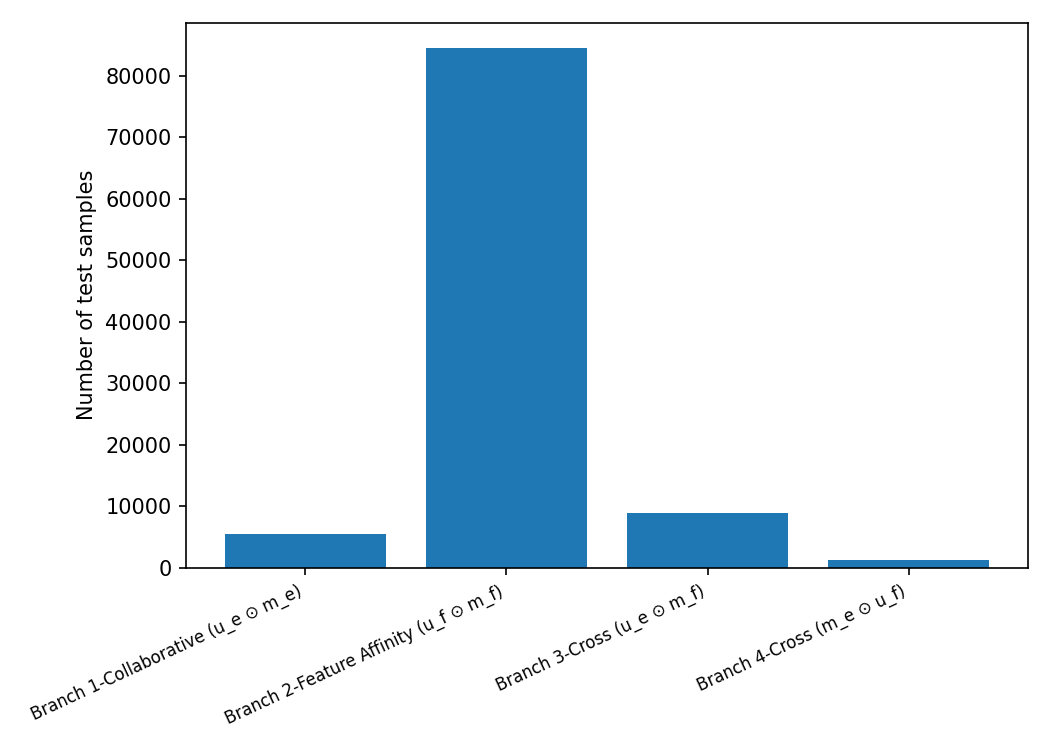}
		\caption{AFIN fine-tuned}
		\label{subfig:afinftbd}
	\end{subfigure}
	\hfill
	\begin{subfigure}[t]{0.49\textwidth}
		\includegraphics[width=\textwidth]{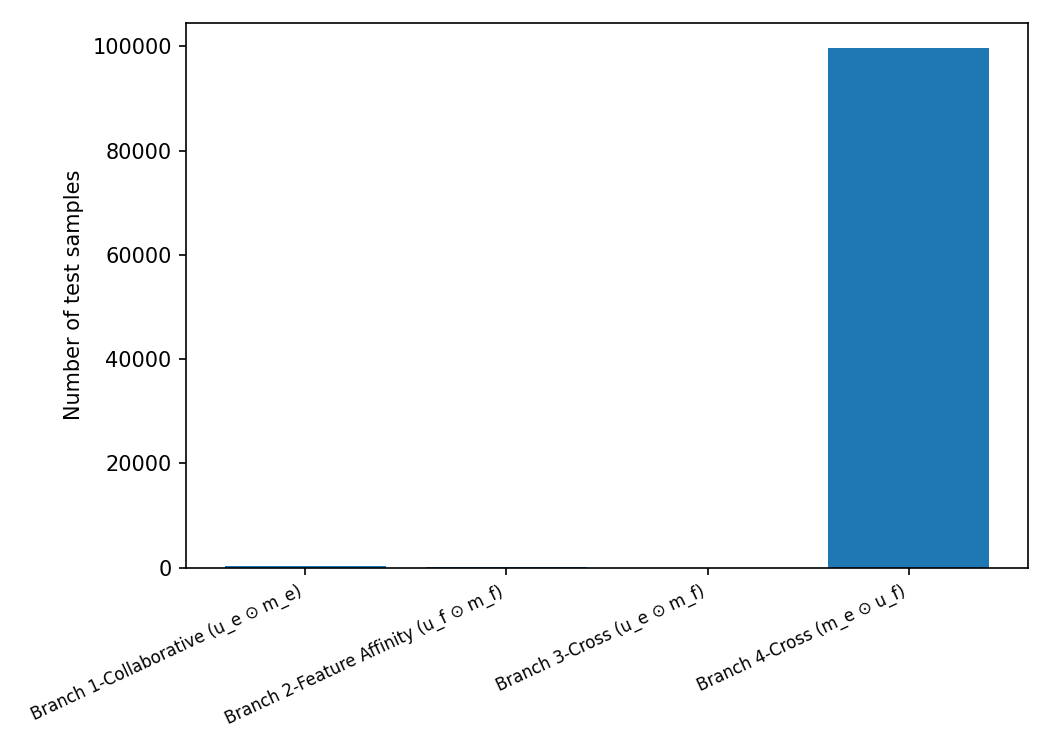}
		\caption{AFINca frozen}
		\label{subfig:afincabd}
	\end{subfigure}
	\hfill
	\begin{subfigure}[t]{0.49\textwidth}
		\includegraphics[width=\textwidth]{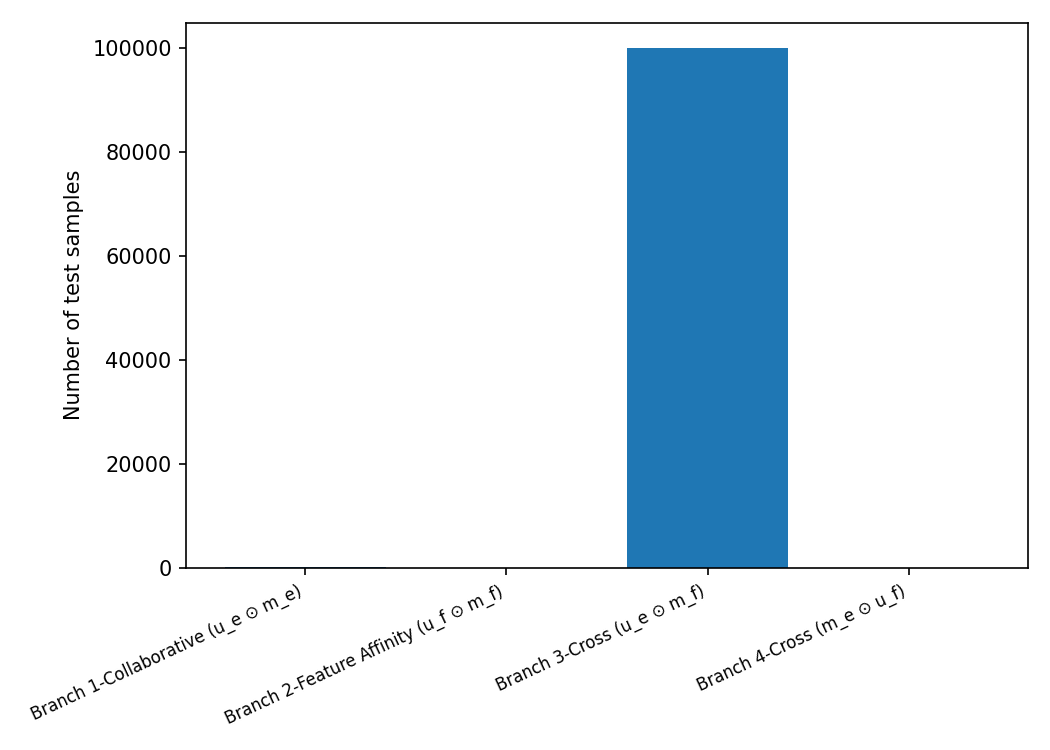}
		\caption{AFINca fine-tuned}
		\label{subfig:afincaftbd}
	\end{subfigure}
	\caption{Branch dominance histogram on the test set: number of test pairs for which each branch has the highest weight, for AFIN and AFINca with frozen and fine-tuned encoders.}
	\label{fig:branchdominance}
\end{figure}

When the encoder is frozen, branch 3 of AFIN (cross, $\mathbf{e}_{u_i} \odot \mathbf{f}_{m_j}$) is dominant for the large majority of test samples (about 80k of 100k), while branches 1, 2, and 4 each win fewer than 10k (Fig.~\ref{subfig:afinbd}). Fine-tuning the encoder moves the majority to branch 2 (feature affinity), which then wins about 85k samples. The count for branch 1 changes little, whereas branch 4 falls to about 1.5k (Fig.~\ref{subfig:afinftbd}). Joint training therefore changes not only the relative strength of each branch but also which branch receives the largest weight.

AFINca shows a more extreme pattern. With the encoder frozen, branch 4 (cross, $\mathbf{e}_{m_j} \odot \mathbf{f}_{u_i}$) is dominant for nearly every test sample, and branches 1 to 3 almost never win (Fig.~\ref{subfig:afincabd}). Fine-tuning does not balance this: branch 3 becomes the dominant branch for nearly every sample, and branch 4 falls to almost zero (Fig.~\ref{subfig:afincaftbd}). AFIN also concentrates on one branch after fine-tuning, but AFINca is more extreme, and fine-tuning only changes which branch it favors.

\subsubsection{Branch weights}
\label{subsubsec:resbranchweights}
\begin{figure}[tbp]
	\centering
	\begin{subfigure}[t]{0.49\textwidth}
		\includegraphics[width=\textwidth]{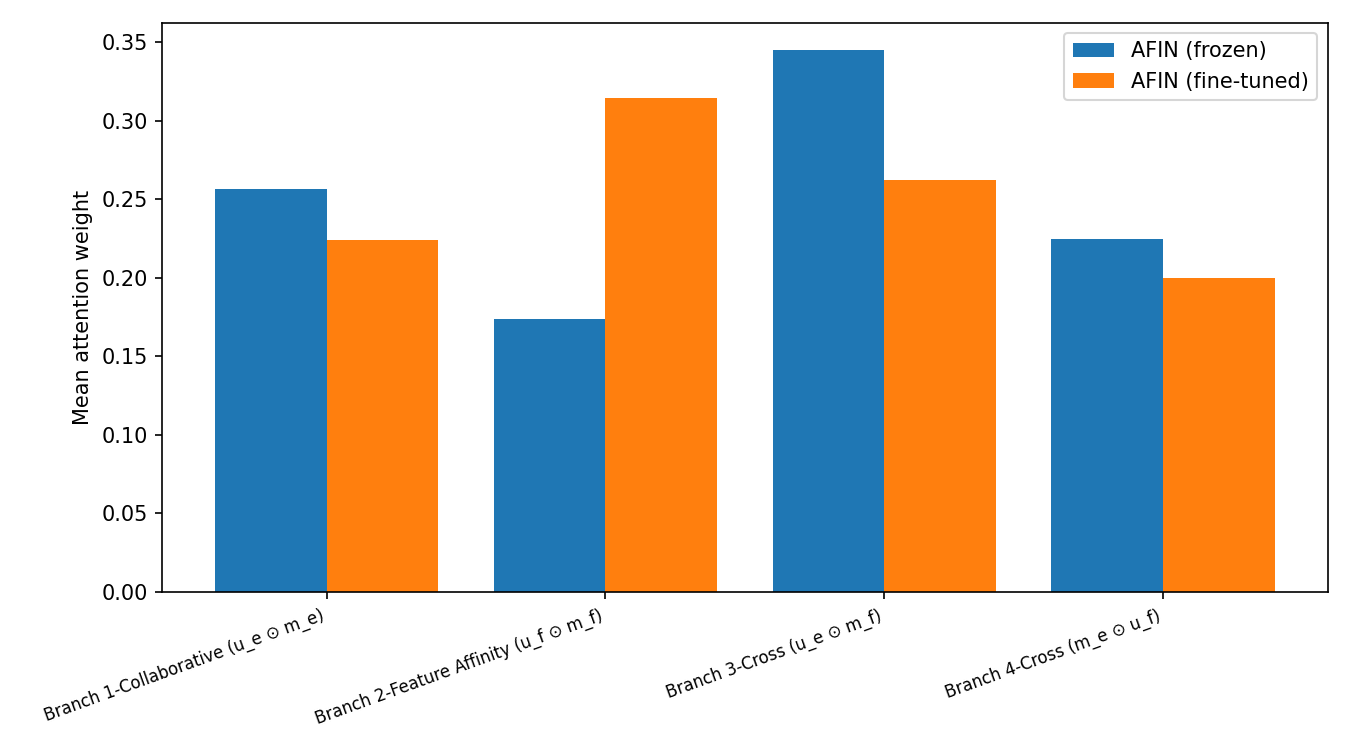}
		\caption{AFIN frozen vs. AFIN fine-tuned }
		\label{subfig:afinfrftbw}
	\end{subfigure}
	\hfill
	\begin{subfigure}[t]{0.49\textwidth}
		\includegraphics[width=\textwidth]{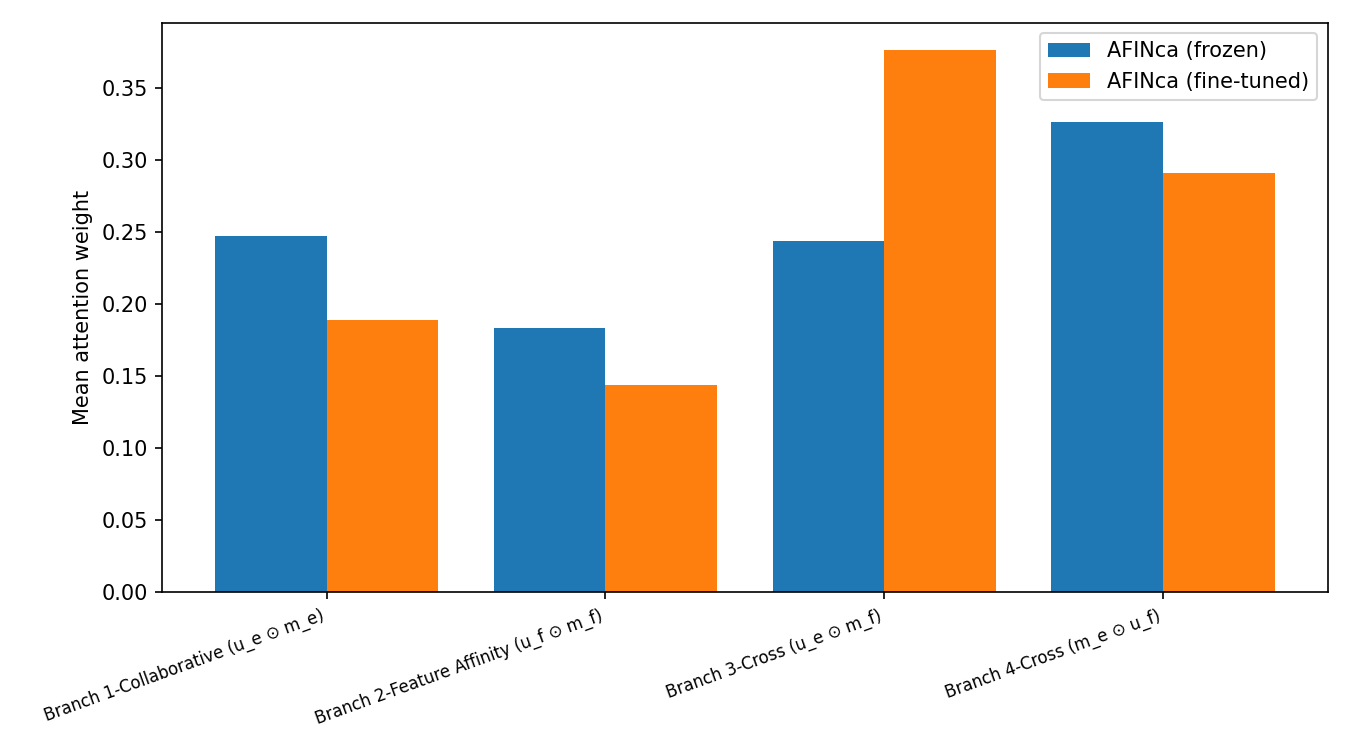}
		\caption{AFINca frozen vs. AFINca fine-tuned}
		\label{subfig:afincafrftbw}
	\end{subfigure}
	\caption{Mean branch weights on the test set, comparing frozen and fine-tuned encoders for AFIN (a) and AFINca (b).}
	\label{fig:branchweight}
\end{figure}

Mean weights show smaller differences than dominance counts, but they remain evident. When AFIN is frozen, branch 3 has the highest average weight (approximately 0.34), followed by branch 1 (0.26), branch 4 (0.22), and branch 2 (0.17). After fine-tuning, branch 2 has the highest mean weight (0.31), branch 3 decreases to 0.26, and branches 1 and 4 decline slightly (Fig.~\ref{subfig:afinfrftbw}). Branch 1's mean weight changes only from 0.256 to 0.223, although the ablation below shows that the model relies on this branch most. The distinction between weight and importance is the main focus of this analysis.

The weights in AFINca are more evenly distributed than the dominance counts suggest. When frozen, branch 4 has the highest mean weight (approximately 0.33), followed by branch 1 (0.25), branch 3 (0.24), and branch 2 (0.18). These values are more closely grouped than in frozen AFIN, even though branch 4 wins nearly every sample. After fine-tuning, branch 3 increases to approximately 0.38, a gain of about 0.13 that is comparable to the largest shift in AFIN (branch 2, from 0.17 to 0.31). Branch 4 decreases to about 0.29, and branches 1 and 2 decline (Fig.~\ref{subfig:afincafrftbw}). The shift in mean weight is thus present but smaller than the shift in dominance. The softmax selects a distinct winner for each sample even though the mean weights are close to one another.

\subsubsection{Weight vs. user activity}
\label{subsubsec:resweightuser}
\begin{figure}[tbp]
	\centering
	\begin{subfigure}[t]{0.49\textwidth}
		\includegraphics[width=\textwidth]{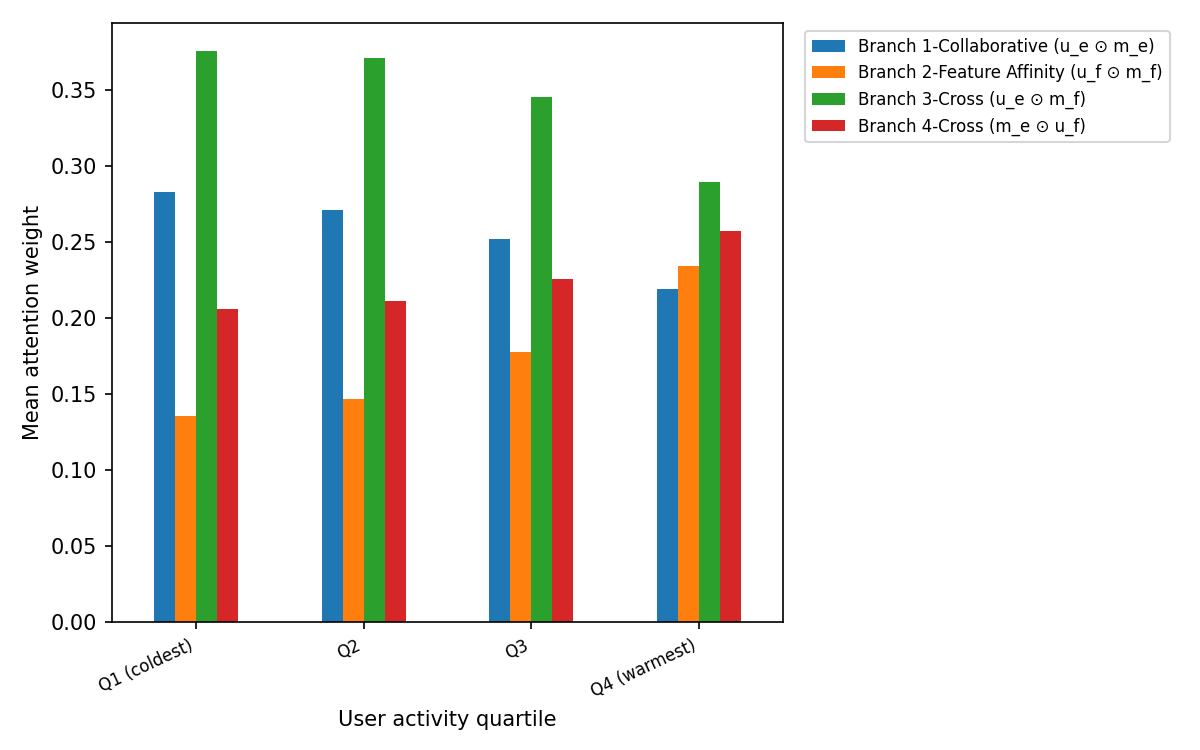}
		\caption{AFIN frozen}
		\label{subfig:afinwu}
	\end{subfigure}
	\hfill
	\begin{subfigure}[t]{0.49\textwidth}
		\includegraphics[width=\textwidth]{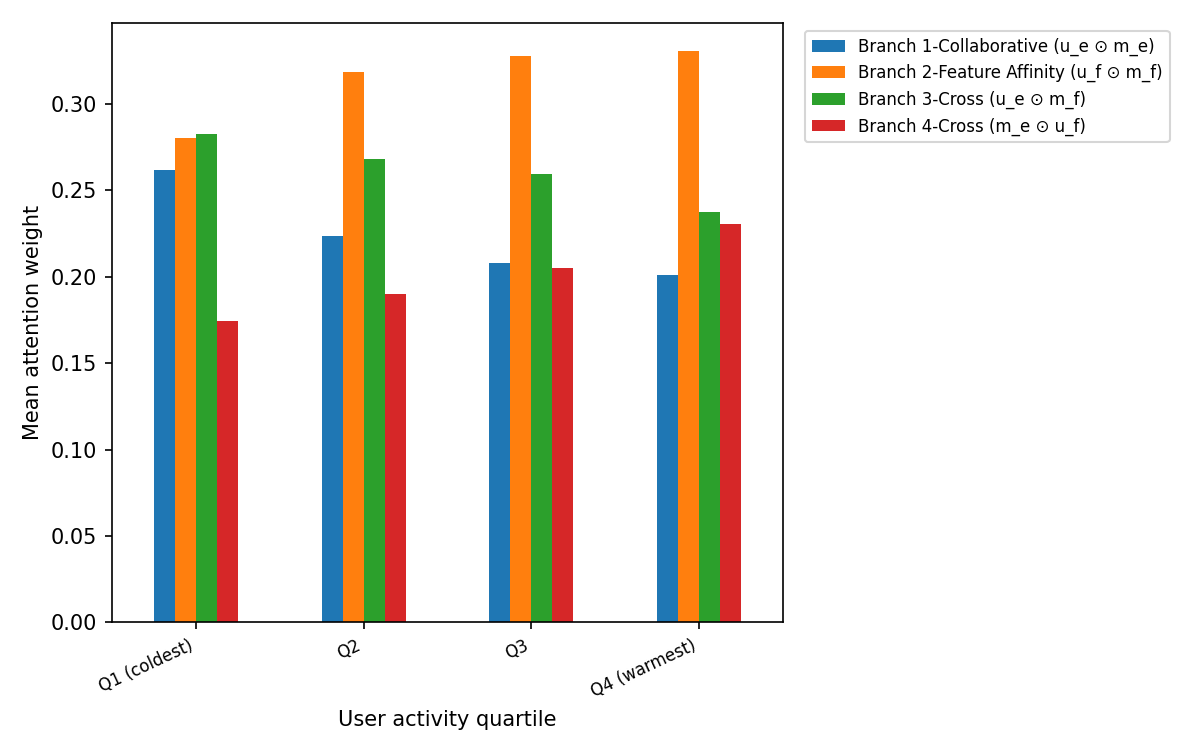}
		\caption{AFIN fine-tuned}
		\label{subfig:afinftwu}
	\end{subfigure}
	\hfill
	\begin{subfigure}[t]{0.49\textwidth}
		\includegraphics[width=\textwidth]{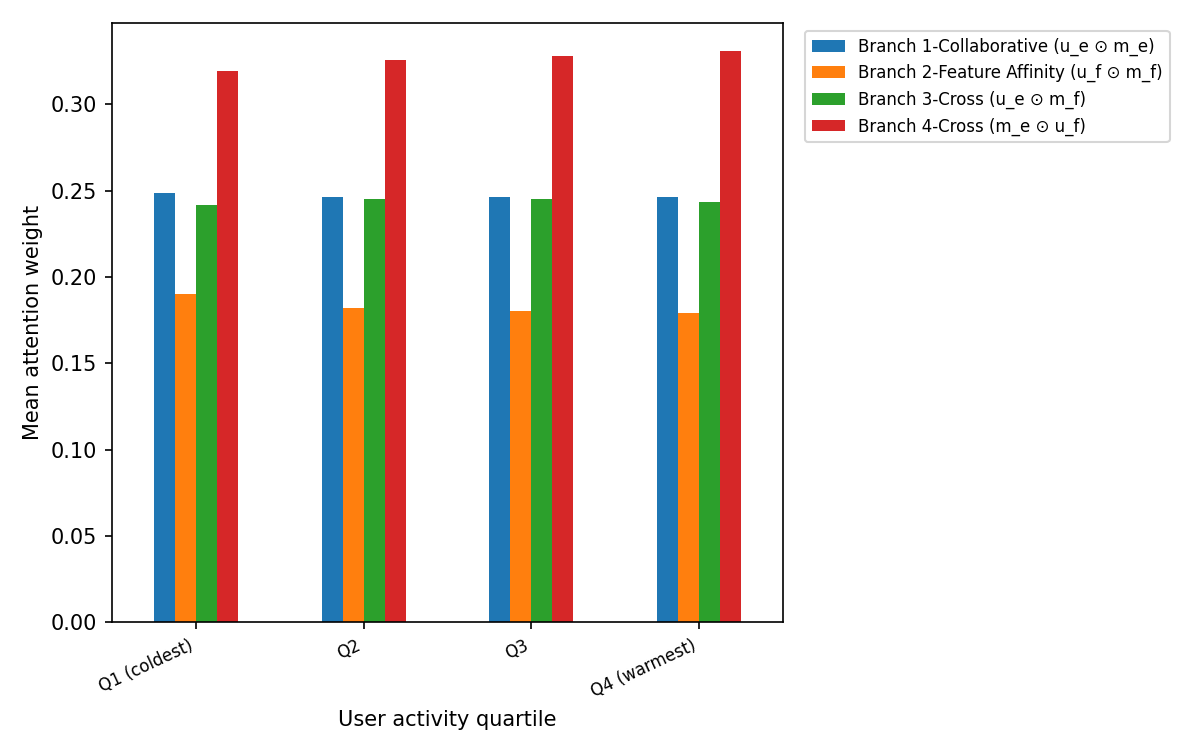}
		\caption{AFINca frozen}
		\label{subfig:afincawu}
	\end{subfigure}
	\hfill
	\begin{subfigure}[t]{0.49\textwidth}
		\includegraphics[width=\textwidth]{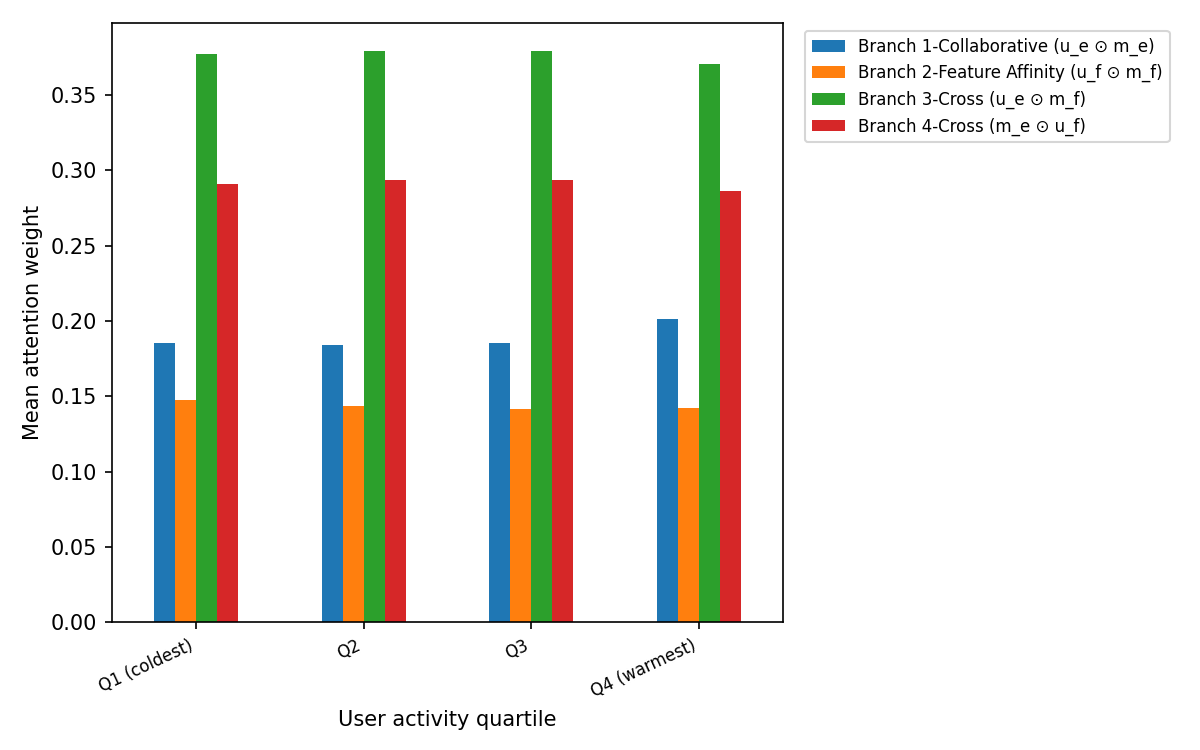}
		\caption{AFINca fine-tuned}
		\label{subfig:afincaftwu}
	\end{subfigure}
	\caption{Mean branch weight by user-activity quartile (coldest to warmest) for AFIN and AFINca with frozen and fine-tuned encoders.}
	\label{fig:weightuser}
\end{figure}

Mean branch weight by user activity quartile shows a distinct pattern in frozen AFIN. Branch 1 (collaborative) receives more weight for cold users and less for warm users. This runs against collaborative filtering theory, which predicts that the collaborative signal becomes more influential as a user accumulates ratings (Fig.~\ref{subfig:afinwu}). Branch 2 shows the opposite trend, with its weight rising as user activity rises. Fine-tuning changes the leading branch: branch 2 matches branch 3 in the coldest quartile and leads from the second quartile onward. The general trends for branches 1 and 2 persist (Fig.~\ref{subfig:afinftwu}), so the activity-dependent reweighting remains after the move from frozen to fine-tuned embeddings.

AFINca does not show this pattern. Across all four activity quartiles, the weights of frozen AFINca stay nearly constant: branch 4 remains around 0.32 to 0.33, branch 1 around 0.25, and the branch order barely changes from the coldest to the warmest users (Fig.~\ref{subfig:afincawu}). Fine-tuning raises branch 3 to the highest rank, and this ranking is again stable across activity levels, with branch 3 near 0.38 for both cold and warm users (Fig.~\ref{subfig:afincaftwu}). AFIN's gate responds to user activity, whereas AFINca's pooling weights are largely unaffected by it. Since user activity is a key determinant of cold-start performance, this is a substantive difference between the two aggregation mechanisms.

\subsubsection{Cold-start shift}
\label{subsubsec:rescolsdstart}
\begin{figure}[tbp]
	\centering
	\begin{subfigure}[t]{0.49\textwidth}
		\includegraphics[width=\textwidth]{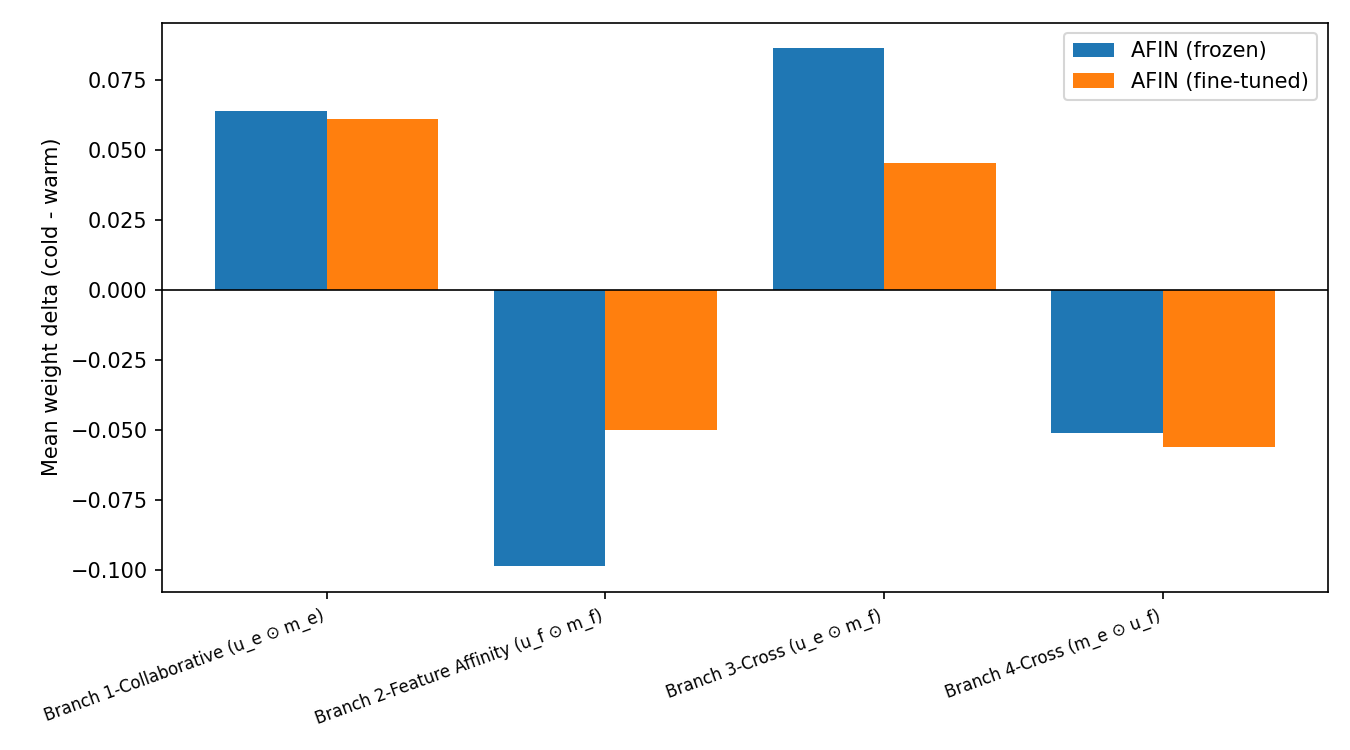}
		\caption{AFIN frozen vs. AFIN fine-tuned }
		\label{subfig:afinfrftcs}
	\end{subfigure}
	\hfill
	\begin{subfigure}[t]{0.49\textwidth}
		\includegraphics[width=\textwidth]{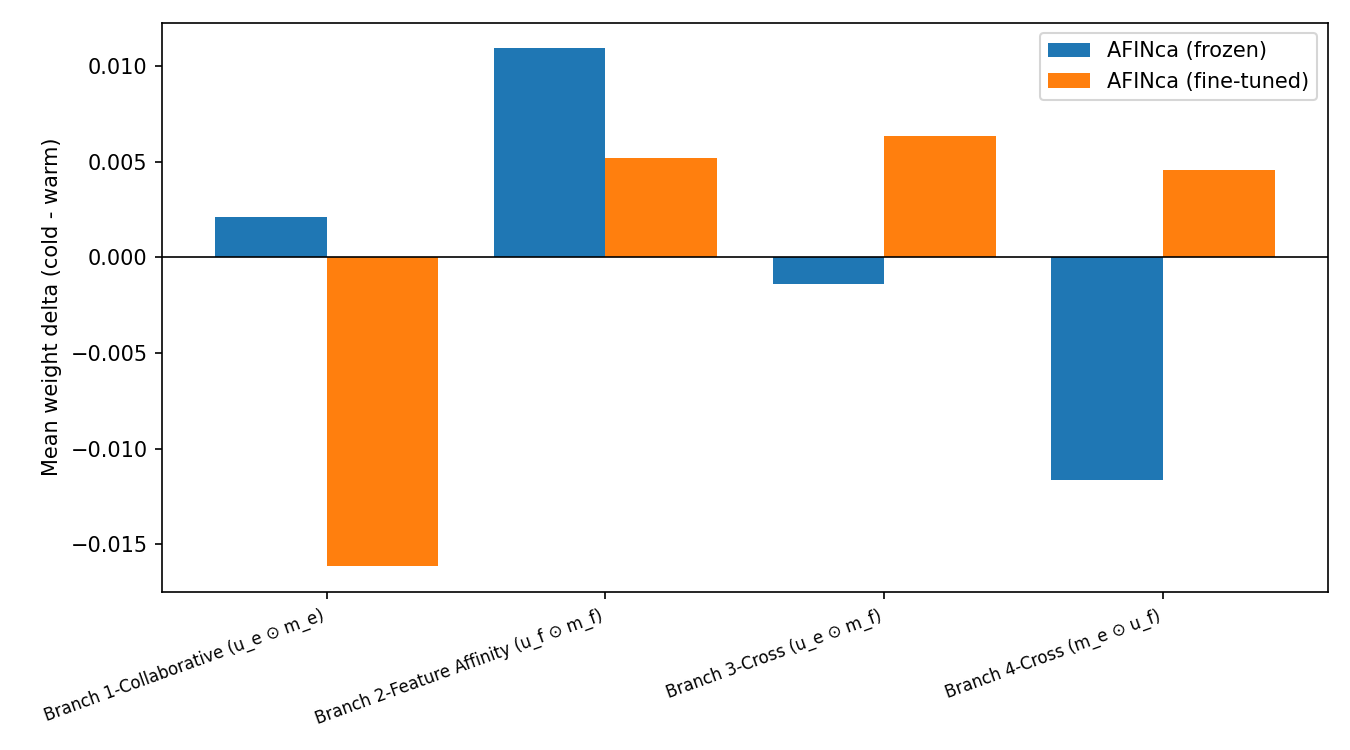}
		\caption{AFINca frozen vs. AFINca fine-tuned}
		\label{subfig:afincafrftcs}
	\end{subfigure}
	\caption{Cold-start shift $\Delta_k^{\mathrm{cold}}$ per branch (coldest minus warmest activity quartile mean weight), comparing frozen and fine-tuned encoders for AFIN (a) and AFINca (b).}
	\label{fig:coldstart}
\end{figure}

Comparing the coldest and warmest user-activity quartiles, branches 1 and 3 of frozen AFIN lean toward cold users (positive deltas of about $+0.064$ and $+0.086$), while branches 2 and 4 lean toward warm users (negative deltas of about $-0.098$ and $-0.050$; Fig.~\ref{subfig:afinfrftcs}). Fine-tuning leaves the direction of every delta unchanged. It roughly halves the deltas for branches 2 and 3, and the deltas for branches 1 and 4 stay nearly the same (still Fig.~\ref{subfig:afinfrftcs}).

The flat activity plots of AFINca appear as near-zero deltas. In the frozen state, the largest cold-warm gaps are for branch 4 ($-0.012$) and branch 2 ($+0.011$), against $-0.098$ for AFIN's branch 2 and $+0.086$ for its branch 3. Fine-tuning produces AFINca's largest shift in this diagnostic: branch 1 moves to approximately $-0.016$, while the other branches stay within $0.007$ of zero (Fig.~\ref{subfig:afincafrftcs}). Fine-tuning thus raises AFINca's sensitivity to cold-start users slightly, but it remains small compared with AFIN.

\subsubsection{Branch-zeroing ablation}
\label{subsubsec:resbranchzeroing}

\begin{table}[tbp]
	\centering
	\caption{Branch-zeroing ablation on the test set. For each model variant, the weight of one branch is set to zero and the remaining weights are renormalized. $\Delta$RMSE and $\Delta$MAE are the increases over the unmodified model, and $\Delta$RMSE (\%) is $\Delta$RMSE as a percentage of the unmodified RMSE.}
	\label{tab:branchzeroingabres}
	\begin{tabular*}{\textwidth}{@{\extracolsep{\fill}}LLLLL@{}}
		\toprule
		{\bf Variant} & {\bf Branch} & {\bf $\Delta$RMSE} & {\bf $\Delta$RMSE (\%)} & {\bf $\Delta$MAE} \\
		\midrule
		AFIN frozen & 1 & 0.1428 & 16.6644 & 0.1240 \\
		AFIN frozen & 2 & 0.0020 & 0.2301 & 0.0004 \\
		AFIN frozen & 3 & 0.0180 & 2.0946 & 0.0122 \\
		AFIN frozen & 4 & 0.0171 & 1.9913 & 0.0175 \\
		AFIN fine-tuned & 1 & 0.1492 & 17.5552 & 0.1252 \\
		AFIN fine-tuned & 2 & 0.0106 & 1.2495 & 0.0036 \\
		AFIN fine-tuned & 3 & 0.0117 & 1.3822 & 0.0121 \\
		AFIN fine-tuned & 4 & 0.0135 & 1.5851 & 0.0165 \\
		AFINca frozen & 1 & 0.0054 & 0.6236 & 0.0067 \\
		AFINca frozen & 2 & 0.0021 & 0.2421 & 0.0010 \\
		AFINca frozen & 3 & 0.0022 & 0.2575 & 0.0004 \\
		AFINca frozen & 4 & 0.0077 & 0.8907 & 0.0054 \\
		AFINca fine-tuned & 1 & 0.0018 & 0.2134 & 0.0012 \\
		AFINca fine-tuned & 2 & 0.0014 & 0.1590 & 0.0007 \\
		AFINca fine-tuned & 3 & 0.0052 & 0.6081 & 0.0039 \\
		AFINca fine-tuned & 4 & 0.0032 & 0.4101 & 0.0030 \\
		\bottomrule
	\end{tabular*}
\end{table}

Here, removing a branch means zeroing its weight and renormalizing the rest, as described in Section~\ref{subsec:intpret}. Table~\ref{tab:branchzeroingabres} gives the changes in RMSE and MAE for each branch across all model variants. In AFIN, branch weight and branch importance clearly disagree. The dominance counts make the gap concrete: branch 3 wins about 80k test samples when frozen, and branch 2 wins about 85k when fine-tuned, yet zeroing either raises RMSE by at most 2.09\%. In the frozen model, removing branch 1 raises RMSE by 0.1428 (16.66\%), while the other branches cause much smaller increases. This persists after fine-tuning: removing branch 1 raises RMSE by 0.1492 (17.56\%), while branches 2 to 4 each cause less than a 1.6\% increase. Branch 1 is therefore the most important branch under intervention, despite a middling average branch weight (0.26 frozen and 0.22 fine-tuned, ranking second and third of four).

The MAE results agree. In frozen AFIN, removing branch 1 raises MAE by 0.1240, whereas removing branches 2, 3, and 4 raises it by 0.0004, 0.0122, and 0.0175. After fine-tuning, branch 1 raises MAE by 0.1252, against 0.0036, 0.0121, and 0.0165 for branches 2 to 4. The discrepancy is therefore not limited to RMSE and does not come from a few large errors. Under both metrics, branch 1 matters far more for prediction than its branch weight implies.

Branch 2 shows the disconnect most clearly. After fine-tuning it has the highest branch weight in AFIN, yet its removal raises RMSE by only 1.25\% and MAE by only 0.0036. Branch 1 remains highly important despite its lower weight.

AFINca shows a different and more even pattern. In both the frozen and the fine-tuned model, removing any branch raises RMSE by less than 1\%. The largest increase is for branch 4 in the frozen model (0.89\%) and the smallest is for branch 2 in the fine-tuned model (0.16\%). MAE changes are also small, from 0.0004 to 0.0067. Unlike AFIN, AFINca has no single branch whose removal greatly hurts performance.

The branch ranking by RMSE effect in AFINca matches the weight ranking in both variants (branches 4, 1, 3, 2 when frozen, and 3, 4, 1, 2 when fine-tuned). Some adjacent ranks differ by less than 0.005 in weight or 0.0005 in RMSE, and the MAE ranking for the frozen model does not match. Branch weights in AFINca thus track branch importance more closely than in AFIN, but the small effects of removing any branch indicate that predictive responsibility is spread across branches, with none indispensable.

Two caveats apply to these comparisons. First, a branch's contribution to the prediction is $\alpha_k\mathbf{b}_k$, so a branch with a small weight but a large norm can still dominate the pooled vector. We did not measure branch norms, so weight and norm effects are confounded in the ablation. Second, for AFINca, zeroing acts on the pooling weights after self-attention, so a small change in RMSE is ambiguous between a branch that is not load-bearing and one whose signal is already present in the other branches' representations. The agreement between AFINca's weights and its ablation effects also rests on effects below 1\%, so it should be read as qualitative.

Overall, the RMSE and MAE results show a pronounced gap between branch weight and functional importance in AFIN. A branch with a low weight may be critical for prediction, while a heavily weighted branch may have limited impact. This is consistent with the claim that attention-style weights alone do not fully explain model behavior \cite{jain2019attention,wiegreffe2019attention}. AFINca's weight rankings correspond more closely to intervention-based importance, but the small effects of removing any branch mean that the weights alone cannot quantify contribution.

The difference between AFIN and AFINca gives a more nuanced view than a simple test of whether attention is explanatory. The relationship between branch weight and functional importance depends on the design of the aggregation mechanism. In AFIN, weight magnitude does not reliably indicate branch necessity. AFINca shows a closer correspondence and appears to distribute predictive responsibility across branches, although the caveats above limit both readings.

\section{Discussion}
\label{sec:disc}

\paragraph{RQ1: Do AFIN's branch weights identify the branches it depends on?}
They do not. With frozen embeddings, branch 3 is dominant for about 80k of the 100k test samples and has the highest mean weight (0.34), but zeroing it raises RMSE by only 2.09\% (Table~\ref{tab:branchzeroingabres}). Branch 1 is dominant for fewer than 10k samples, yet zeroing it raises RMSE by 0.1428 (16.66\%). After fine-tuning the mismatch is sharper. Branch 2 becomes dominant for about 85k samples and has the highest mean weight (0.31), but zeroing it raises RMSE by only 1.25\%, the smallest effect of the four, while branch 1 again causes a 17.56\% increase. The same pattern holds for MAE (0.1252 for branch 1 against 0.0036 for branch 2), so it is not an artifact of a few large errors. The weights are not uninformative: in the frozen model, branches 2 and 4 rank lowest on both measures. They simply do not point to the branch the predictor needs.

A plausible reason is that the gate scales branch vectors without accounting for their magnitude or content, so a small weight on a high-variance branch can carry more of the final signal than a large weight on a low-variance one. Branch 1 is also the only branch built from two collaborative embeddings, so it may be where pair-specific collaborative evidence enters the MLP most directly. We did not test either explanation. The result is consistent with the attention-as-explanation debate \cite{jain2019attention, wiegreffe2019attention}: a gate readout describes allocation, not necessity.

\paragraph{RQ2: Does the mismatch carry over to AFINca?}
Not in the same form. In AFINca, removing any single branch changes RMSE by less than 1\%, and the ranking of branches by RMSE effect matches the weight ranking in both variants. Weight rank and zeroing rank therefore agree more closely than in AFIN. This is weak evidence of faithfulness. Some adjacent ranks are nearly tied, the MAE ranking for the frozen model does not match, one branch wins for nearly every test sample, and every zeroing effect is below 1\%. Cross-attention also runs before the pooling gate, so a zeroed branch may already have passed its information to the other tokens. A small RMSE change then means either that the branch is not needed or that its signal is encoded elsewhere, and our procedure cannot separate the two. What we can say is that predictive dependence in AFINca is spread across tokens, whereas in AFIN it is concentrated in one branch the gate does not highlight. The activity analysis agrees: AFIN's weights shift with user activity, while AFINca's stay nearly flat, which suggests its pooling gate carries little per-sample information once the tokens are mixed.

\paragraph{RQ3: Does self-attention across branches improve rating prediction over independent gating?}
No. With frozen encoders AFIN is better than AFINca, by 0.0010 mean RMSE on ML-1M and 0.0054 on ML-100K (Table~\ref{tab:seedres}). After fine-tuning the two are tied on ML-1M (0.8513 for both) and AFINca is ahead by 0.0004 on ML-100K, well within the seed standard deviations of 0.0019 and 0.0021. AFINca therefore never shows a gain beyond seed variation, so we do not claim that independent gating is generally superior. No significance test was run. With four branch tokens built from the same inputs, there is little for attention to route between, and the extra parameters give no measurable gain. This does not rule out a benefit for models with more or more heterogeneous branches.

\paragraph{RQ4: Does fine-tuning the pretrained encoder help?}
Modestly, and consistently. Fine-tuning lowers AFIN's mean RMSE by 0.0066 on ML-1M and 0.0067 on ML-100K, and AFINca's by 0.0076 and 0.0125, which exceeds the seed standard deviations in every case (Table~\ref{tab:seedres}). MAE moves in the same direction in every case. On ML-1M, fine-tuning moves the best AFIN run from 0.8559 (frozen) to 0.8497, which crosses below $\Gamma$UICF (0.8506) in Table~\ref{tab:basecompresults}. Fine-tuning also changes where the gate puts its weight. AFIN's dominant branch moves from branch 3 to branch 2 and AFINca's from branch 4 to branch 3, while branch 1 stays the most important branch under zeroing in AFIN. Higher accuracy therefore does not come with a stable internal account of the prediction, which reinforces the caution in RQ1.

\section{Conclusion and Limitations}
\label{sec:concl}
This study proposes an Attentive Factorization Interaction Network (AFIN) and an Attentive Factorization Interaction Network with Cross-Attention (AFINca) to improve performance of rating prediction over sparse data sets (MovieLenz100K and MovieLenz1M), while examining how four complementary interaction branches (collaborative, feature affinity, and two cross branches) are combined in a collaborative filtering rating prediction pipeline. In the first stage, a Collaborative Attention Autoencoder (CAAE) learns dense user and movie embeddings by reconstructing the sparse rating matrix, with an attention step over side features that lets each embedding borrow strength from demographically and collaboratively similar users or movies. In the second stage, a predictor builds the four branches from these embeddings and the side features. Two aggregation mechanisms have been compared: an independent softmax gate (AFIN) and self-attention across branches before pooling (AFINca), each with frozen and fine-tuned CAAE encoders, on MovieLens 1M and 100K.

On the predictive side, the pipeline achieves strong accuracy. Fine-tuned AFIN reaches a mean RMSE of 0.8513 and a mean MAE of 0.6685 over 11 seeds on ML-1M, and a mean RMSE of 0.9109 and a mean MAE of 0.7151 on ML-100K. Fine-tuned AFINca is tied with it in RMSE on both datasets. Against ten published collaborative filtering baselines, the best run of fine-tuned AFIN attains the lowest RMSE on both datasets and the lowest MAE on ML-100K, and it ranks second in MAE on ML-1M. Fine-tuning the pretrained encoder with the rating loss improves the mean RMSE of both architectures in every comparison, which indicates that embeddings shaped jointly by reconstruction and rating prediction are better suited to the downstream task than embeddings trained for reconstruction alone. In contrast, cross-branch attention does not improve accuracy over the simpler gate beyond seed variation, so the additional modeling capacity of AFINca does not translate into lower error in this setting.

On the explainability side, the four-branch design makes each prediction decomposable into named interaction types, and the per-sample branch weights provide a direct readout of how the model allocates its reliance across them. The analyses of branch dominance, mean weights, and user activity describe how this allocation changes with fine-tuning and with the amount of available user history. The branch-zeroing ablation then tests whether this allocation reflects dependence. In AFIN, it does not: the branches that receive the largest weights have small effects when removed, whereas zeroing the collaborative branch increases RMSE by about 17\%, and this holds for both frozen and fine-tuned encoders and for both RMSE and MAE. In AFINca, the weight ranking agrees more closely with the zeroing ranking, but every zeroing effect is below 1\%, and the attention step may already have transferred the information of a removed branch to the remaining tokens. The practical consequence is that the proposed analysis provides a way to verify, rather than assume, what a branch-level gate reveals. Branch weights therefore require an intervention-based check before they are explained as importance, and the combination of competitive accuracy with this check yields a model whose internal account is examined and not merely displayed.

The findings are subject to several limitations. Both datasets require a minimum of 20 ratings per user, so the evaluation does not cover unseen users. The baseline figures are taken from the cited publications and may reflect different splits and preprocessing, which makes the comparison indicative rather than decisive. AFIN and AFINca employ different interaction dimensions (1536 and 256) and are not matched in parameter count. In the frozen setting, validation ratings form part of the encoder input during model selection, and in both settings a training pair's own rating is part of the encoder input during training. Branch zeroing is a sensitivity analysis rather than a causal attribution, and it produces gate configurations that the models have not encountered during training. Branch norms are not controlled for, so the effects of weight and magnitude remain confounded. Future work includes evaluating users and movies without rating history, matching model capacity across architectures, and measuring branch contributions in a way that accounts for branch norms.

\section*{Data availability}
The MovieLens 100K and 1M datasets are publicly available from GroupLens at \url{https://grouplens.org/datasets/movielens/}. The source code to reproduce all experiments is available at \url{https://github.com/ComputationIASBS/RecSys}.

\bibliographystyle{unsrtnat}
\bibliography{refs}

\end{document}